\documentclass[iop,apj,tighten]{emulateapj}

\usepackage{apjfonts}
\usepackage{graphicx}
\usepackage{dcolumn}
\usepackage{bm}
\usepackage{epsfig}

\usepackage{float}
\usepackage{amsmath,amstext}
\usepackage{floatflt}
\usepackage{subfigure}
\usepackage[usenames]{color}
\usepackage{ulem}
\usepackage{ragged2e}

\usepackage[breaklinks,colorlinks,citecolor=blue,linkcolor=magenta]{hyperref}
\usepackage[all]{hypcap} 

\shorttitle{Tilted flat and untilted nonflat $\phi$CDM models}
\shortauthors{Park \& Ratra}

\begin{document}


\title{Observational constraints on the tilted spatially-flat and the untilted nonflat $\phi$CDM dynamical dark energy inflation models}

\author{
Chan-Gyung Park\altaffilmark{1, 2} and
Bharat Ratra\altaffilmark{2}
}

\altaffiltext{1}{Division of Science Education and Institute of Fusion
                 Science, Chonbuk National University, Jeonju 54896, South Korea;
                 e-mail: park.chan.gyung@gmail.com}
\altaffiltext{2}{Department of Physics, Kansas State University, 116 Cardwell Hall,
                 Manhattan, KS 66506, USA
                 }

\date{\today}



%
%
\begin{abstract}
We constrain spatially-flat tilted and nonflat untilted scalar field ($\phi$)
dynamical dark energy inflation ($\phi$CDM) models by using Planck 2015 
cosmic microwave background (CMB) anisotropy measurements and recent baryonic 
acoustic oscillation distance observations, Type Ia supernovae apparent 
magnitude data, Hubble parameter measurements, and growth rate data. We assume 
an inverse power-law scalar field potential energy density 
$V(\phi)=V_0 \phi^{-\alpha}$. We find that the combination of the CMB data 
with the four non-CMB data sets significantly improves parameter constraints 
and strengthens the evidence for nonflatness in the nonflat untilted 
$\phi$CDM case from $1.8\sigma$ for the CMB measurements only to more than 
$3.1\sigma$ for the combined data. In the nonflat untilted $\phi$CDM model
current observations favor a spatially closed universe with spatial 
curvature contributing about two-thirds of a percent of the present 
cosmological energy budget. The flat tilted $\phi$CDM model is a 
0.4$\sigma$ better fit to the data than is the standard flat tilted 
$\Lambda$CDM model: current data allow for the possibility that dark 
energy is dynamical.
The nonflat tilted $\phi$CDM model is in better accord with the 
Dark Energy Survey bounds on the rms amplitude 
of mass fluctuations now ($\sigma_8$) as a function of 
the nonrelativistic matter density parameter now ($\Omega_m$) but it does not 
provide as good a fit to the larger-multipole Planck 2015 CMB anisotropy 
data as does the standard flat tilted $\Lambda$CDM model. A few cosmological 
parameter value measurements differ significantly when determined using the 
tilted flat and the untilted nonflat $\phi$CDM models, including the cold dark
matter density parameter and the reionization optical depth.
\end{abstract}

\maketitle

%
%

\section{Introduction}

In the standard flat $\Lambda$CDM cosmogony \citep{Peebles1984} 
the cosmological energy budget is currently dominated by the cosmological 
constant $\Lambda$, which is responsible for powering the currently 
accelerated cosmological expansion.\footnote{For reviews of the standard 
model see \citet{RatraVogeley2008}, \citet{Martin2012}, \citet{Brax2018}, and
\citet{Lukovicetal2018}. In this model, cold dark matter (CDM) and 
baryonic matter, both nonrelativistic, are the second and third largest 
contributors to the current cosmological energy budget; earlier they dominated 
over $\Lambda$ and were responsible for decelerating the cosmological 
expansion.}
This standard $\Lambda$CDM model is consistent with most 
observational constraints, including CMB anisotropy measurements 
\citep{PlanckCollaboration2016}, baryonic acoustic oscillations (BAO) 
distance observations \citep{Alametal2017, Ryanetal2018}, Type Ia 
supernova (SNIa) apparent magnitude data \citep{Scolnicetal2017},
and Hubble parameter measurements \citep{Farooqetal2017, Yuetal2018},

The standard flat $\Lambda$CDM inflation cosmogony is characterized 
by six cosmological parameters usually picked to be: $\Omega_{c} h^2$ and
$\Omega_{b} h^2$, the current values of the cold dark matter and baryonic
matter density parameters multiplied by the square of the 
Hubble constant $H_0$ (in units of 100 km s$^{-1}$  Mpc$^{-1}$); $A_{s}$ and 
$n_{s}$, the amplitude and spectral index of the primordial 
fractional energy density inhomogeneity power-law power spectrum;
$\theta_{\rm MC}$, the angular size of the sound 
horizon at recombination; and $\tau$, the reionization optical depth.

While the standard $\Lambda$CDM model assumes flat spatial geometry, 
current observational data allow for slightly curved spatial hypersurfaces.
Current measurements also allow a dark energy density that decreases
slowly in time (and so also varies weakly spatially) and do not require a 
space- and time-independent $\Lambda$. Theoretically, it seems easier to 
accommodate dynamical dark energy than a $\Lambda$.

XCDM is a simple and widely used dynamical dark energy parameterization.
Here the equation of state relating the dark energy fluid pressure and energy 
density is $p_X = w \rho_X$ where $w$ is the equation of state parameter and 
the additional seventh cosmological parameter. XCDM does not provide a 
consistent description of 
the evolution of energy density spatial inhomogeneities and so
is not a physically consistent description of dark energy. The simplest 
physically consistent dynamical dark energy model is $\phi$CDM 
\citep{PeeblesRatra1988, RatraPeebles1988}. In this model the dynamical
dark energy is a scalar field $\phi$ with potential energy density 
$V(\phi) \propto \phi^{-\alpha}$ and $\alpha > 0$ is the additional seventh 
cosmological parameter.\footnote{Many cosmological data sets have been used to 
place constraints on the $\phi$CDM model \citep[see, e.g.,][and references therein]{Samushiaetal2007, Yasharetal2009, SamushiaRatra2010, ChenRatra2011b, Campanellietal2012, Avsajanishvilietal2015, Solaetal2017b, Solaetal2017c, Zhaietal2017, Sangwanetal2018}.}

There have been a number of suggestions that some measurements favor 
dynamical dark energy over a $\Lambda$ \citep{Sahnietal2014, Dingetal2015, Solaetal2015, Zhengetal2016, Solaetal2017a, Solaetal2018, Solaetal2017b, Zhaoetal2017, Solaetal2017c, Zhangetal2017a, Solaetal2017d, GomezValentSola2017, Caoetal2018, GomezValentSola2018}. These analyses made a number of simplifying 
assumptions, either ignoring CMB anisotropy data, or only approximately 
accounting for it, or using it in the context of a generalized XCDM 
parameterization of dynamical dark energy.  Some of these analyses also 
include a high $H_0$ value determined from the local expansion 
rate in the data collections they use to investigate dark energy 
dynamics.\footnote{We exclude this high local $H_0$ value from the data we use 
here to constrain cosmological model parameters, as it is inconsistent with 
the other data sets we utilize for this purpose, in the models we study.}

\citet{Oobaetal2018c} have more exactly analyzed the Planck CMB data (as 
well as a few BAO distance measurements) by using the seven parameter 
spatially-flat XCDM and $\phi$CDM dynamical dark energy tilted inflation
models and discovered that both were weakly favored by the data, compared 
to the standard six parameter flat $\Lambda$CDM model, by 1.1$\sigma$ and 
1.3$\sigma$ for the XCDM and $\phi$CDM cases.\footnote{\citet{ParkRatra2018b} 
used a much larger compilation of non-CMB data in an analysis of the tilted 
flat XCDM parameterization, confirming the \cite{Oobaetal2018c} findings, 
but at a lower level of significance, 0.3$\sigma$ instead of 1.1$\sigma$.}
These are not significant improvements over the standard flat $\Lambda$CDM 
case, but current data allow for the possibility that dark energy is 
dynamical. Furthermore, 
both dynamical dark energy models decrease the tension between 
the Planck CMB and the weak lensing observational 
bounds on $\sigma_8$, the current value of rms fractional energy density 
inhomogeneity averaged over 8$h^{-1}$ Mpc radius spheres.

Nonflat models have a characteristic length set by the non-vanishing spatial 
curvature and an 
energy density inhomogeneity power spectrum in a nonflat model that does not 
fully account for this spatial curvature length scale \citep[as was done in the 
analyses of nonflat models by][]{PlanckCollaboration2016} is not 
physically consistent. Nonflat
cosmological inflation models are the only known way of defining 
physically consistent fractional energy density inhomogeneity power spectra 
in nonflat models. For open geometries the open-bubble inflation model 
\citep{Gott1982} is used to derive the non-power-law power spectrum 
\citep{RatraPeebles1994, RatraPeebles1995}. For closed geometries 
the Hawking prescription for the initial state of the universe 
\citep{Hawking1984, Ratra1985} defines a closed 
inflation model that is used to compute the non-power-law power spectrum
\citep{Ratra2017}. Unlike in the flat inflation case, there is no simple
way to also accommodate tilt in nonflat inflation models. In the nonflat 
case $n_s$ is no longer a free parameter but is instead replaced by the 
current spatial curvature density parameter $\Omega_k$.

\citet{Oobaetal2018a} used this physically consistent nonflat untilted 
model non-power-law power spectrum of energy density spatial inhomogeneities 
in analyses of the Planck 2015 CMB anisotropy measurements 
\citep{PlanckCollaboration2016} and found that these data do not require flat 
spatial geometry in the six parameter nonflat untilted $\Lambda$CDM inflation 
model.\footnote{Non-CMB observations do not provide tight constraints on 
spatial curvature \citep{Farooqetal2015, Chenetal2016, YuWang2016, LHuillierShafieloo2017, Farooqetal2017, Lietal2016, WeiWu2017, Ranaetal2017, Yuetal2018, Mitraetal2018, Ryanetal2018}, with the recent exception of a collection of all of 
the most recent BAO, Hubble parameter, and SNIa data, which (weakly) favors 
closed spatial geometry \citep{ParkRatra2018c}, as well as a recent collection 
of deuterium abundances that favor flat spatial hypersurfaces \citep{Pentonetal2018}.}
\citet{ParkRatra2018a, ParkRatra2018b} confirmed the results of 
\citet{Oobaetal2018a} by using the largest compilation of reliable 
observational data to study the nonflat untilted $\Lambda$CDM 
inflation model, and found stronger evidence for non-flatness, 5.2$\sigma$, 
favoring a very slightly closed model. The Planck 2015 CMB anisotropy data 
also do not require flat spatial surfaces in the seven parameter nonflat 
untilted XCDM dynamical dark energy inflation parameterization 
\citep{Oobaetal2017}. 
In the XCDM parameterization $w$ is the seventh cosmological parameter with
$n_s$ again replaced 
by $\Omega_k$. Using a much larger compilation of non-CMB data,
\citet{ParkRatra2018b} confirmed the \citet{Oobaetal2017} results with 
higher significance: in the untilted nonflat XCDM case the data favor a 
closed model at 3.4$\sigma$ significance and favor dynamical dark 
energy over a cosmological constant at 1.2$\sigma$ significance.     
In the seven parameter nonflat untilted $\phi$CDM dynamical dark 
energy inflation model 
\citep{Pavlovetal2013} --- with $\alpha$ as the seventh cosmological
parameter --- \citet{Oobaetal2018b} again discovered that Planck 2015 
CMB anisotropy data do not demand flat spatial hypersurfaces.
In both the XCDM and $\phi$CDM dynamical dark energy inflation cases 
the data again favor a very mildly closed model. All three 
closed models are more compatible with weak lensing $\sigma_8$ constraints 
but do not fit the 
higher-$\ell$ $C_\ell$ data as well as the flat models do.

In this paper we determine observational limits on parameters of the seven 
parameter flat tilted $\phi$CDM and the seven parameter nonflat untilted
$\phi$CDM dynamical dark energy inflation models. For this purpose, we use 
the same observational data in as \citet{ParkRatra2018b}, the Planck CMB 
anisotropy,
the Pantheon collection of 1048 SNIa apparent magnitudes 
\citep{Scolnicetal2017}, and a collection of BAO distances, Hubble 
parameters, and growth rates (see \citealt{ParkRatra2018a,ParkRatra2018b} 
for the data compilation and update).
  
We find that the seven parameter flat tilted $\phi$CDM inflation model
provides a better fit to these data than does the six parameter
standard flat tilted $\Lambda$CDM model. However, for the larger 
compilation of data here the $\phi$CDM dynamical dark energy inflation
model is only 0.40$\sigma$ better than the standard $\Lambda$CDM model
(compared to the 1.3$\sigma$ \citealt{Oobaetal2018c} found with their 
smaller data collection). While not a significant improvement over the 
standard model, the $\phi$CDM model cannot be ruled out. In agreement with 
\citet{Oobaetal2018c} we also do not detect a deviation from $\alpha = 0$ 
(a cosmological constant) for the flat $\phi$CDM model.\footnote{These 
conclusions do not agree with those from earlier approximate analyses, 
based on less, as well as less reliable, data 
\citep{Solaetal2017b, Solaetal2017c}, that favor the flat $\phi$CDM model 
over the flat $\Lambda$CDM one by more than 3$\sigma$ and find $\alpha$ deviating from $0$ by more than 2$\sigma$.} 

Our results for the nonflat untilted $\phi$CDM inflation model, 
derived using many more non-CMB observations, are consistent with and 
strengthen the \citet{Oobaetal2018b} conclusions. For the full data 
collection we use here we find a more than 3.1$\sigma$ deviation from 
spatial flatness. The nonflat untilted $\phi$CDM model better fits 
the weak lensing $\sigma_8$--$\Omega_m$ bound. For the full data 
collection we use here (including CMB lensing data), 
the best-fit nonflat untilted $\phi$CDM model has a reduced low-$\ell$ CMB 
temperature anisotropy multipole number ($\ell$) power spectrum $C_\ell$ 
and is more compatible with the observations. However, overall
the standard tilted flat $\Lambda$CDM model better fits the CMB 
data.\footnote{As discussed elsewhere and below, the number of degrees of 
freedom of the Planck 2015 data are ambiguous and the nonflat untilted 
$\phi$CDM model and the flat tilted $\Lambda$CDM model are not nested, 
thus it is impossible to translate the 
$\Delta \chi^2$'s we compute here to quantitative goodness of fit 
probabilities, consequently a large number of our statements about 
goodness of fit are qualitative. See below and see 
\citet{ParkRatra2018a,ParkRatra2018b} for more details about this issue.}
  
These data determine $H_0$ in an almost model-independent way with a 
value that is compatible with most other estimates. As found in 
\citet{ParkRatra2018a, ParkRatra2018b}, however, $\Omega_c h^2$ and
$\tau$ differ significantly between the tilted flat and the untilted 
nonflat models and so care must be taken when utilizing cosmological 
measurements of such parameters. 

In Sec.\ 2 we summarize the data sets we use in our analyses. 
In Sec.\ 3 we summarize the methods we use in our analyses here. 
Observational constraints following from these data for the flat 
tilted $\phi$CDM and the nonflat untilted $\phi$CDM inflation models 
are presented and discussed in Sec.\ 4. We summarize our main results 
in Sec.\ 5.

\section{Data}

Following \citet{ParkRatra2018a, ParkRatra2018b} we utilize the Planck 2015 
TT + lowP and TT + lowP + lensing CMB anisotropy measurements 
\citep{PlanckCollaboration2016} to set bounds on the parameters of the 
$\phi$CDM dynamical dark energy model. Here TT is
the low-$\ell$ ($2 \le \ell \le 29$) and high-$\ell$ ($30 \le \ell \le 2508$; 
PlikTT) Planck temperature-only $C_\ell^{TT}$ angular power spectrum observations
and lowP is the low-$\ell$ polarization $C_\ell^{TE}$, $C_\ell^{EE}$, and 
$C_\ell^{BB}$ angular power
spectra measurements at $2 \le \ell \le 29$. The collection of
low-$\ell$ CMB temperature and polarization power spectra is called lowTEB.
The  CMB lensing data we use is the measured Planck lensing potential 
power spectrum. The abbreviations TT + lowP and TT + lowP + lensing are 
used for the CMB data without and with CMB lensing data, respectively.
The Planck collaboration recommends using the TT + lowP + lensing data combination as a conservative choice for parameter estimation (see the footnote to Table 4 of \citealt{PlanckCollaboration2016}).

The Type Ia supernova data set we use is the Pantheon set of 1048
SNIa apparent magnitude observations over the wide redshift ($z$) range of
$0.01 < z < 2.3$ \citep{Scolnicetal2017}. This data set includes 276 SNIa
($0.03 < z < 0.65$ from the Pan-STARRS1 Medium Deep Survey and
SNIa distance measurements from the SDSS, SNLS and low-$z$ HST collections. 
We use the abbreviation SN to refer to the Pantheon sample.

We use the compilation of BAO data given in Table 1 of \citet{ParkRatra2018a}. 
As in \citet{ParkRatra2018b}, we use the updated BAO data point, 
$D_V (r_{d,\textrm{fid}} / r_d)=3843\pm147$ Mpc of \cite{Ataetal2018}, instead 
of the old value. See Sec.\ 2.3 of \citet{ParkRatra2018a} for more details.
We note that the BAO data from BOSS DR12 
\citep{Alametal2017} include growth rate ($f\sigma_8$) and radial BAO 
$H(z)$ data that are correlated with the other BOSS DR12 BAO measurements.
We use the abbreviation BAO to refer to this BAO data compilation.

We also use the Hubble parameter, $H(z)$ (with 31 data points in 
total),\footnote{Hubble parameters have been measured over a wide range of 
redshift, from the present epoch to well beyond the cosmological 
deceleration-acceleration transition redshift. They provide evidence 
that this transition occurred and they have been used to measure the redshift 
of this transition at roughly the value expected in standard $\Lambda$CDM 
and other dark energy models \citep{FarooqRatra2013, Farooqetal2013, Capozzielloetal2014, Morescoetal2016, Farooqetal2017, Yuetal2018, Jesusetal2018, Haridasuetal2018b}.}
and growth rate (with 10 points in total), $f(z) \sigma_8 (z)$, observations of 
Tables 2 and 3 of \citet{ParkRatra2018a}.

\section{Methods}

In the $\phi$CDM model we study here, the minimally coupled dark energy scalar 
field $\phi$ has an inverse power-law potential energy density
\begin{equation}
    V(\phi)=V_0 \phi^{-\alpha}
\end{equation}
with $\alpha> 0$ being a constant parameter and $V_0$ is determined in 
terms of 
$\alpha$ \citep{PeeblesRatra1988}. When $\alpha$ goes to zero, the 
dark energy behaves like the cosmological constant $\Lambda$.

We evolve a system of multiple components including radiation, neutrinos, 
matter, and the scalar field (that only directly couples to the gravitational
field). The evolution equations for the spatially homogeneous background
and the spatial inhomogeneity linear perturbation variables are summarized 
in \citet{HwangNoh2001, HwangNoh2002}. For the homogeneous background scalar 
field we use the initial conditions of \citet{PeeblesRatra1988} at
scale factor $a_i=10^{-10}$. This places the homogeneous background scalar field 
on the attractor/tracker solution \citep{PeeblesRatra1988, RatraPeebles1988, Pavlovetal2013}.\footnote{As a consequence of there being an 
attractor/tracker solution of the background scalar field nonlinear equation
of motion (coupled to the Friedmann equation), the long-term time evolution is 
independent of the chosen initial conditions. However, there can be differences
caused by different approaches from different initial conditions to the 
attractor/tracker solution and future data might require a more careful study 
of initial conditions effects.}
As initial conditions of the spatially inhomogeneous scalar field perturbation 
and its time derivative, we take them to vanish in the CDM-comoving gauge 
(this is synchronous gauge without gauge modes) at $a_i=10^{-10}$.\footnote{The 
evolution of spatially inhomogeneous scalar field quantities linearly perturbed 
about the background attractor solution also show tracking behavior and 
so are largely independent of the choice of initial conditions \citep{RatraPeebles1988, Braxetal2000}.}

For background evolution, we numerically solve the equation of motion
of the scalar field,
\begin{equation}
   \phi'' + \left( 3 + \frac{\dot{H}}{H^2} \right) \phi'
      - \hat{V}_0 \alpha \phi^{-\alpha -1}
      \left(\frac{H_0}{H}\right)^2
    = 0,
\end{equation}
where $\phi' \equiv d\phi / d\ln a$, $H=\dot{a} / a$,
$\hat{V}_0 \equiv V_0 / H_0^2$,
and an  overdot denotes the time derivative $d/dt$.
For the matter and dark energy dominated epochs, the normalized Hubble
parameter $H(a)$ can be written as
\begin{equation}
   \left(\frac{H}{H_0}\right)^2
     = \frac{1}{1-\frac{1}{6}\left(\phi' \right)^2} 
       \left[ \Omega_m a^{-3} + \Omega_k a^{-2}
       + \frac{1}{3}\hat{V}_0 \phi^{-\alpha}\right],
\label{eq:hubble}
\end{equation}
where $\Omega_m$ and $\Omega_k$ are present values of the matter and curvature
density parameters, respectively, and we have chosen units such that 
$8\pi G \equiv 1$. In actual calculations of Eq.\ (\ref{eq:hubble}), we 
have taken
into account the contribution of photons as well as massless and massive 
neutrinos. Given cosmological parameters and initial conditions for the
scalar field, we adjust the value of $\hat{V}_0$ to satisfy the condition
$H/H_0=1$ at the present epoch ($a_0=1$) by applying the bisection method.

\begin{figure*}
\centering
\mbox{\includegraphics[width=120mm,bb=30 240 600 720]{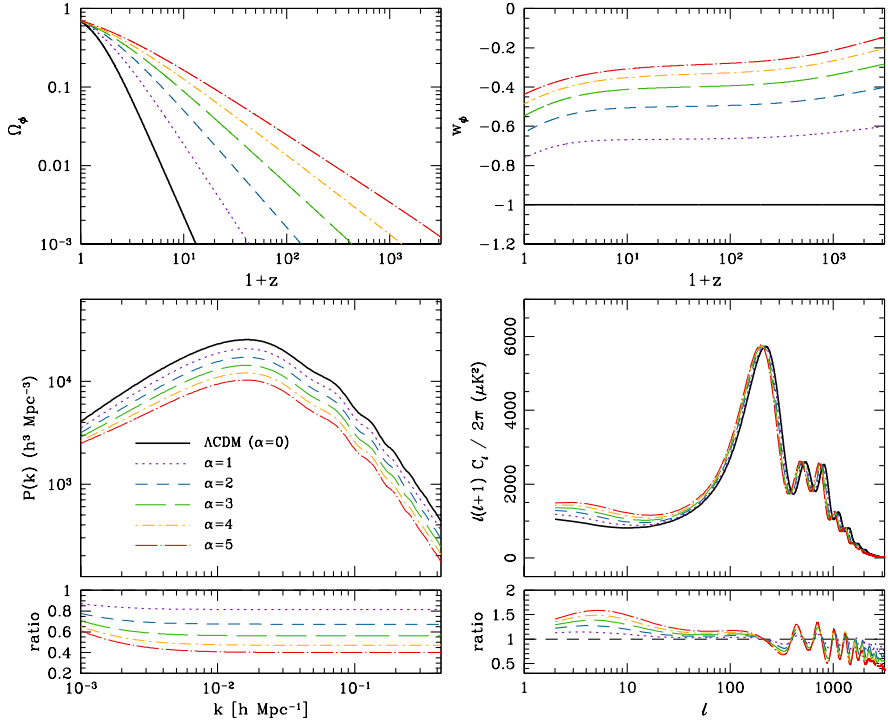}}
\caption{Top: Evolution of the dark energy scalar field density parameter ($\Omega_\phi$) and equation of state parameter ($w_\phi$) in the tilted flat $\phi$CDM model for integer values of $\alpha$ from 1 to 5. The black solid curve is for $\Lambda$CDM which corresponds to $\phi$CDM with $\alpha = 0$. For these illustrations all other cosmological parameters are fixed to the mean values of the $\Lambda$CDM model parameters constrained by using the Planck 2015 CMB (TT + lowP) and the four non-CMB data sets (see Table 5 bottom-right panel of \citealt{ParkRatra2018a}). Bottom: Theoretical predictions for matter density and CMB temperature anisotropy angular power spectra in the $\phi$CDM model at the corresponding $\alpha$ values. The ratios of the $\phi$CDM model power spectra relative to the $\Lambda$CDM one are shown in the lower panels.} 
\label{fig:bgpert_flat}
\end{figure*}

To estimate the likelihood distributions of $\phi$CDM model parameters, 
we use the CAMB/COSMOMC package (Nov.\ 2016 version) 
\citep{ChallinorLasenby1999, Lewisetal2000, LewisBridle2002}. CAMB is used 
to compute the theoretical CMB temperature anisotropy, polarization,
and lensing potential power spectra, as well as the matter density power 
spectrum, 
by solving for the evolution of the cosmological spatial inhomogeneity linear
perturbations. COSMOMC determines model parameter values that are favored by 
the observational data by using the Markov chain Monte Carlo (MCMC) method. 
Since the current version of CAMB/COSMOMC package cannot be applied to scalar 
field dynamical dark energy models, we generalized CAMB by including the 
dynamical equations of motion for the spatially homogeneous background and 
spatial inhomogeneity linear perturbation quantities for the scalar field 
inverse power-law potential energy density model. CAMB uses the RECFAST routine
to compute the recombination history of the universe \citep{Seageretal1999, Wongetal2008}.
We modified RECFAST to use the background evolution of the $\phi$CDM model.
We also altered the COSMOMC parameter interface to use the scalar field potential energy density 
parameter $\alpha$ as a new free parameter, in place of the constant equation 
of state parameter $w$ of the XCDM model.

Unlike in the $\Lambda$CDM and XCDM analyses of \citet{ParkRatra2018a, ParkRatra2018b},
here we use the Hubble constant $H_0$ as a new free parameter, instead of 
$\theta_\textrm{MC}$ (a default free parameter used in COSMOMC). There are 
two reasons for this change. First, $\theta_\textrm{MC}$, the approximate 
angular size of the sound horizon at the decoupling epoch, is based on the 
fitting 
formula of the sound horizon size given in \citet{HuSugiyama1996} and is 
appropriate for models with a negligible level of dark energy in the early 
universe. In general, however, scalar field dark energy can be non-negligible 
at early times, depending on the scalar field potential energy density 
parameters and the initial conditions (e.g., see \citealt{Parketal2014} for 
episodic domination of scalar field dark energy in the early universe).
In the $\phi$CDM model we study here a large value of $\alpha$ can result 
in a significant amount
of dark energy at early times. Thus, a more accurate model parameterization 
is needed. Second, using the angular size of the sound horizon ($\theta$) as a 
free parameter is less suitable in the presence of scalar field dark energy. 
Scalar field dark energy has its own dynamical equation that needs to be 
numerically evolved and so it is a matter of practical difficulty to adjust 
other cosmological parameter values along with the potential parameters
of the scalar field to reproduce $\theta$, a quantity that is obtained from an 
integration of the spatially homogeneous background equations of motion.
The drawback of choosing the Hubble constant as a free parameter is that 
this makes it difficult to achieve MCMC convergence as the Hubble constant 
has degeneracy with spatial curvature and with the dark energy parameter 
$\alpha$ resulting in likelihood distributions that are degenerate and 
non-Gaussian.

Figure \ref{fig:bgpert_flat} shows the evolution of the scalar field dark
energy density parameter ($\Omega_\phi$) and equation of state parameter
($w_\phi = p_\phi/\rho_\phi$, where $p_\phi$ and $\rho_\phi$ are the pressure 
and energy density of the scalar field) as well as theoretical predictions 
for matter density and CMB temperature 
anisotropy angular power spectra in the spatially-flat $\phi$CDM model 
for some $\alpha$ values. The other cosmological parameters are fixed
to the mean $\Lambda$CDM model parameters obtained by using the Planck 
2015 CMB (TT + lowP) and the four non-CMB data sets 
(see the bottom-right panel of Table 5 in \citealt{ParkRatra2018a}).
We can expect that the spatially-flat $\phi$CDM model with large $\alpha$ 
can be excluded by CMB data alone. However, we will see that the 
nonflat $\phi$CDM model with large values of $\alpha$ can be consistent 
with Planck CMB data.\footnote{However, from the bottom-left panels of Figs.\ \ref{fig:bgpert_flat} and \ref{fig:bgpert_nonflat}, we see that matter power spectrum measurements over a wide range of wavenumbers, such as those shown in Fig.\ 19 of \citet{PlanckCollaboration2018}, exclude large $\alpha$ values.}

The primordial fractional energy density spatial inhomogeneity power spectrum 
in the tilted flat $\phi$CDM 
inflation model \citep{LucchinMatarrese1985, Ratra1992, Ratra1989} is
\begin{equation}
   P(k)=A_s \left(\frac{k}{k_0} \right)^{n_s},
\end{equation}
where $k$ is wavenumber and $A_s$ is the amplitude of the power spectrum 
at the pivot scale wavenumber 
$k_0=0.05~\textrm{Mpc}^{-1}$. The corresponding power spectrum in the 
nonflat untilted $\phi$CDM inflation model 
\citep{RatraPeebles1995, Ratra2017} is
\begin{equation}
   P(q) \propto \frac{(q^2-4K)^2}{q(q^2-K)},
\end{equation}
which becomes the $n_s=1$ spectrum in the flat limit (when $K=0$).
For scalar-type perturbations, $q=\sqrt{k^2 + K}$ is the wavenumber 
where spatial curvature $K=-(H_0^2 / c^2 ) \Omega_k$ and $c$ is the speed 
of light. For the negative $\Omega_k$ closed model, normal modes 
are characterized by positive integers $\nu=q K^{-1/2}=3,4,5,\cdots$. 
For the nonflat model, we use $P(q)$ as the
initial perturbation power spectrum and 
normalize its amplitude at $k_0$ to $A_s$. 

Our analyses methods are those described in Sec.\ 3.2
of \citet{ParkRatra2018a} and Sec.\ 3 of \citet{ParkRatra2018b}.

\section{Observational Constraints}

We constrain the tilted flat $\phi$CDM model with seven cosmological parameters
($\Omega_b h^2$, $\Omega_c h^2$, $H_0$, $\tau$, $A_s$, $n_s$, and $\alpha$) and
the untilted nonflat $\phi$CDM model with seven parameters 
($\Omega_b h^2$, $\Omega_c h^2$, $H_0$, $\Omega_k$, $\tau$, $A_s$, and $\alpha$).
The calibration and foreground model parameters of the Planck data are also 
constrained as nuisance parameters by the COSMOMC program. In all parameter 
constraint tables presented in this work we also list three derived 
parameters, $\theta_{\textrm{MC}}$, $\Omega_m$ (present value of the 
nonrelativistic matter density parameter), and $\sigma_8$. 

We use the COSMOMC settings adopted by the Planck team 
\citep{PlanckCollaboration2016} and the same priors on the model parameters 
as well as the same values of the present CMB temperature 
($T_0=2.7255~\textrm{K}$), the effective number of neutrino species 
($N_\textrm{eff}=3.046$), and one massive neutrino species (with mass 
$m_\nu=0.06~\textrm{eV}$) as used in \citet{ParkRatra2018a, ParkRatra2018b}.
We set tophat priors on the scalar field potential energy density 
parameter $0 < \alpha < 10$ and on the Hubble constant $0.2 \le h \le 1.0$. 
However, as detailed below, constraining the nonflat $\phi$CDM models 
using the Planck CMB data alone is a complicated task due to the 
highly degenerate and non-Gaussian likelihood distributions of 
$H_0$, $\Omega_k$, and $\alpha$, that make it difficult for the MCMC chains 
to converge. In this case (for only the CMB TT + lowP and TT + lowP + lensing 
data alone analyses), we apply a more restrictive tophat prior on the 
Hubble constant, $0.45 \le h \le 1.0$, to achieve convergence of 
the MCMC chains in a reasonable amount of time (given our computational 
resources).

\begin{figure*}
\centering
\mbox{\includegraphics[width=87mm]{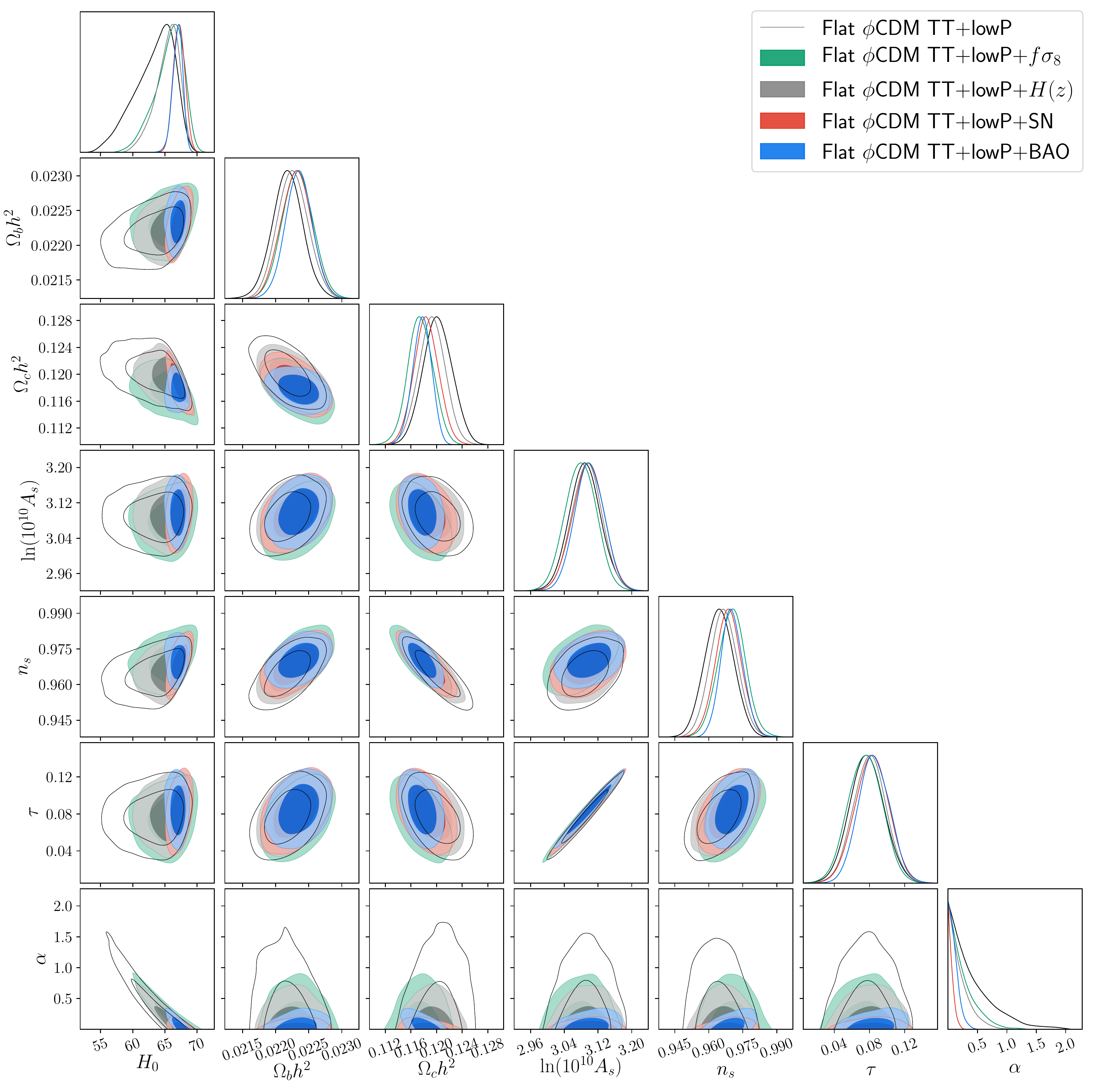}}
\mbox{\includegraphics[width=87mm]{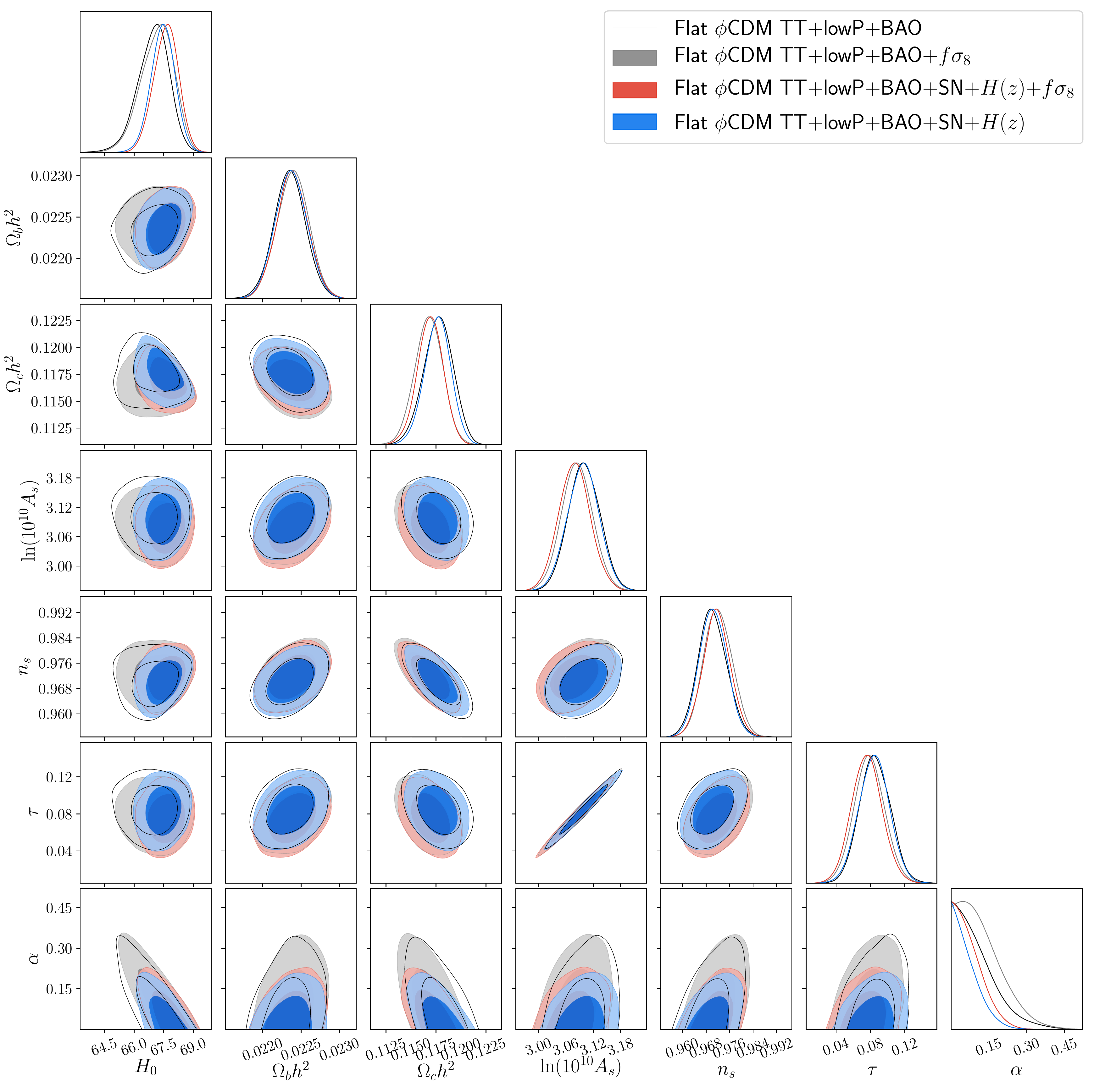}}
\caption{Likelihood distributions of the tilted flat $\phi\textrm{CDM}$
model parameters constrained by using Planck CMB TT + lowP, SN, BAO, $H(z)$,
and $f\sigma_8$ data. Two-dimensional marginalized likelihood constraint 
contours and one-dimensional likelihoods are plotted for when each set of 
non-CMB data is combined with the Planck TT + lowP measurements
(left panel) and when the growth rate, Hubble parameter, and SN data, as well
as their combination, are combined with the TT + lowP + BAO data (right panel).
For viewing clarity, the cases of TT + lowP (left) and TT + lowP + BAO
(right panel) are shown with solid black curves.
}
\label{fig:para_flat}
\end{figure*}

\begin{figure*}
\centering
\mbox{\includegraphics[width=87mm]{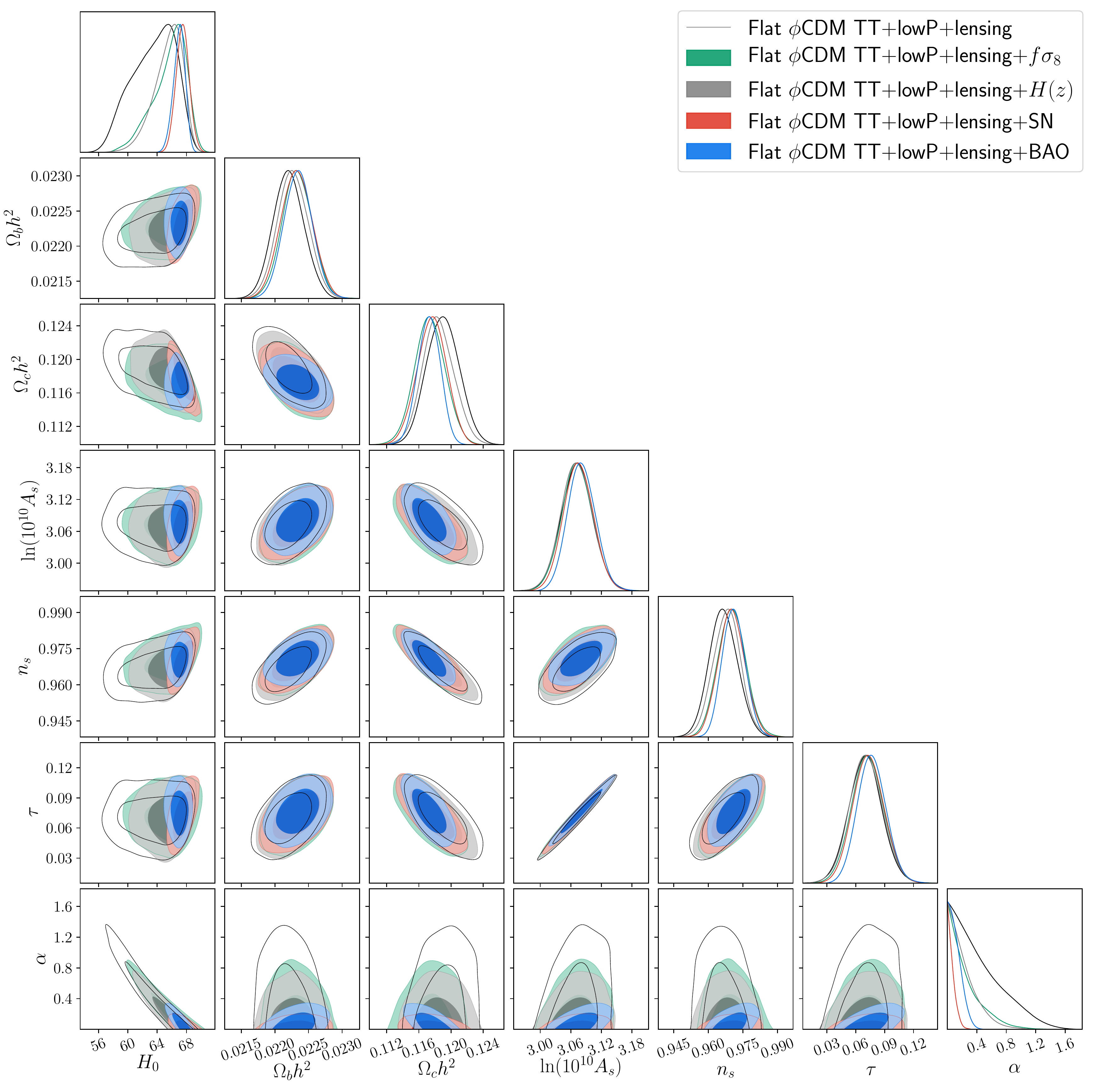}}
\mbox{\includegraphics[width=87mm]{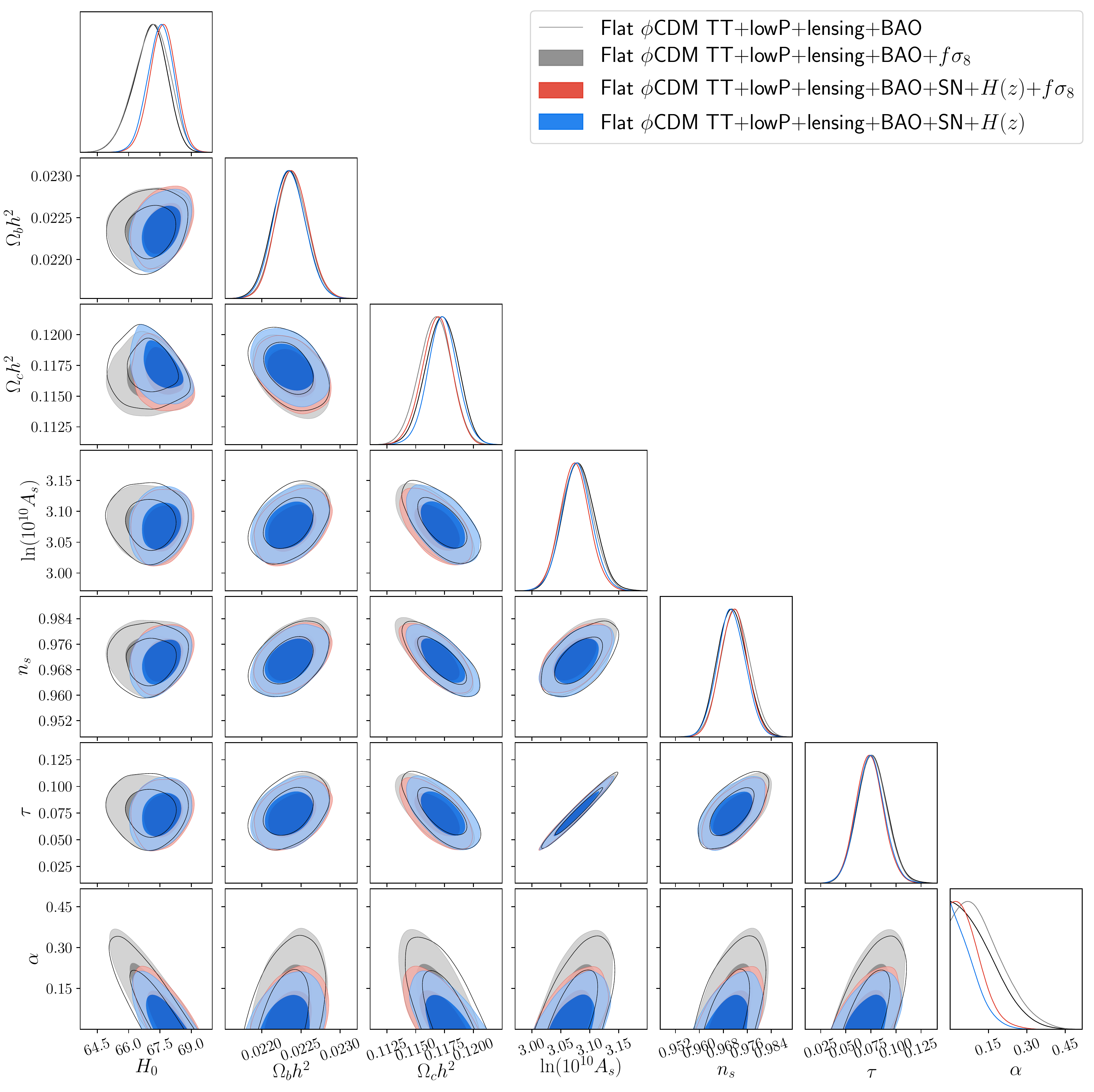}}
\caption{Same as Fig.\ \ref{fig:para_flat} but now also accounting for
the Planck CMB lensing data.
}
\label{fig:para_flat_lensing}
\end{figure*}

Our results for the flat tilted $\phi$CDM model 
are given in Figs.\ \ref{fig:para_flat} and \ref{fig:para_flat_lensing}
and Tables \ref{tab:para_flat} and \ref{tab:para_flat_lensing}.
The likelihood distributions for the TT + lowP (+ lensing) + SN + BAO data
combination (ignoring or accounting for the CMB lensing data) are omitted 
in the figures since they are very similar to those for the TT + lowP 
(+ lensing) + SN + BAO + $H(z)$ combination.
 
The results for the flat tilted $\phi$CDM model in the TT + lowP 
panel of Table \ref{tab:para_flat} and in the TT + lowP + lensing panel 
in Table \ref{tab:para_flat_lensing} agree well with the corresponding 
entries in Table 2 of \citet{Oobaetal2018c}, except for 
the 2$\sigma$ upper limit on $\alpha$ for the TT + lowP case where we find
$\alpha < 1.49$ while \cite{Oobaetal2018c} give $\alpha < 1.1$. 
\citet{Oobaetal2018c} use CLASS \citep{Blasetal2011} for computing the
$C_\ell$'s and Monte Python \citep{Audrenetal2013} for the MCMC analyses,
so it is comforting that both our results agree well.

Tables \ref{tab:para_flat} and \ref{tab:para_flat_lensing} show that, 
when added to the Planck anisotropy data, for the flat tilted 
$\phi$CDM cosmogony, the BAO distance observations are largely more 
constraining than the $f \sigma_8$, SN, or $H(z)$ data, except for 
$\alpha$ when the SN measurements are more restrictive than the BAO ones 
and for $\sigma_8$ in the TT + lowP + lensing case when again the SN 
measurements are more restrictive than the BAO ones. This is very similar 
to the results of the XCDM analyses \citep{ParkRatra2018b}. As in the XCDM 
case, each of the four non-CMB data sets used with the CMB data provide
approximately equally tight bounds on $\Omega_b h^2$, $\tau$, 
and $A_s$. We also note that the full combination of CMB and non-CMB 
data sets gives a somewhat worse constraint on the potential energy density 
parameter 
$\alpha$ than does the CMB + SN case, because the BAO, $H(z)$, and $f\sigma_8$ 
data favor a wider range of $\alpha$ and so weaken the $\alpha$ constraint.
For a similar reason, the combination of the CMB and SN + BAO +
$H(z)$ data constrains $\alpha$ tighter than does the full data combination.

\begin{table*}
\caption{Tilted flat $\phi\textrm{CDM}$ model parameters constrained by using 
Planck TT + lowP, SN, BAO, $H(z)$, and $f\sigma_8$ data (mean and 68.3\% confidence limits).}
\begin{ruledtabular}
\begin{tabular}{lccc}
  Parameter                & TT+lowP                & TT+lowP+SN               &  TT+lowP+BAO   \\[+0mm]
 \hline \\[-2mm]
  $\Omega_b h^2$           & $0.02218 \pm 0.00023$  & $0.02231 \pm 0.00023$    &  $0.02235  \pm 0.00021$    \\[+1mm]
  $\Omega_c h^2$           & $0.1202  \pm 0.0023$   & $0.1184  \pm 0.0019$     &  $0.1177   \pm 0.0015$       \\[+1mm]
  $H_0$ [km s$^{-1}$ Mpc$^{-1}$] & $63.3 \pm 3.1$   & $67.20  \pm 0.91$        &  $66.97    \pm 0.80$          \\[+1mm]
  $\tau$                   & $0.077 \pm 0.019$      & $0.083   \pm 0.019$      &  $0.085    \pm 0.018$         \\[+1mm]
  $\ln(10^{10} A_s)$       & $3.089 \pm 0.037$      & $3.097   \pm 0.037$      &  $3.100    \pm 0.035$        \\[+1mm]
  $n_s$                    & $0.9646 \pm 0.0064$    & $0.9686  \pm 0.0058$     &  $0.9702   \pm 0.0049$     \\[+1mm]
  $\alpha$~[95.4\%~C.L.]   & $<1.49$                & $<0.19$                  &  $<0.32$                   \\[+1mm]
 \hline \\[-2mm]
  $100\theta_\textrm{MC}$  & $1.04063 \pm 0.00049$  & $1.04084 \pm 0.00046$    &  $1.04095  \pm 0.00043$     \\[+1mm]
  $\Omega_m$               & $0.359 \pm 0.039$      & $0.313 \pm 0.012$        &  $0.3139   \pm 0.0085$      \\[+1mm]
  $\sigma_8$               & $0.791 \pm 0.032$      & $0.822 \pm 0.016$        &  $0.816    \pm 0.016$        \\[+1mm]
  \hline \hline \\[-2mm]
    Parameter              & TT+lowP+$H(z)$         &  TT+lowP+SN+BAO          &  TT+lowP+SN+BAO+$H(z)$  \\[+0mm]
 \hline \\[-2mm]
  $\Omega_b h^2$           & $0.02226 \pm 0.00023$  &  $0.02236 \pm 0.00020$   &  $0.02237 \pm 0.00020$    \\[+1mm]
  $\Omega_c h^2$           & $0.1193  \pm 0.0021$   &  $0.1177 \pm 0.0013$     &  $0.1177 \pm 0.0013$     \\[+1mm]
  $H_0$ [km s$^{-1}$ Mpc$^{-1}$] & $65.4 \pm 1.9$   &  $67.44 \pm 0.59$        &  $67.45 \pm 0.61$       \\[+1mm]
  $\tau$                   & $0.080 \pm 0.019$      &  $0.084 \pm 0.017$       &  $0.084 \pm 0.018$      \\[+1mm]
  $\ln(10^{10} A_s)$       & $3.094 \pm 0.036$      &  $3.098 \pm 0.034$       &  $3.098 \pm 0.035$      \\[+1mm]
  $n_s$                    & $0.9666 \pm 0.0061$    &  $0.9704 \pm 0.0045$     &  $0.9704 \pm 0.0046$     \\[+1mm]
  $\alpha$~[95.4\%~C.L.]   & $<0.68$                &  $<0.19$                 &  $<0.20$     \\[+1mm]
 \hline \\[-2mm]
  $100\theta_\textrm{MC}$  & $1.04075 \pm 0.00047$  &  $1.04093 \pm 0.00043$   &  $1.04096 \pm 0.00042$    \\[+1mm]
  $\Omega_m$               & $0.334 \pm 0.022$      &  $0.3094 \pm 0.0069$     &  $0.3093 \pm 0.0070$     \\[+1mm]
  $\sigma_8$               & $0.808 \pm 0.022$      &  $0.820 \pm 0.014$       &  $0.819 \pm 0.015$      \\[+1mm]
    \hline \hline \\[-2mm]
    Parameter              & TT+lowP+$f\sigma_8$    & TT+lowP+BAO+$f\sigma_8$  &  TT+lowP+SN+BAO+$H(z)$+$f\sigma_8$  \\[+0mm]
 \hline \\[-2mm]
  $\Omega_b h^2$           & $0.02233 \pm 0.00023$  & $0.02239 \pm 0.00020$    &  $0.02238 \pm 0.00020$    \\[+1mm]
  $\Omega_c h^2$           & $0.1175  \pm 0.0020$   & $0.1168 \pm 0.0014$      &  $0.1169 \pm 0.0013$     \\[+1mm]
  $H_0$ [km s$^{-1}$ Mpc$^{-1}$] & $65.6 \pm 2.3$   & $67.17 \pm 0.83$         &  $67.61 \pm 0.62$       \\[+1mm]
  $\tau$                   & $0.075 \pm 0.019$      & $0.079 \pm 0.018$        &  $0.076 \pm 0.018$      \\[+1mm]
  $\ln(10^{10} A_s)$       & $3.079 \pm 0.037$      & $3.084 \pm 0.034$        &  $3.079 \pm 0.035$      \\[+1mm]
  $n_s$                    & $0.9701 \pm 0.0060$    & $0.9721 \pm 0.0048$      &  $0.9716 \pm 0.0046$     \\[+1mm]
  $\alpha$~[95.4\%~C.L.]   & $<0.85$                & $<0.33$                  &  $<0.21$     \\[+1mm]
 \hline \\[-2mm]
  $100\theta_\textrm{MC}$  & $1.04091 \pm 0.00047$  & $1.04103 \pm 0.00042$    &  $1.04101 \pm 0.00042$    \\[+1mm]
  $\Omega_m$               & $0.328 \pm 0.026$      & $0.3101 \pm 0.0084$      &  $0.3062 \pm 0.0069$     \\[+1mm]
  $\sigma_8$               & $0.792 \pm 0.024$      & $0.805 \pm 0.015$        &  $0.808 \pm 0.014$       \\[+0mm]
\end{tabular}
\end{ruledtabular}
\label{tab:para_flat}
\end{table*}


\begin{table*}
\caption{Tilted flat $\phi\textrm{CDM}$ model parameters constrained by using Planck TT + lowP + lensing, SN, BAO, $H(z)$, and $f\sigma_8$ data (mean and 68.3\% confidence limits).}
\begin{ruledtabular}
\begin{tabular}{lccc}
  Parameter                & TT+lowP+lensing        & TT+lowP+lensing+SN       &  TT+lowP+lensing+BAO   \\[+0mm]
 \hline \\[-2mm]
  $\Omega_b h^2$           & $0.02221 \pm 0.00023$  & $0.02232 \pm 0.00023$    &  $0.02235  \pm 0.00021$    \\[+1mm]
  $\Omega_c h^2$           & $0.1190  \pm 0.0020$   & $0.1176  \pm 0.0018$     &  $0.1173   \pm 0.0014$     \\[+1mm]
  $H_0$ [km s$^{-1}$ Mpc$^{-1}$] & $63.6 \pm 2.9$   & $67.48  \pm 0.91$        &  $67.01    \pm 0.80$       \\[+1mm]
  $\tau$                   & $0.070 \pm 0.017$      & $0.073   \pm 0.016$      &  $0.077    \pm 0.015$      \\[+1mm]
  $\ln(10^{10} A_s)$       & $3.072 \pm 0.031$      & $3.075   \pm 0.029$      &  $3.081    \pm 0.027$      \\[+1mm]
  $n_s$                    & $0.9667 \pm 0.0062$    & $0.9701  \pm 0.0058$     &  $0.9710   \pm 0.0049$     \\[+1mm]
  $\alpha$~[95.4\%~C.L.]   & $<1.20$                & $<0.19$                  &  $<0.32$                   \\[+1mm]
 \hline \\[-2mm]
  $100\theta_\textrm{MC}$  & $1.04076 \pm 0.00047$  & $1.04095 \pm 0.00045$    &  $1.04100  \pm 0.00042$    \\[+1mm]
  $\Omega_m$               & $0.353 \pm 0.035$      & $0.309 \pm 0.011$        &  $0.3125   \pm 0.0082$     \\[+1mm]
  $\sigma_8$               & $0.780 \pm 0.026$      & $0.810 \pm 0.010$        &  $0.805    \pm 0.011$      \\[+1mm]
  \hline \hline \\[-2mm]
    Parameter              & TT+lowP+lensing+$H(z)$ &  TT+lowP+lensing+SN+BAO  &  TT+lowP+lensing+SN+BAO+$H(z)$  \\[+0mm]
 \hline \\[-2mm]
  $\Omega_b h^2$           & $0.02227 \pm 0.00023$  &  $0.02234 \pm 0.00021$   &  $0.02235 \pm 0.00020$  \\[+1mm]
  $\Omega_c h^2$           & $0.1184  \pm 0.0020$   &  $0.1174 \pm 0.0013$     &  $0.1173 \pm 0.0013$    \\[+1mm]
  $H_0$ [km s$^{-1}$ Mpc$^{-1}$] & $65.5 \pm 2.0$   &  $67.52 \pm 0.61$        &  $67.53 \pm 0.62$       \\[+1mm]
  $\tau$                   & $0.071 \pm 0.017$      &  $0.074 \pm 0.013$       &  $0.075 \pm 0.014$      \\[+1mm]
  $\ln(10^{10} A_s)$       & $3.072 \pm 0.031$      &  $3.076 \pm 0.024$       &  $3.078 \pm 0.026$      \\[+1mm]
  $n_s$                    & $0.9681 \pm 0.0060$    &  $0.9706 \pm 0.0047$     &  $0.9708 \pm 0.0046$    \\[+1mm]
  $\alpha$~[95.4\%~C.L.]   & $<0.73$                &  $<0.20$                 &  $<0.20$     \\[+1mm]
 \hline \\[-2mm]
  $100\theta_\textrm{MC}$  & $1.04085 \pm 0.00045$  &  $1.04098 \pm 0.00041$   &  $1.04098 \pm 0.00043$   \\[+1mm]
  $\Omega_m$               & $0.331 \pm 0.023$      &  $0.3080 \pm 0.0069$     &  $0.3078 \pm 0.0071$     \\[+1mm]
  $\sigma_8$               & $0.794 \pm 0.018$      &  $0.8092 \pm 0.0097$     &  $0.8096 \pm 0.0098$     \\[+1mm]
    \hline \hline \\[-2mm]
    Parameter              & TT+lowP+lensing+$f\sigma_8$  & TT+lowP+lensing+BAO+$f\sigma_8$  &  TT+lowP+lensing+SN+BAO+$H(z)$+$f\sigma_8$  \\[+0mm]
 \hline \\[-2mm]
  $\Omega_b h^2$           & $0.02233 \pm 0.00022$  & $0.02238 \pm 0.00020$    &  $0.02238 \pm 0.00020$   \\[+1mm]
  $\Omega_c h^2$           & $0.1174  \pm 0.0019$   & $0.1167 \pm 0.0014$      &  $0.1168 \pm 0.0013$     \\[+1mm]
  $H_0$ [km s$^{-1}$ Mpc$^{-1}$] & $65.7 \pm 2.3$   & $67.09 \pm 0.84$         &  $67.63 \pm 0.62$        \\[+1mm]
  $\tau$                   & $0.072 \pm 0.017$      & $0.076 \pm 0.015$        &  $0.074 \pm 0.014$       \\[+1mm]
  $\ln(10^{10} A_s)$       & $3.073 \pm 0.030$      & $3.078 \pm 0.027$        &  $3.074 \pm 0.025$       \\[+1mm]
  $n_s$                    & $0.9702 \pm 0.0060$    & $0.9720 \pm 0.0049$      &  $0.9715 \pm 0.0045$     \\[+1mm]
  $\alpha$~[95.4\%~C.L.]   & $<0.85$                & $<0.34$                  &  $<0.22$     \\[+1mm]
 \hline \\[-2mm]
  $100\theta_\textrm{MC}$  & $1.04092 \pm 0.00046$  & $1.04101 \pm 0.00042$    &  $1.04101 \pm 0.00042$   \\[+1mm]
  $\Omega_m$               & $0.327 \pm 0.026$      & $0.3106 \pm 0.0084$      &  $0.3059 \pm 0.0068$     \\[+1mm]
  $\sigma_8$               & $0.789 \pm 0.021$      & $0.801 \pm 0.011$        &  $0.8055 \pm 0.0098$       \\[+0mm]
\end{tabular}
\end{ruledtabular}
\label{tab:para_flat_lensing}
\end{table*}

Next, we use the same observational data to explore and constrain the 
parameter space of untilted nonflat $\phi$CDM models. For these models 
the MCMC parameter search using only the Planck 
CMB data (either TT + lowP or TT + lowP + lensing) is very slow because of 
the highly degenerate and non-Gaussian shape of the likelihood distributions 
of $H_0$, $\Omega_k$, and $\alpha$. The overall shape of the likelihood 
function in the three dimensional space of these three parameters can 
be described as a sheet of bent paper. Thus the full likelihood distribution 
is not well approximated by a simple multivariate Gaussian function. In 
practical terms the problem is that the MCMC random walks in the parameter 
space that are usually determined by the square root of the covariance matrix 
of model parameters multiplied by a random number vector drawn from a 
Gaussian distribution does not properly propagate throughout the whole space 
but stays within a local maximum of the likelihood distribution. 

\begin{figure*}
\centering
\mbox{\includegraphics[width=120mm,bb=30 240 600 720]{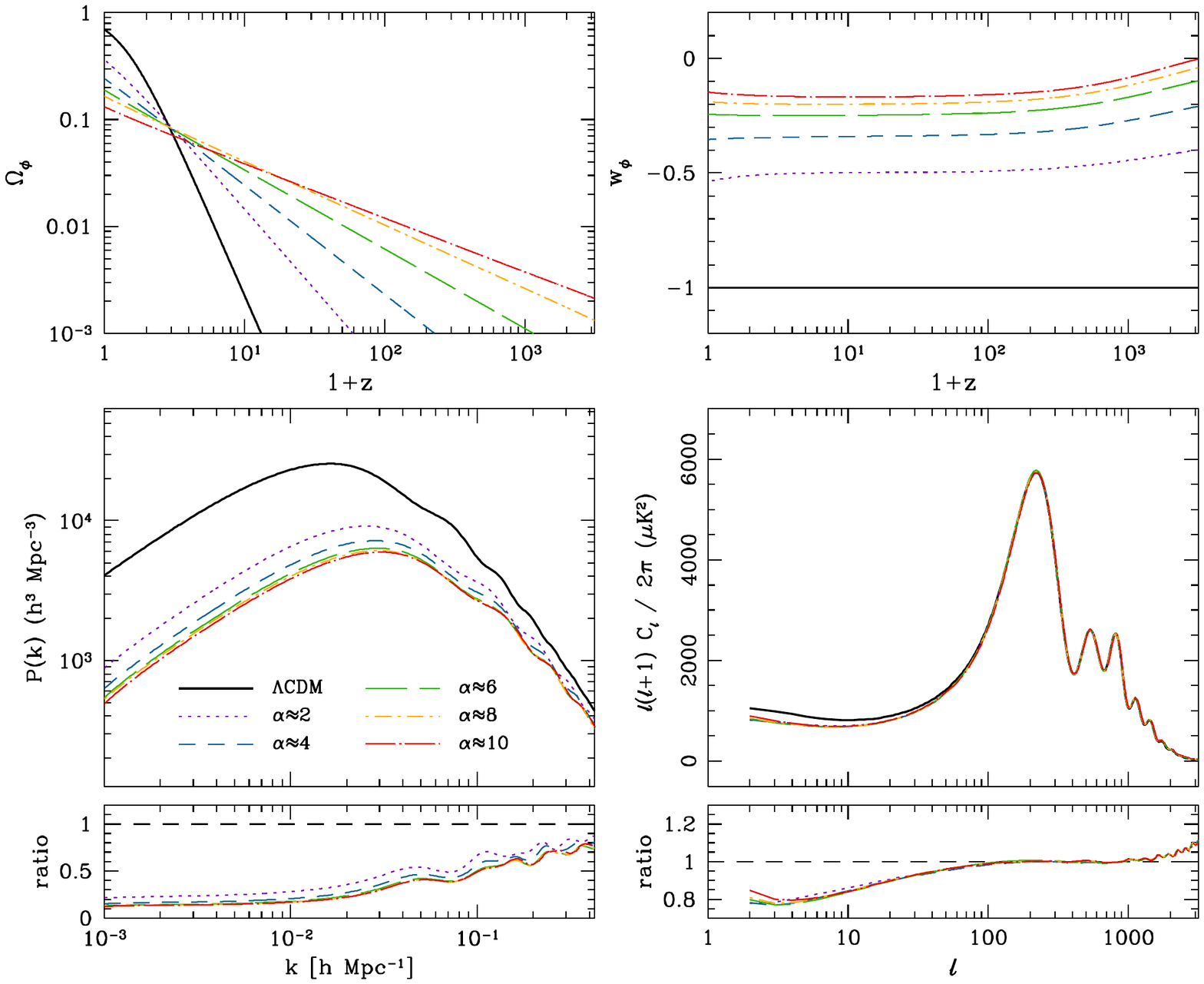}}
\caption{Top: Evolution of the dark energy scalar field 
density parameter ($\Omega_\phi$) and equation of state parameter ($w_\phi$) for
the untilted nonflat $\phi$CDM model for various $\alpha$ values
corresponding to the green dots in the bottom panels of Fig.\ 
\ref{fig:para_nonflat_tt_lowp_comparison}. The black solid curve is for
$\Lambda$CDM, whose model parameters are the same as in Fig.\ \ref{fig:bgpert_flat}.
The $\phi$CDM models presented here were selected from the unconverged MCMC chains and are 
consistent with Planck observations. Bottom: Theoretical predictions for 
matter density and CMB temperature anisotropy angular power spectra
for the $\phi$CDM models. }
\label{fig:bgpert_nonflat}
\end{figure*}

The most dramatic feature of the nonflat $\phi$CDM analyses is that for this 
model the CMB data poorly constrains $\alpha$ and is also consistent with 
large values of $\alpha$, unlike in the spatially-flat $\phi$CDM 
model.\footnote{$\alpha$ governs the dynamics of dark energy and when spatial
curvature is allowed to vary this weakens the constraints on dark energy 
dynamics from any data set \citep[see, e.g.,][and references therein]{Farooqetal2017}.} 
This phenomenon can be understood more easily by comparing CMB data parameter 
constraints for the nonflat $\phi$CDM and XCDM models. Figure 
\ref{fig:bgpert_nonflat} shows several examples of the $\phi$CDM model with 
large $\alpha$'s that are consistent with the Planck CMB observations. Here 
the parameters of the nonflat $\phi$CDM models were chosen from the 
unconverged MCMC output determined using the Planck TT + lowP data, for 
$\alpha$'s near 2, 4, 6, 8, and 10 (shown in the three-dimensional view of the
MCMC chains in the $H_0$-$\Omega_k$-$\alpha$ space of Fig.\ 
\ref{fig:para_nonflat_tt_lowp_comparison} below).
The individual cosmological parameters of these five models are very 
different. All the models have small Hubble constant 
($H_0 < 45$ $\textrm{km~s}^{-1}\textrm{Mpc}^{-1}$) and are highly 
positively curved ($\Omega_k < 0$).\footnote{For example, for the nonflat 
$\phi$CDM model with $\alpha \approx 10$, which corresponds
to the red curves in Fig.\ \ref{fig:bgpert_nonflat}, the parameters are
$\Omega_b h^2=0.0022946$, $\Omega_c h^2=0.10990$, $\Omega_\nu h^2=6.4514\times 10^{-4}$,
$\Omega_k=-0.2023$, $\alpha=9.795$, $\tau=0.1123$, and $A_s=2.2921\times 10^{-9}$ 
(at $k_0=0.05~\textrm{Mpc}^{-1}$).} 
From Fig.\ \ref{fig:bgpert_nonflat}, we see that as $\alpha$ becomes larger 
($\alpha \gtrsim 6$), the behavior of the equation of state parameter $w_\phi$ 
becomes less sensitive to the variation of $\alpha$, and $\phi$CDM 
asymptotically behaves like the $w \approx -0.15$ XCDM parameterization for 
$z < 1000$. As shown in Fig.\ \ref{fig:para_nonflat_tt_lowp_comparison} 
(top-right panel), untilted nonflat XCDM parameterizations 
with $w \approx -0.15$ are 
consistent with Planck TT + lowP data for small Hubble constant and highly 
negative $\Omega_k$. A similar thing happens in the nonflat $\phi$CDM model 
for large values of $\alpha$, and $\phi$CDM models with very large $\alpha$ 
around 10 still have CMB power spectra that are consistent with the 
$\Lambda$CDM prediction (black curve in Fig.\ \ref{fig:bgpert_nonflat}) 
within observational precision. However, as shown in the matter power 
spectra panel of Fig.\ \ref{fig:bgpert_nonflat}, these $\phi$CDM models are 
excluded by large-scale structure observations.

\begin{figure*}
\centering
\mbox{\includegraphics[width=87mm]{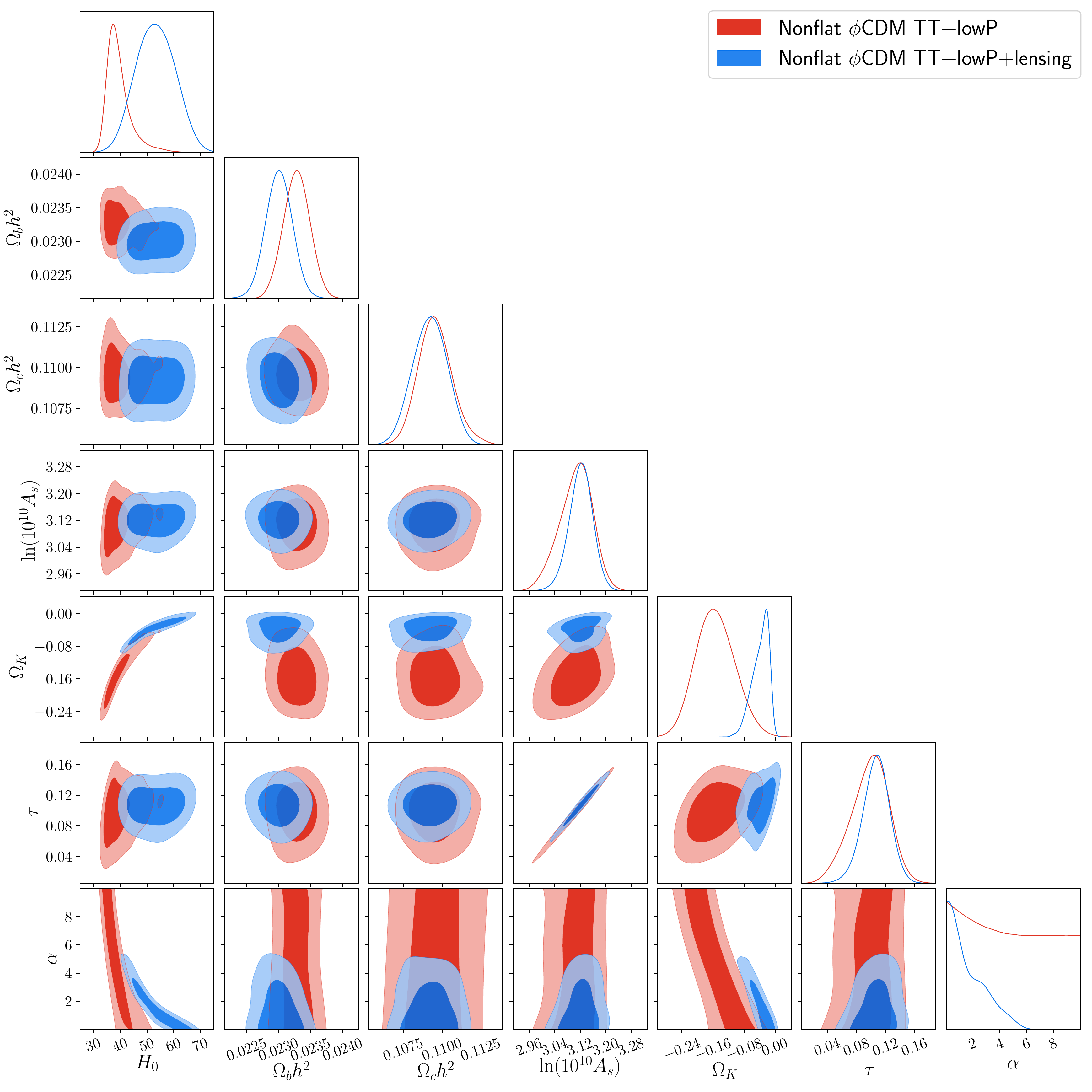}}
\mbox{\includegraphics[width=87mm]{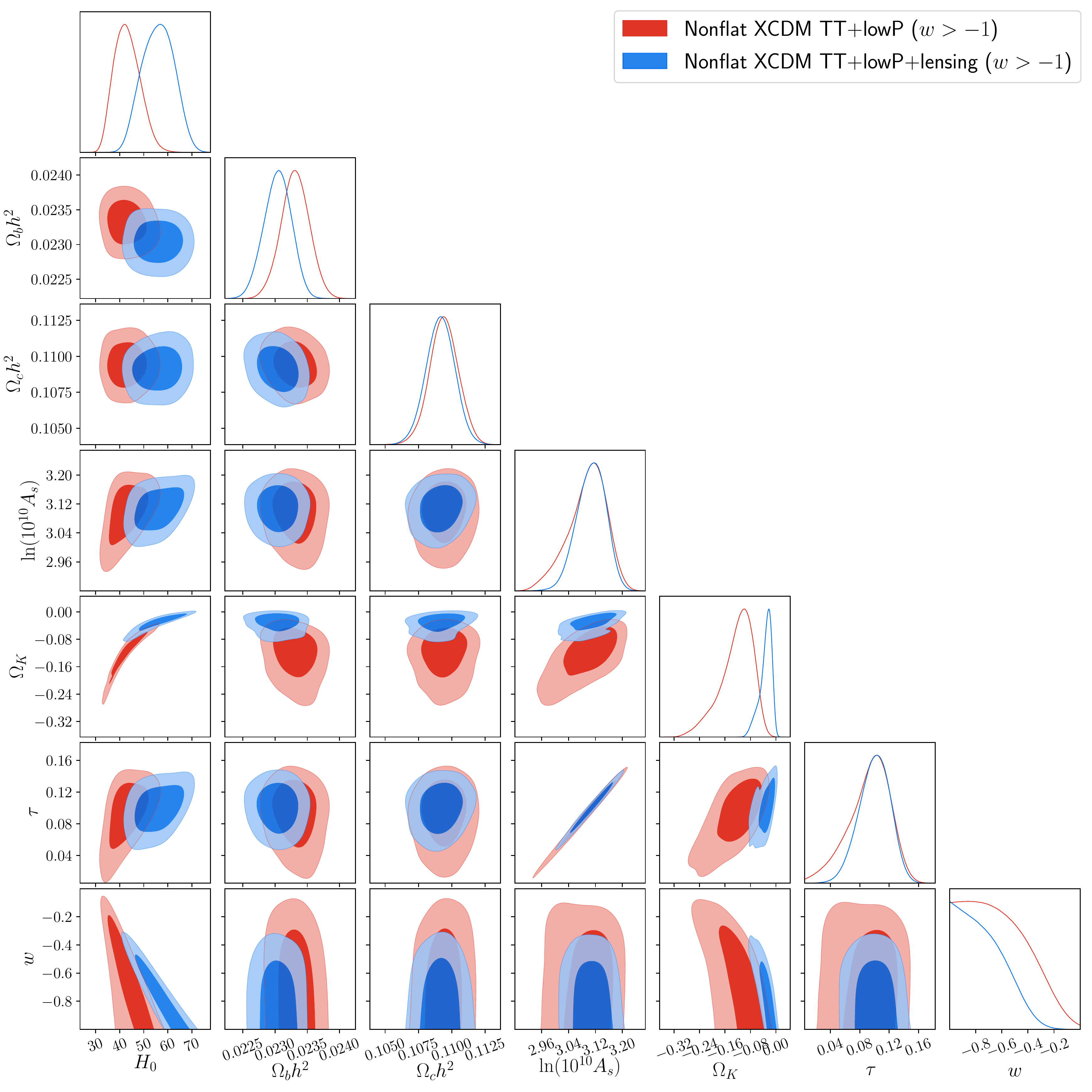}}\\
\mbox{\includegraphics[width=82mm]{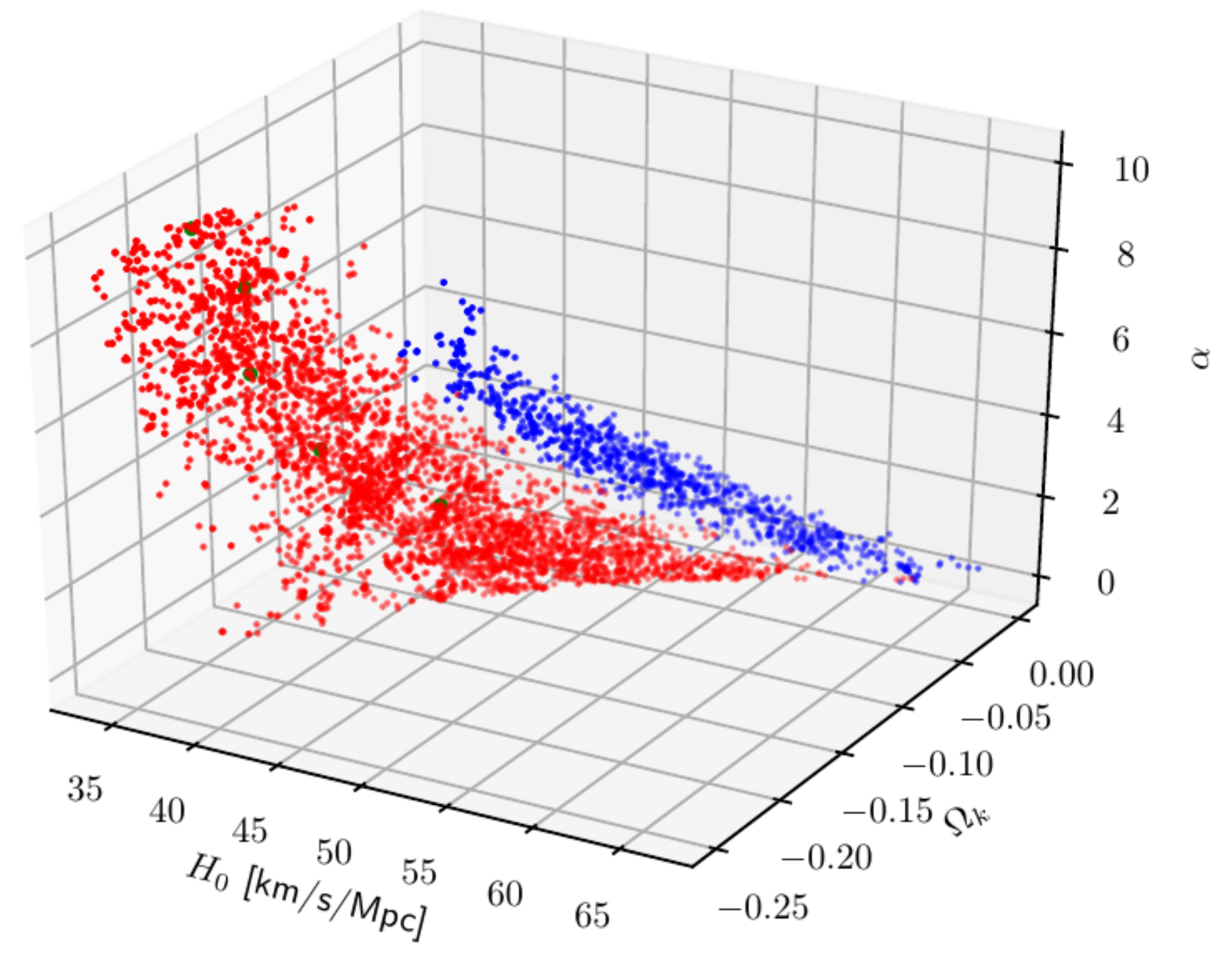}}
\mbox{\includegraphics[width=75mm]{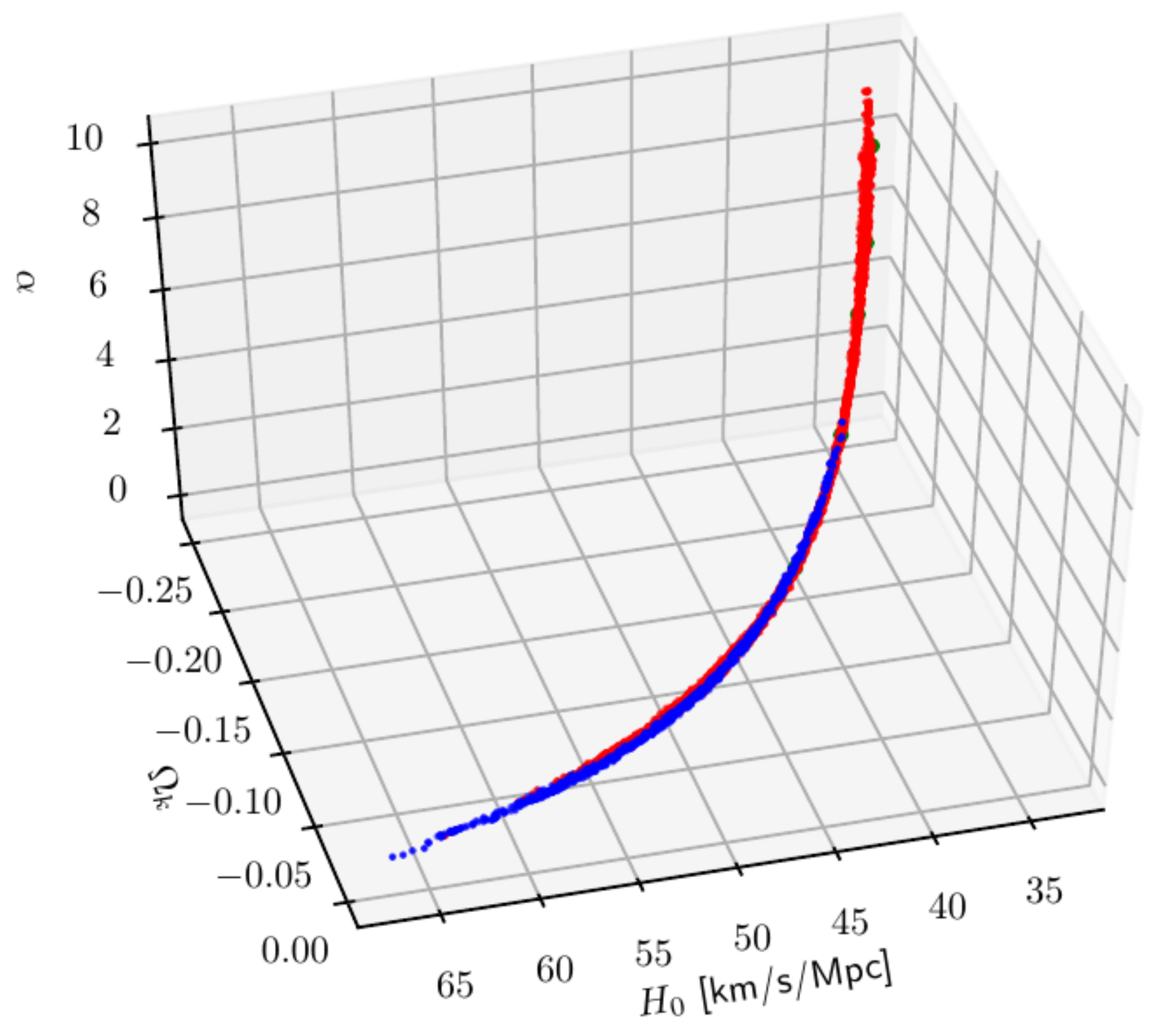}}
\caption{
Top left: Comparison of likelihood distributions of the untilted nonflat $\phi\textrm{CDM}$ model parameters constrained using Planck CMB TT + lowP and TT + lowP + lensing data. Note that these likelihood distributions have been obtained from unconverged MCMC chains. Top right: Similar plot as in top-left panel but now for the untilted nonflat XCDM model \citep{ParkRatra2018b}. Here the likelihood distributions have been derived from MCMC chain elements computed for equation of state parameter $w > -1$. 
Bottom panels: Three-dimensional views of some selected untilted nonflat $\phi$CDM model MCMC chain elements with $\Delta\chi^2 < 18$ relative to the minimum value (see Table \ref{tab:chi2_phiCDM} below), constrained using Planck TT + lowP (red dots) and TT + lowP + lensing (blue dots) data. The five green dots for the TT + lowP data indicate the position of the five untilted nonflat $\phi$CDM models presented in Fig.\ \ref{fig:bgpert_nonflat}.
}
\label{fig:para_nonflat_tt_lowp_comparison}
\end{figure*}

\begin{figure*}
\centering
\mbox{\includegraphics[width=87mm]{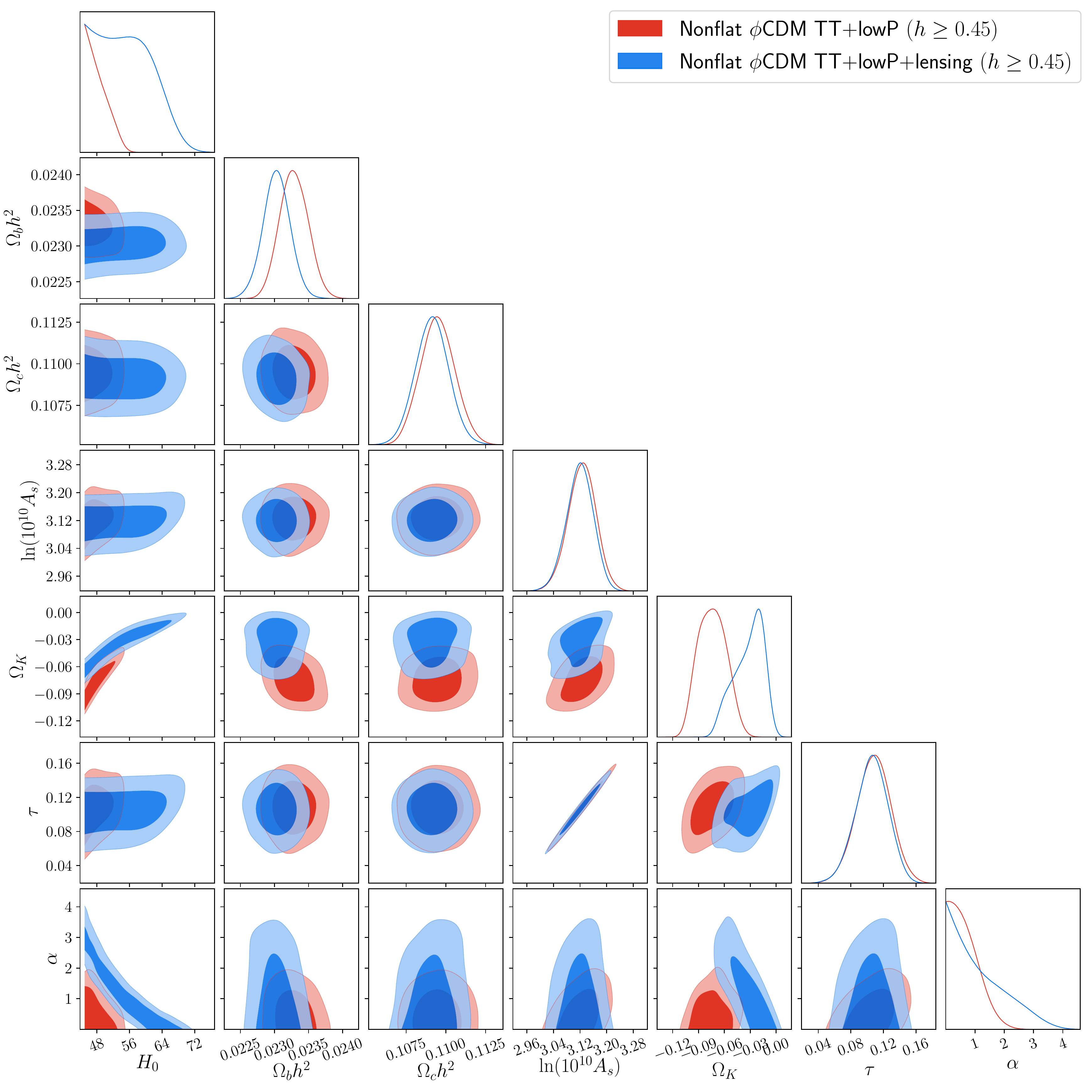}}
\caption{
Likelihood distributions of the untilted nonflat $\phi\textrm{CDM}$ model
parameters for the Planck CMB TT + lowP and TT + lowP + lensing data. Here we have used the more restrictive Hubble constant prior, $h \ge 0.45$. The MCMC processes have statistically converged.}
\label{fig:para_nonflat_tt_lowp_comparison_H45}
\end{figure*}

Figure \ref{fig:para_nonflat_tt_lowp_comparison} (top-left panel) shows the 
likelihood distributions of the untilted nonflat $\phi$CDM model parameters 
constrained by using the Planck TT + lowP (red) and TT + lowP + lensing (blue)
data. These are approximate estimates based on several sets of unconverged 
MCMC chains. For comparison we present results for the untilted nonflat 
XCDM parameterization parameters in the top-right panel. Here
we use MCMC chain elements computed for dark energy equation of state 
parameter $w > -1$ to derive likelihood contours of nonflat XCDM model 
parameters.
Thus the resulting likelihood distributions differ from those obtained from
the full MCMC outputs presented in \citet{ParkRatra2018b}.
The bottom panels show three-dimensional views of some selected untilted nonflat
$\phi$CDM model MCMC chains corresponding to parameter values that are 
consistent with the Planck TT + lowP (red dots) and TT + lowP + lensing 
(blue dots) data. The likelihood isosurfaces in the $H_0$-$\Omega_k$-$\alpha$ 
space appear to be long, thin, and curved, which means that the three 
parameters are highly degenerate and the likelihood functions are 
non-Gaussian. Since the nonflat $\phi$CDM model with small Hubble constant 
and large $\alpha$ will be excluded by other cosmological observations, from 
here on we set a more restrictive prior for the Hubble constant, $h \ge 0.45$,
to guarantee reasonably rapid convergence of the MCMC chains (given our 
computational resources) for the CMB data alone analyses. Likelihood 
distributions from the CMB data alone analyses for the more restrictive 
Hubble constant prior 
are shown in Fig.\ \ref{fig:para_nonflat_tt_lowp_comparison_H45}. Note that 
for the Planck TT + lowP data the constraint on $\alpha$ seems tighter than 
for the case of the TT + lowP + lensing data. This is because the region of 
large $\alpha$ but small Hubble constant favored by TT + lowP data is 
excluded by the more restrictive Hubble constant prior.

\begin{figure*}
\centering
\mbox{\includegraphics[width=87mm]{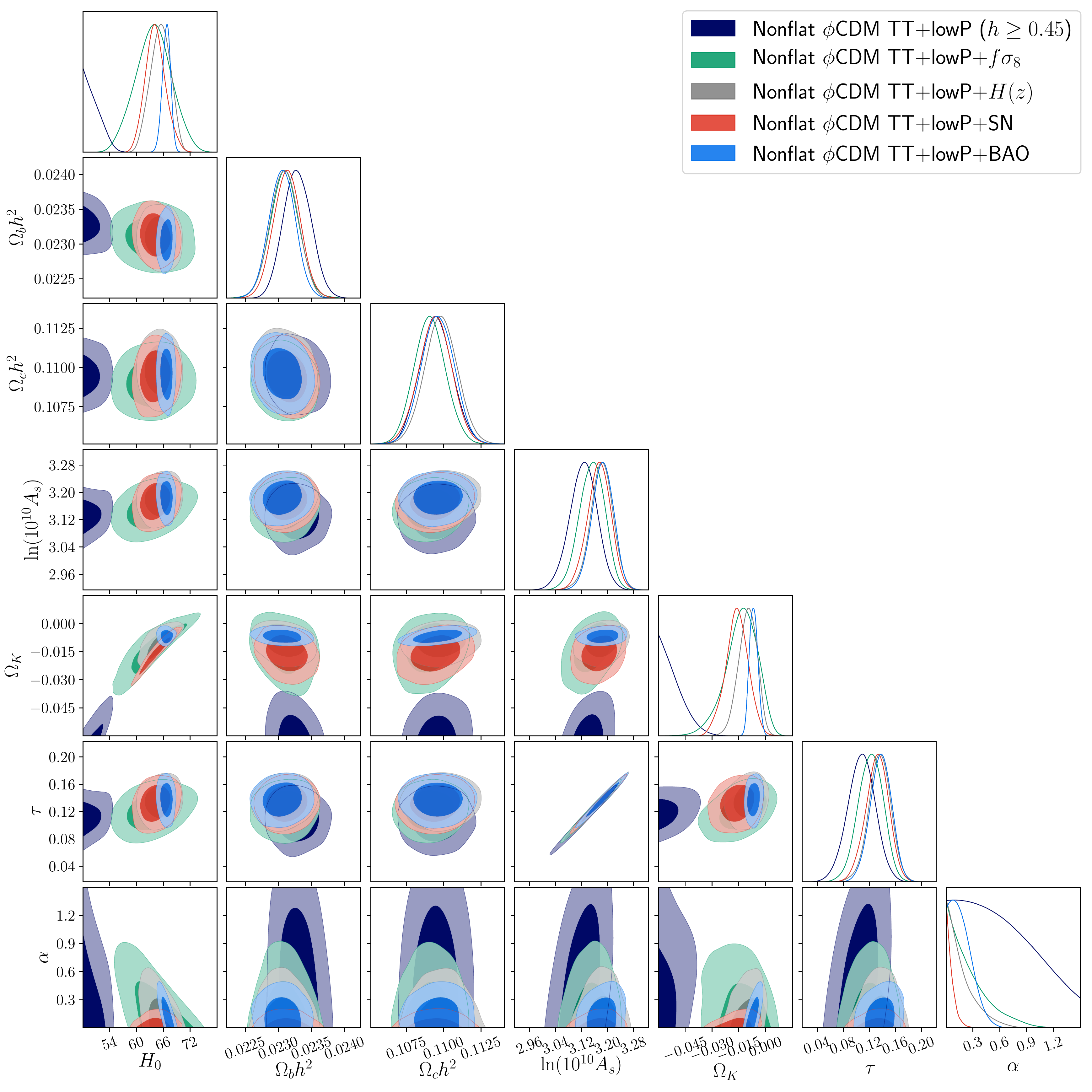}}
\mbox{\includegraphics[width=87mm]{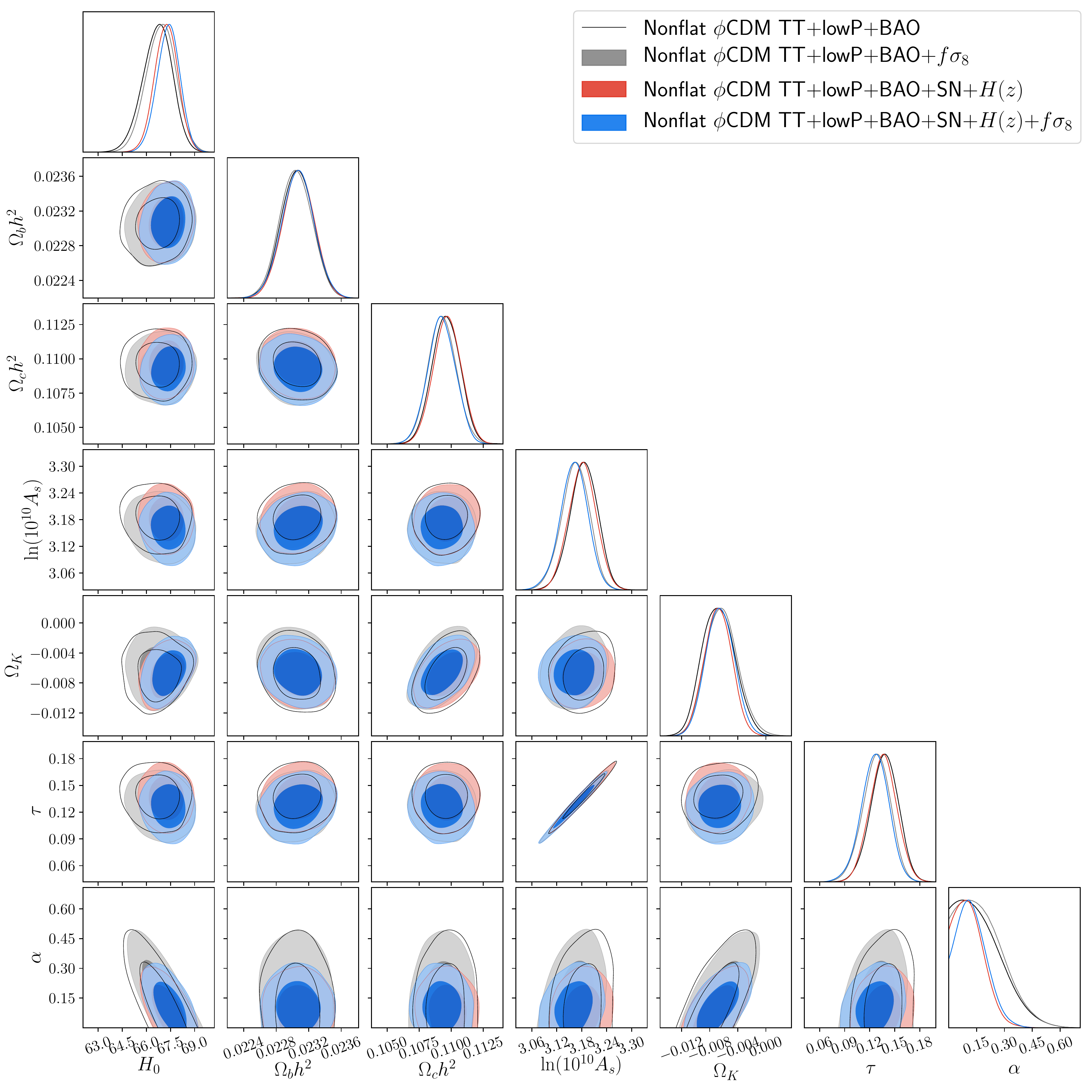}}
\caption{
Likelihood distributions of the untilted nonflat $\phi\textrm{CDM}$ model
parameters constrained using Planck CMB TT + lowP, SN, BAO, $H(z)$, and
$f\sigma_8$ data sets. Two-dimensional marginalized likelihood constraint 
contours and one-dimensional likelihood distributions are plotted for when 
each set of non-CMB data is combined with  the Planck TT + lowP data
(left panel) and when the growth rate, Hubble parameter, and SN data, as well
as their combination, are added to TT + lowP + BAO data
(right panel). For clarity of viewing, the result of TT + lowP + BAO
is shown with solid black curves in the right panel. The TT + lowP CMB data alone
contours are derived using the more restrictive $h \ge 0.45$ prior.
}
\label{fig:para_nonflat}
\end{figure*}

\begin{figure*}
\centering
\mbox{\includegraphics[width=87mm]{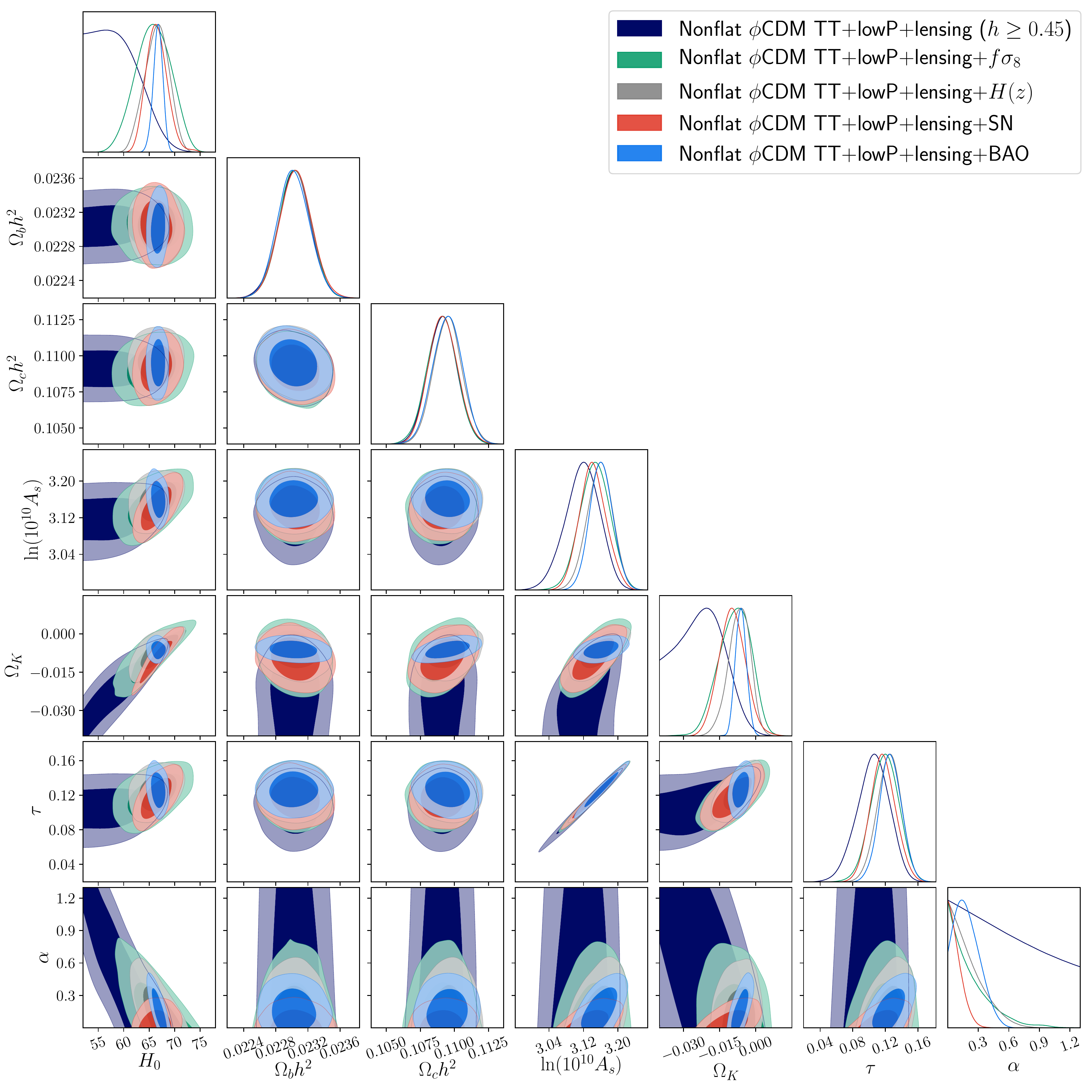}}
\mbox{\includegraphics[width=87mm]{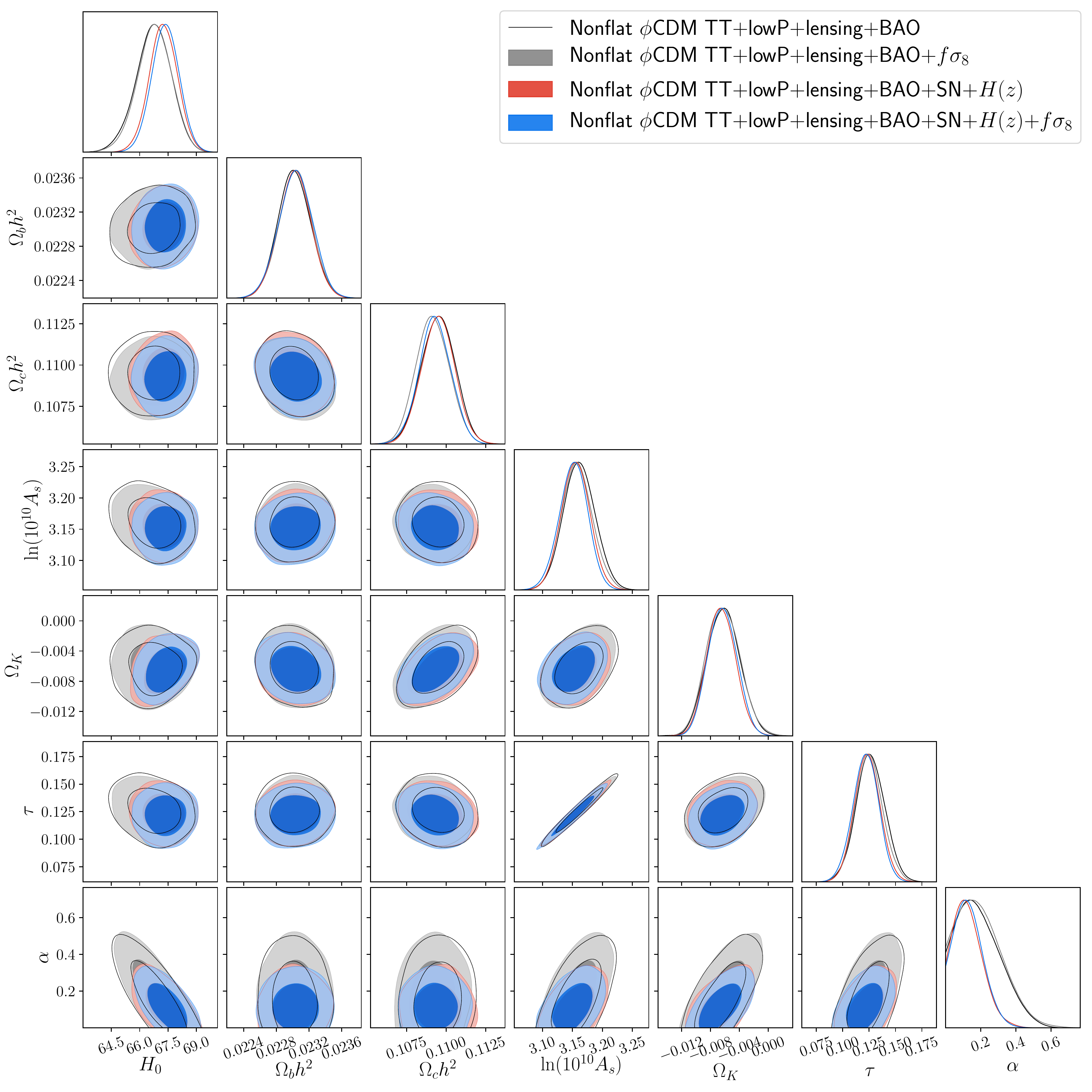}}
\caption{Same as Fig.\ \ref{fig:para_nonflat} but now also accounting for 
the Planck CMB lensing data. The TT + lowP + lensing CMB data alone
contours are derived using the more restrictive $h \ge 0.45$ prior.
}
\label{fig:para_nonflat_lensing}
\end{figure*}

Our results for the nonflat untilted $\phi$CDM model 
are presented in Figs.\ \ref{fig:para_nonflat} and 
\ref{fig:para_nonflat_lensing} and Tables \ref{tab:para_nonflat} and 
\ref{tab:para_nonflat_lensing}. As in the flat tilted $\phi$CDM models,
the likelihood distributions for 
TT + lowP (+ lensing) + SN + BAO data (ignoring or accounting for the 
CMB lensing data) are omitted in the figures since they are very 
similar to those for TT + lowP (+ lensing) + SN + BAO + $H(z)$ data.

The entries for the nonflat untilted $\phi$CDM model in Table 
\ref{tab:para_nonflat} (TT + lowP panel) and those in the Table 
\ref{tab:para_nonflat_lensing} 
(TT + lowP + lensing panel) are very consistent 
with the corresponding entries in Table 1 of 
\citet{Oobaetal2018b}.\footnote{The TT + lowP and TT + lowP + lensing 
entries in the original version of \citet{Oobaetal2018b} were incorrect 
because of a numerical error in their initial computation. Our comparison 
here is made to the corrected \citet{Oobaetal2018b} results.}
\citet{Oobaetal2018b} computed the $C_\ell$'s by using CLASS 
\citep{Blasetal2011} and used the Monte Python software package 
\citep{Audrenetal2013} for the 
MCMC analyses; it is reassuring that both sets of results agree well.

The parameter constraints are more interesting 
in the nonflat untilted $\phi$CDM model than in the flat tilted case.
The general behavior of the cosmological parameter constraints
are similar to those in the XCDM model \citep{ParkRatra2018b}.
When CMB lensing data are accounted for, Table \ref{tab:para_nonflat_lensing},
Planck CMB data with either $H(z)$, BAO, SN, or $f \sigma_8$ data,
provide roughly equally tight constraints on $\Omega_b h^2$,
$\Omega_c h^2$, and $\theta_\textrm{MC}$, while CMB + BAO measurements provide
the tightest limits on $H_0$, $\tau$, $A_s$, $\Omega_k$, 
$\Omega_m$, 
and $\sigma_8$, with CMB + SN setting the tightest limits on 
$\alpha$. We note that the full combination of CMB and non-CMB data results 
in somewhat weaker constraints on $\alpha$ (compared to the CMB + SN case) 
and on $\Omega_b h^2$ (compared to the CMB + $f \sigma_8$ case). 

Let us focus on the results for CMB TT + lowP + lensing data, presented
in Figs.\ \ref{fig:para_flat_lensing} and \ref{fig:para_nonflat_lensing}
and Tables \ref{tab:para_flat_lensing}
and \ref{tab:para_nonflat_lensing}, where we see that adding in turn each 
of the four sets of non-CMB measurements to the CMB measurements
(left panels in the two figures) result in likelihood contours
that are quite compatible with each other, as well as with
the CMB alone contours, for both the flat tilted 
and the nonflat untilted $\phi$CDM model. It might be significant 
that the four sets of non-CMB observations do not pull the CMB only 
contours in very different directions. This is also true 
for the flat tilted $\phi$CDM model when CMB lensing 
data are ignored (left panel of Fig.\ \ref{fig:para_flat}).
However, in the nonflat untilted $\phi$CDM cosmogony without the lensing data
each of the four sets of non-CMB data added to the CMB data
(left panel of Fig.\ \ref{fig:para_nonflat}) push the results towards a 
smaller $|\Omega_k|$ (closer to flat space) and larger $H_0$ as well as 
slightly larger $\tau$ and $A_s$ and slightly smaller $\Omega_b h^2$ than is 
preferred by 
the CMB data alone, but all five constraint contour sets are 
largely mutually compatible, except for the $H_0$ and $\Omega_k$ constraints
where the TT + lowP data alone results differ from those derived using TT + lowP
in combination with any one of the four non-CMB data sets.

Although augmenting the CMB data with the BAO data typically results 
in the largest difference, each of the other three sets of non-CMB data
also contribute. Considering the TT + lowP + lensing data, we see from Table 
\ref{tab:para_flat_lensing} for the flat tilted $\phi$CDM model
that the $H_0$ error bar is reduced the most by the 
full compilation of measurements relative to the CMB + BAO observations 
compilation, 
followed by the $\Omega_m$ error bar decrease compared to the 
CMB + BAO data collection. For the nonflat untilted $\phi$CDM model, 
from Table \ref{tab:para_nonflat_lensing}, the error bars that reduce 
the most when CMB (accounting for lensing) data are used in combination with 
the four sets of non-CMB data are those on $\Omega_m$ and $H_0$ (relative to 
the CMB + BAO combination).

Focusing again on the TT + lowP + lensing data, Tables   
\ref{tab:para_flat_lensing} and \ref{tab:para_nonflat_lensing},
for the flat tilted $\phi$CDM model, we see that augmenting the 
CMB data with the four non-CMB data sets most affects $H_0$, 
$\Omega_m$, $\Omega_c h^2$, and $\sigma_8$,
with the  $H_0$ and $\sigma_8$ central values moving up by 1.4$\sigma$ 
and 0.98$\sigma$ and the $\Omega_m$ and $\Omega_c h^2$ central
values moving down by 1.3$\sigma$ and 1.1$\sigma$, all of the CMB data only
error bars; $\ln (10^{10} A_s)$ is not much affected by including
the four non-CMB sets of data, changing by only 0.065$\sigma$. The 
situation for the nonflat untilted $\phi$CDM model is a little more
dramatic, with $H_0$ and $\sigma_8$ central values moving up by 1.9$\sigma$ 
and 1.7$\sigma$ of the CMB data only error bars, $\Omega_m$ decreasing 
by 1.6$\sigma$, and 
$\Omega_k$ more closely approaching flatness by 1.5$\sigma$; in this case 
the $\Omega_b h^2$ central value is not affected.
 
\begin{figure*}
\centering
\mbox{\includegraphics[width=65mm]{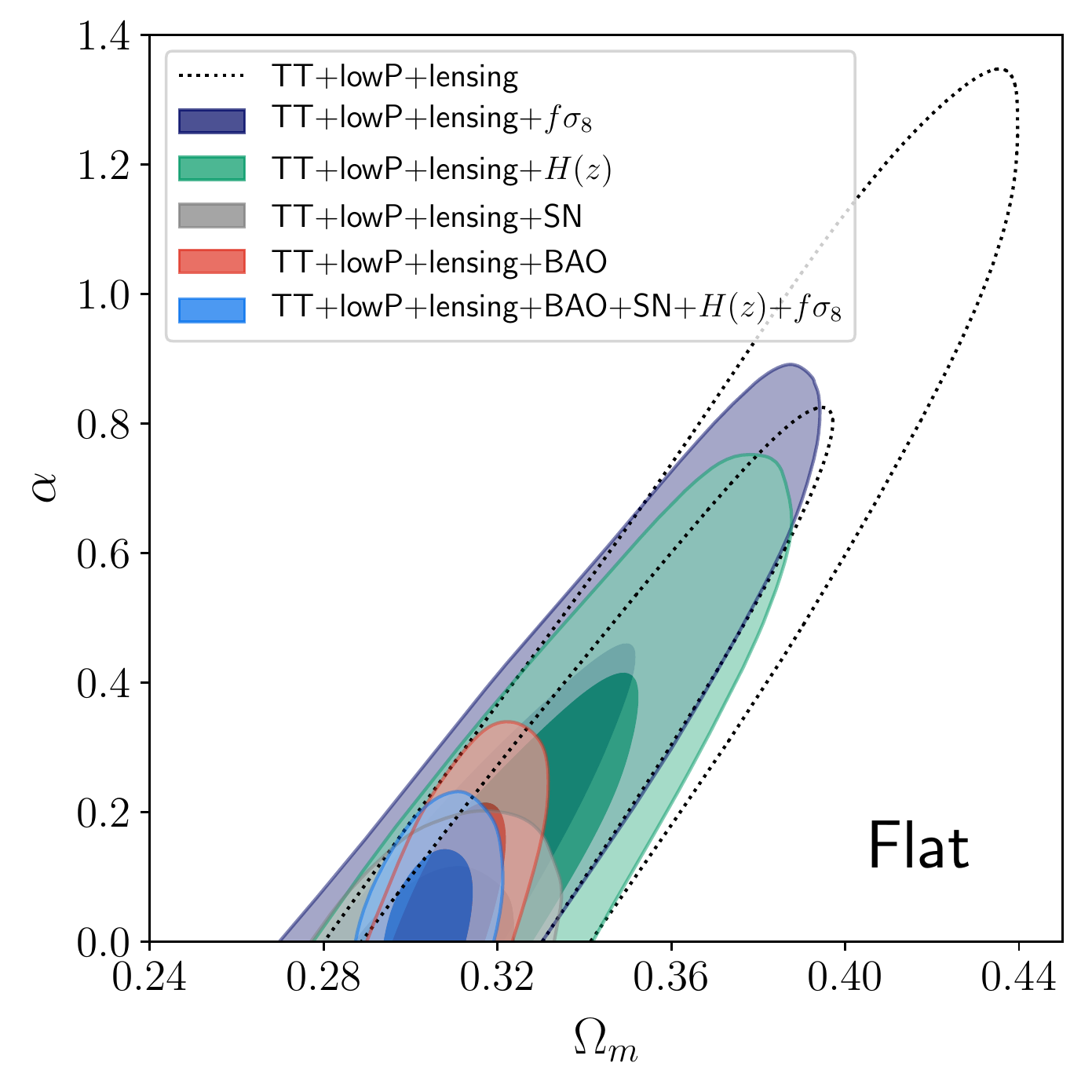}}
\mbox{\includegraphics[width=70mm]{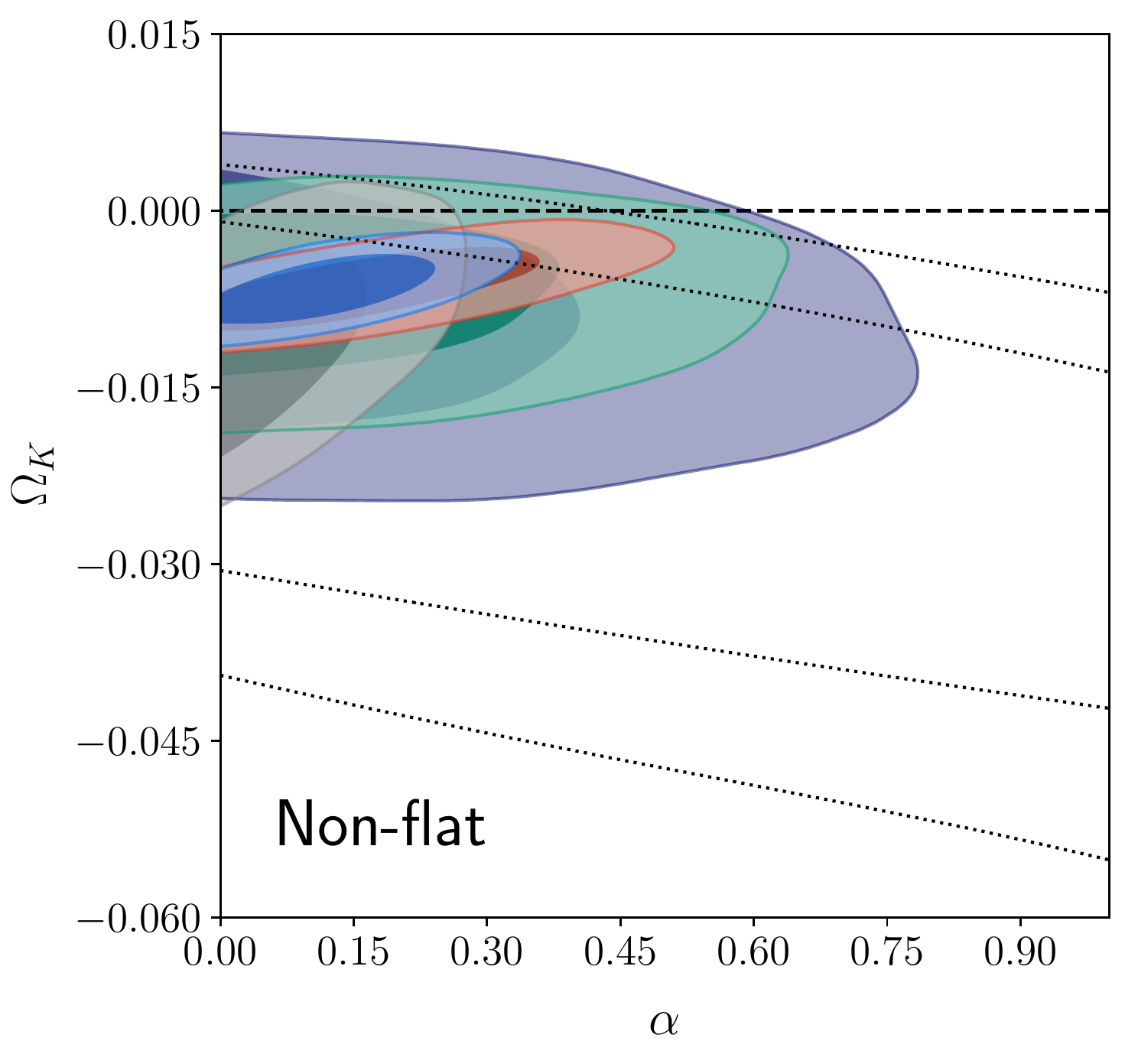}}
\caption{1$\sigma$ and 2$\sigma$ likelihood contours in the $\Omega_m$--$\alpha$ plane for the tilted flat $\phi\textrm{CDM}$ model (left panel) 
and in the $\alpha$--$\Omega_k$ plane for the untilted nonflat 
$\phi\textrm{CDM}$ model (right panel), constrained by using Planck CMB TT + lowP + lensing and non-CMB data sets.
The horizontal dashed line in the right panel indicates $\Omega_k=0$.
The color scheme shown in the left panel governs the contours in both panels.
However, for nonflat $\phi$CDM model constrained with TT + lowP + lensing, the prior $h\ge 0.45$ has been applied.
}
\label{fig:omm_alpha_omk}
\end{figure*}

Figure \ref{fig:omm_alpha_omk} shows marginalized likelihood contours in 
the $\Omega_m$--$\alpha$ plane for the flat tilted $\phi$CDM model
and in $\alpha$--$\Omega_k$ plane for the nonflat untilted $\phi$CDM case. For 
CMB TT + lowP + lensing measurements combined with the non-CMB observations, 
the flat $\phi$CDM model prefers $\alpha=0$, favoring the cosmological constant 
over dynamical dark energy. However, the nonflat $\phi$CDM model, 
when constrained using all the data, prefers closed spatial hypersurfaces 
and also mildly prefers dynamical dark energy with scalar field potential 
energy density parameter $\alpha > 0$.
Estimating 68.3\% and 95.4\% confidence limits of $\alpha$ using the 
information on the right-hand side with respect to the peak value (mode) 
in the marginalized 1-dimensional likelihood distribution, the mode 
$\pm$ $1\sigma$ ($2\sigma$) values for $\alpha$ are $0.113 \pm 0.094~(0.19)$.

More precisely, including the four sets of non-CMB data, we discover 
in the tilted flat $\phi$CDM model (bottom right panel of Table 
\ref{tab:para_flat_lensing}) that $\alpha < 0.22$ (at 2$\sigma$), which is
 more tightly restricted to $\alpha = 0$ and a cosmological constant than 
is the original \citet{Oobaetal2018c} finding of $\alpha < 0.28$ (at 2$\sigma$, 
the last column of their Table 2).\footnote{These results do not agree with 
those of earlier approximate analyses that used less reliable, and less, 
data, and indicated evidence for $\alpha$ deviating from $0$ by more than 
2$\sigma$ \citep{Solaetal2017b, Solaetal2017c}.}

However, perhaps the most interesting consequence of adding
in the four non-CMB data sets is the significant improvement of the 
evidence for nonflatness in the nonflat untilted $\phi$CDM model, 
with it increasing to $\Omega_k = - 0.0063 \pm 0.0020$, a more than 
3.1$\sigma$ deviation from flatness now, for the total data compilation in the 
bottom right panel of Table \ref{tab:para_nonflat_lensing}, compared to 
the 1.8$\sigma$ away from flatness for the CMB only case. This is 
now accompanied by very mild evidence favoring dynamical 
dark energy, see the right panel of Fig.\ \ref{fig:omm_alpha_omk}.
This result is compatible with and strengthens that of 
\cite{Oobaetal2017} who found $\Omega_k = - 0.006 \pm 0.003$ from 
Planck CMB data combined with a few BAO 
measurements. In favoring a closed geometry, the BAO measurements are the 
most important of the four non-CMB data sets.

From the total data combination (also accounting for CMB lensing data) in 
Tables \ref{tab:para_flat_lensing} and \ref{tab:para_nonflat_lensing}, 
$H_0$ measured using the flat tilted and the nonflat untilted 
$\phi$CDM models, $67.63 \pm 0.62$ and $67.36 \pm 0.72$ 
km s$^{-1}$ Mpc$^{-1}$, are quite consistent with each other to within
0.28$\sigma$ (of the quadrature sum of both the error bars).
These values agree with the median statistics measurement
$H_0=68\pm2.8$ km s$^{-1}$ Mpc$^{-1}$ 
\citep{ChenRatra2011a}, which agrees with earlier median statistics 
estimates \citep{Gottetal2001, Chenetal2003}. They are also
compatible with many recent
estimates of $H_0$ \citep{LHuillierShafieloo2017,Chenetal2017,Lukovicetal2016,Wangetal2017,LinIshak2017,DESCollaboration2017b,Yuetal2018,Haridasuetal2018a,Zhangetal2018,GomezValentAmendola2018,Haridasuetal2018b,daSilvaCavalcanti2018,Zhang2018}, but, as is well known, they are lower than the local expansion rate 
estimate of $H_0 = 73.48\pm1.66$ km s$^{-1}$ Mpc$^{-1}$
\citep{Riessetal2018}.\footnote{This local expansion rate estimate 
is 3.3$\sigma$ (3.4$\sigma$), of the quadrature sum of both the error bars,
larger than the $H_0$ value measured here using the flat tilted $\phi$CDM 
(nonflat untilted  $\phi$CDM) model. However, some other local expansion rate 
determinations find somewhat lower $H_0$'s with somewhat larger error bars
\citep{Rigaultetal2015, Zhangetal2017b, Dhawanetal2017, FernandezArenasetal2018}; for related discussions see \citet{Romanetal2017}, \citet{Kimetal2018}, and 
\citet{Jonesetal2018}.}

We find that $H_0$ and $\sigma_8$ (see discussion below) are the only 
measured parameters whose values are almost independent of the cosmological 
model (spatial curvature and tilt) used in the analysis. Measurements of other 
parameters determined by using the two $\phi$CDM models 
differ more significantly. Specifically, measurements determined using
the total data (also including CMB lensing) of $\Omega_m$, 
$\theta_\textrm{MC}$, $\ln(10^{10} A_s)$, $\Omega_b h^2$, $\tau$, and  
$\Omega_c h^2$, differ by 1.4$\sigma$, 1.9$\sigma$, 2.3$\sigma$, 
2.3$\sigma$, 2.6$\sigma$, and 4.6$\sigma$ (of the quadrature 
sum of both the error bars). For some parameters, especially 
$\Omega_c h^2$ as well as possibly $\tau$, $\Omega_b h^2$, and $A_s$, the 
model dependence of the value results in a much larger uncertainty than
that due to the statistical uncertainty in the given cosmological model. 
This was first detected when comparing measurements made using the    
flat tilted $\Lambda$CDM and the nonflat untilted $\Lambda$CDM model 
\citep{ParkRatra2018a} and is also present in the XCDM case 
\citep{ParkRatra2018b}. From Tables \ref{tab:para_flat_lensing} and 
\ref{tab:para_nonflat_lensing}, for the total data collection (also including 
CMB lensing), we find in the tilted flat $\phi$CDM (untilted nonflat 
$\phi$CDM) case $0.046 \le \tau \le 0.102$ ($0.098 \le \tau \le 0.146$) and
$0.02198 \le \Omega_b h^2 \le 0.02278$ ($0.02264 \le \Omega_b h^2 \le 0.02344$)
at 2$\sigma$, which are almost disjoint, and
$0.1142 \le \Omega_c h^2 \le 0.1194$ ($0.1073 \le \Omega_c h^2 \le 0.1113$),
which are completely separated from each other.
Current cosmological data cannot be used to measure $\Omega_c h^2$ or $\tau$
(and possibly some of the other cosmological parameters also) 
in a model independent way.

\begin{figure*}
\centering
\mbox{\includegraphics[width=65mm]{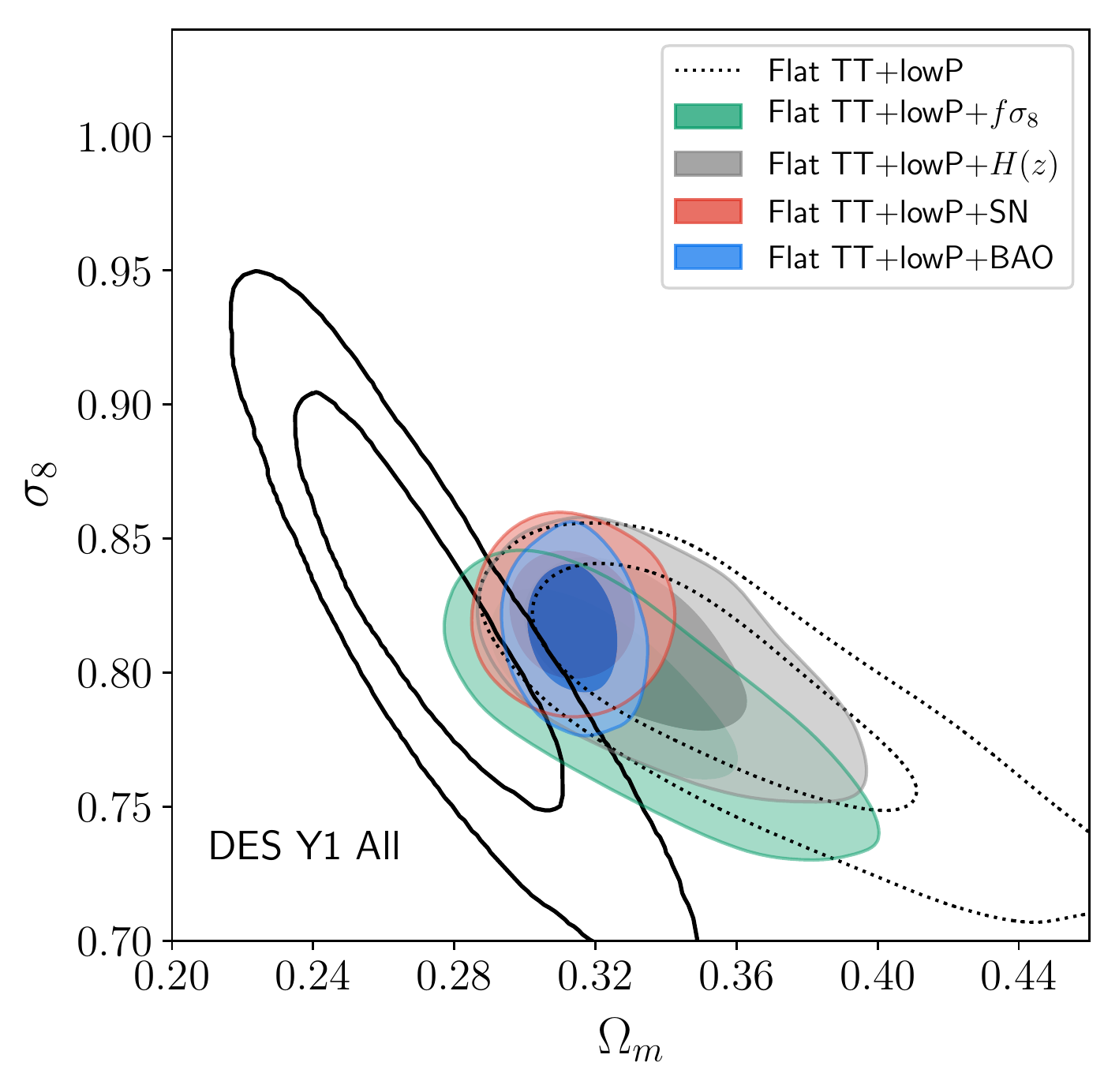}}
\mbox{\includegraphics[width=65mm]{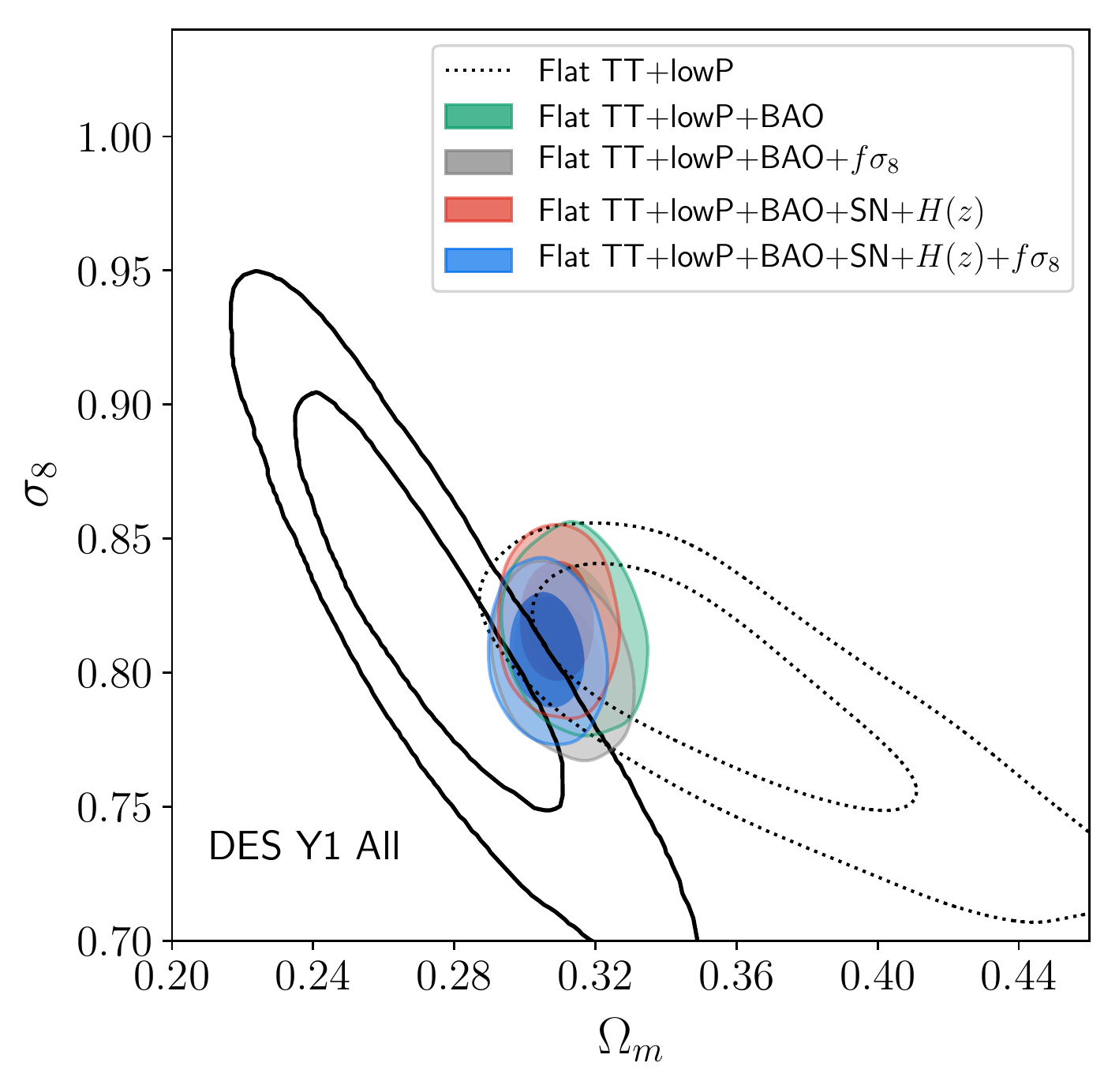}} \\
\mbox{\includegraphics[width=65mm]{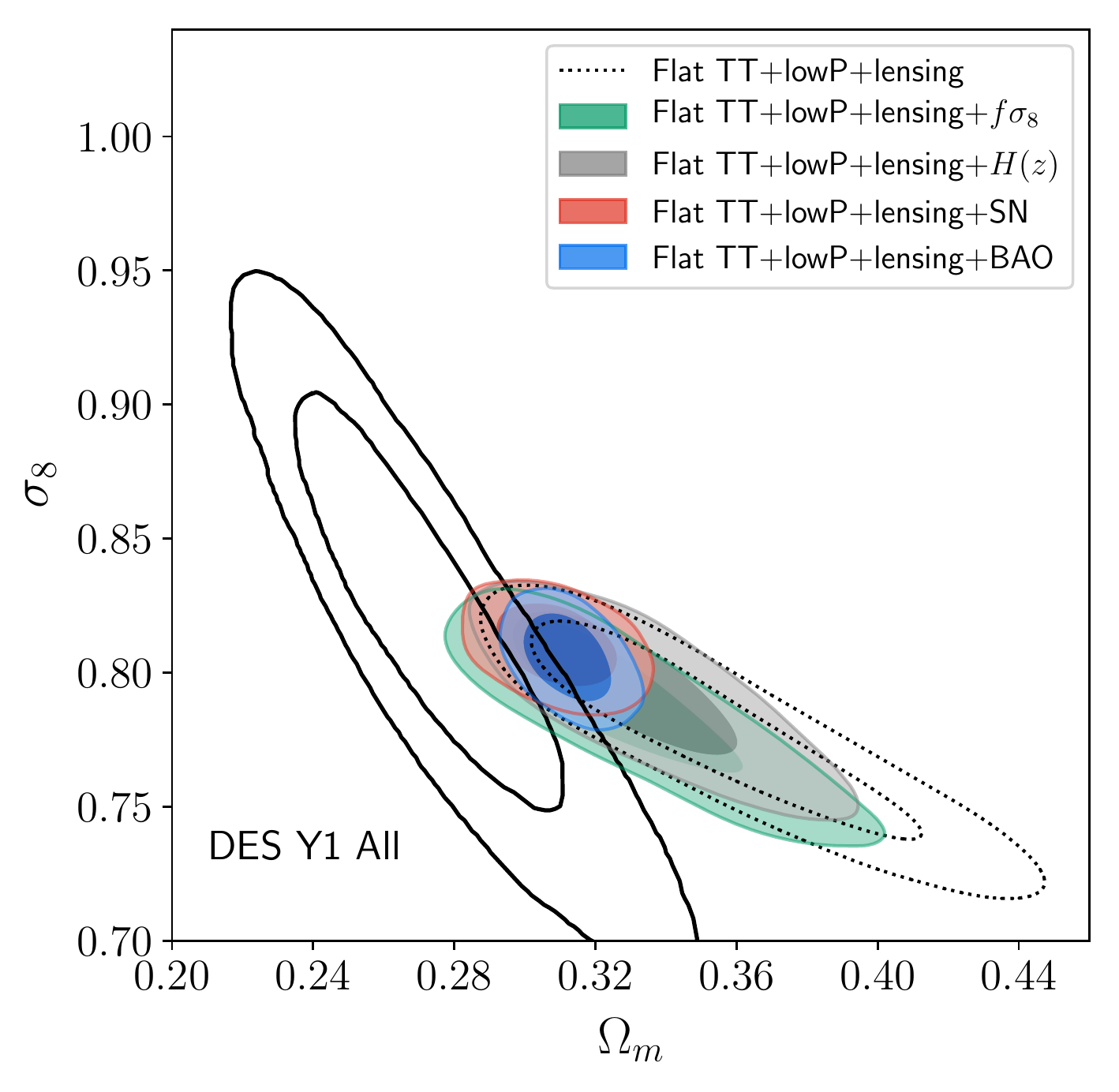}}
\mbox{\includegraphics[width=65mm]{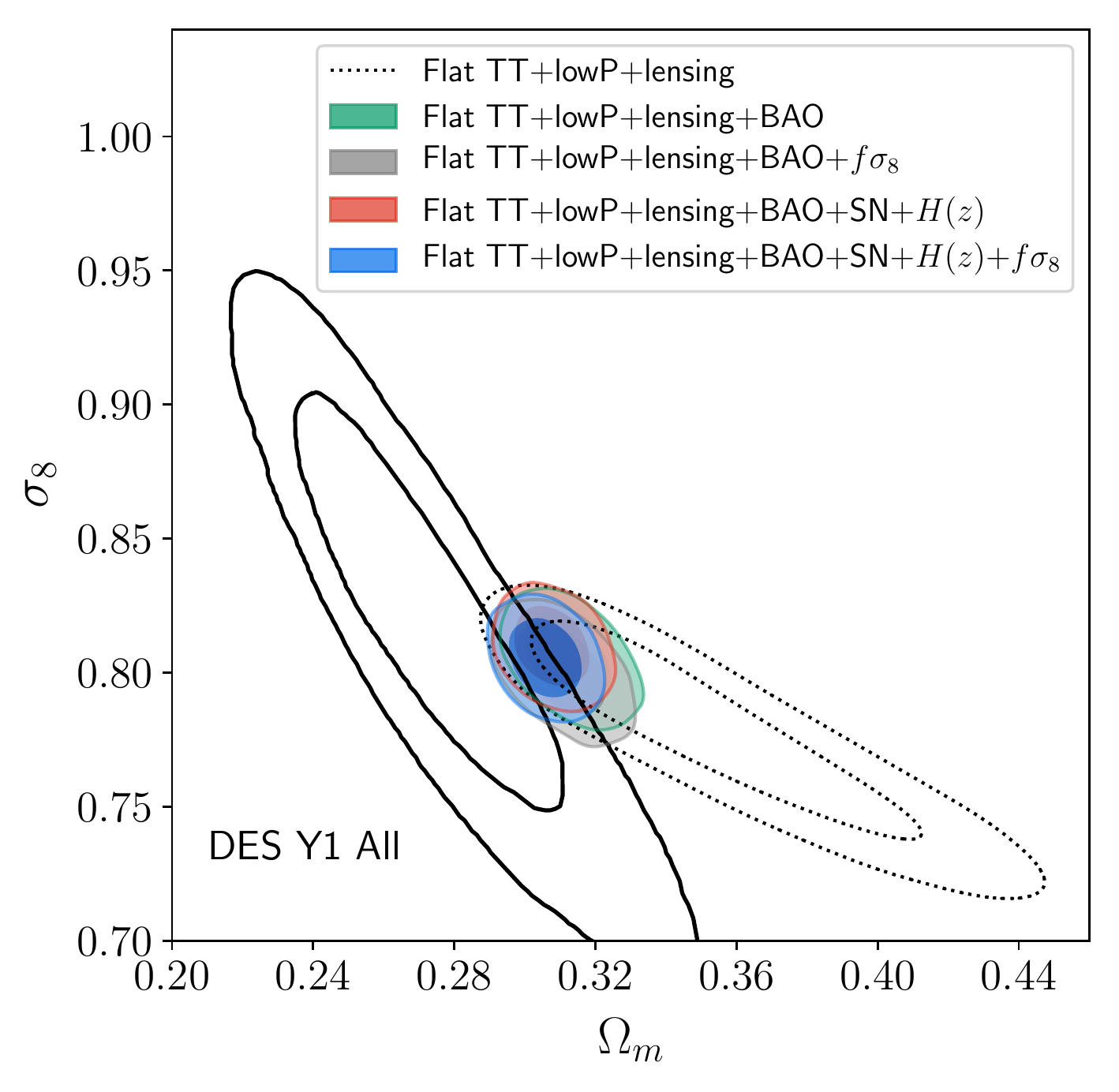}}
\caption{1$\sigma$ and 2$\sigma$ likelihood contours in the $\Omega_m$--$\sigma_8$ plane for
the tilted flat $\phi\textrm{CDM}$ model parameters constrained by using Planck CMB TT + lowP (+lensing), SN, BAO, $H(z)$, and $f\sigma_8$ data. For comparison,
in each panel the $\Lambda\textrm{CDM}$ model 1$\sigma$ and 2$\sigma$ 
constraint contours from the first-year results of the 
Dark Energy Survey (DES Y1 All) 
\citep{DESCollaboration2017a} are plotted as thick solid curves.
}
\label{fig:omm_sig8_flat}
\end{figure*}

\begin{figure*}
\centering
\mbox{\includegraphics[width=65mm]{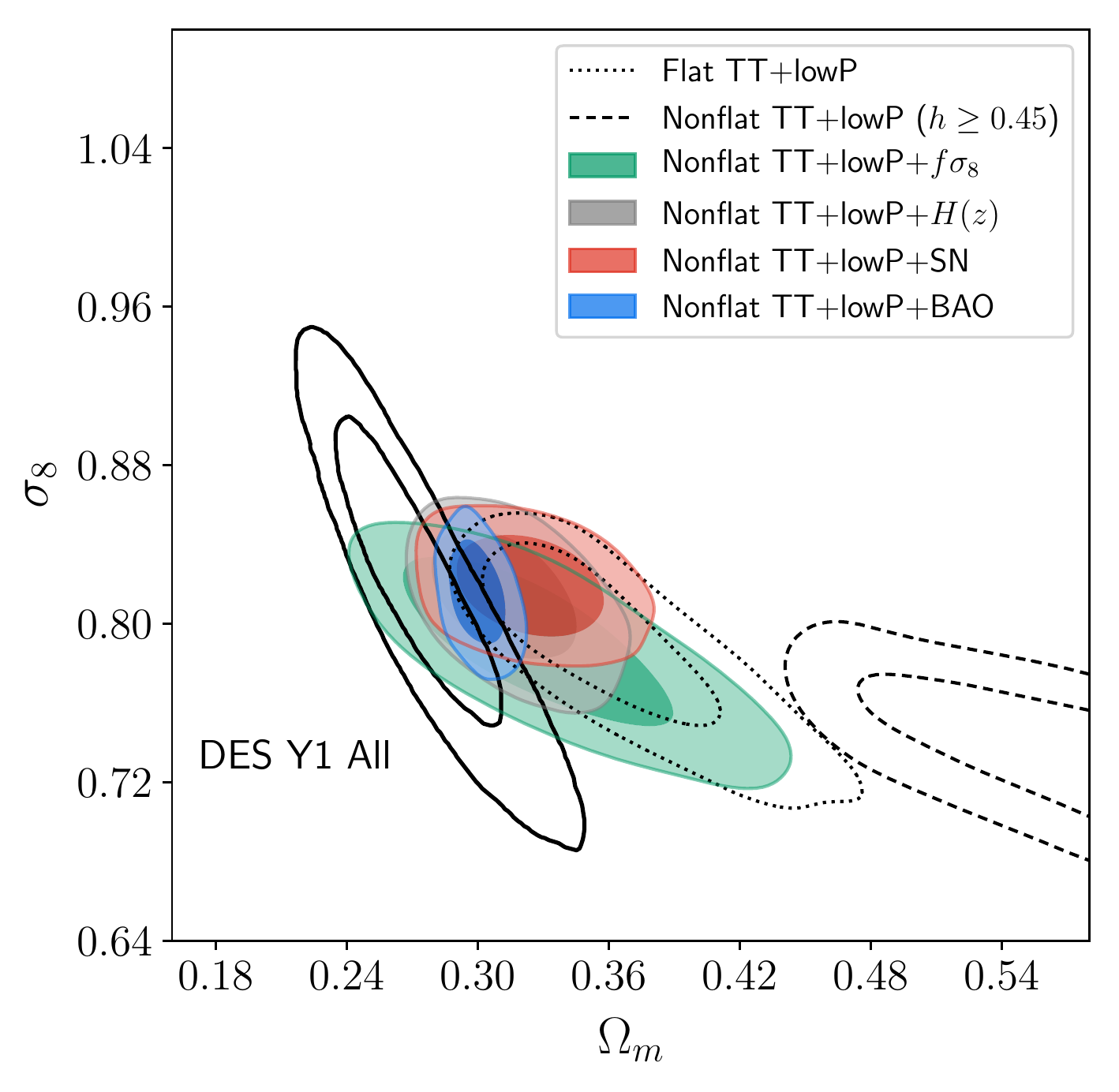}}
\mbox{\includegraphics[width=65mm]{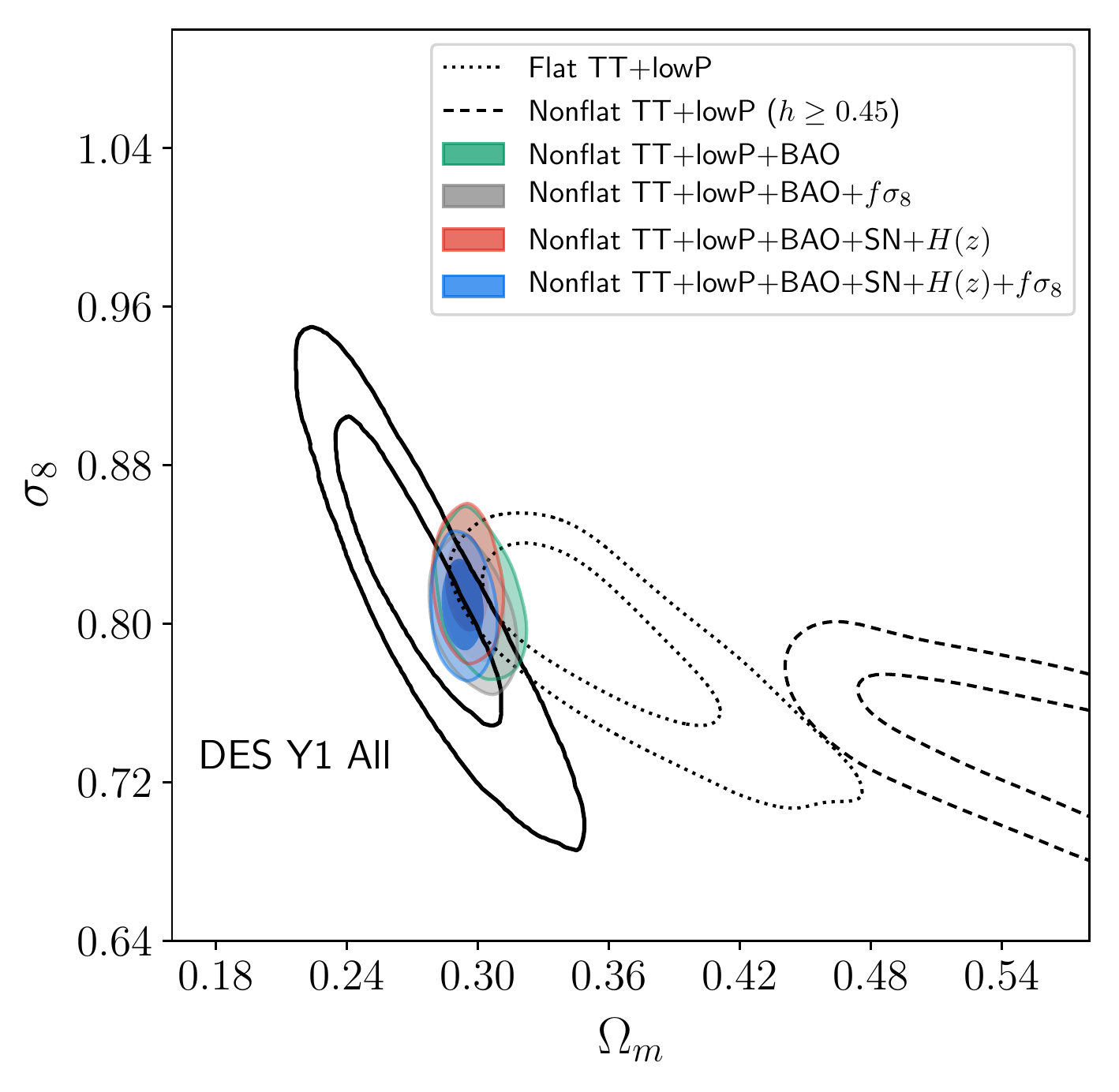}} \\
\mbox{\includegraphics[width=65mm]{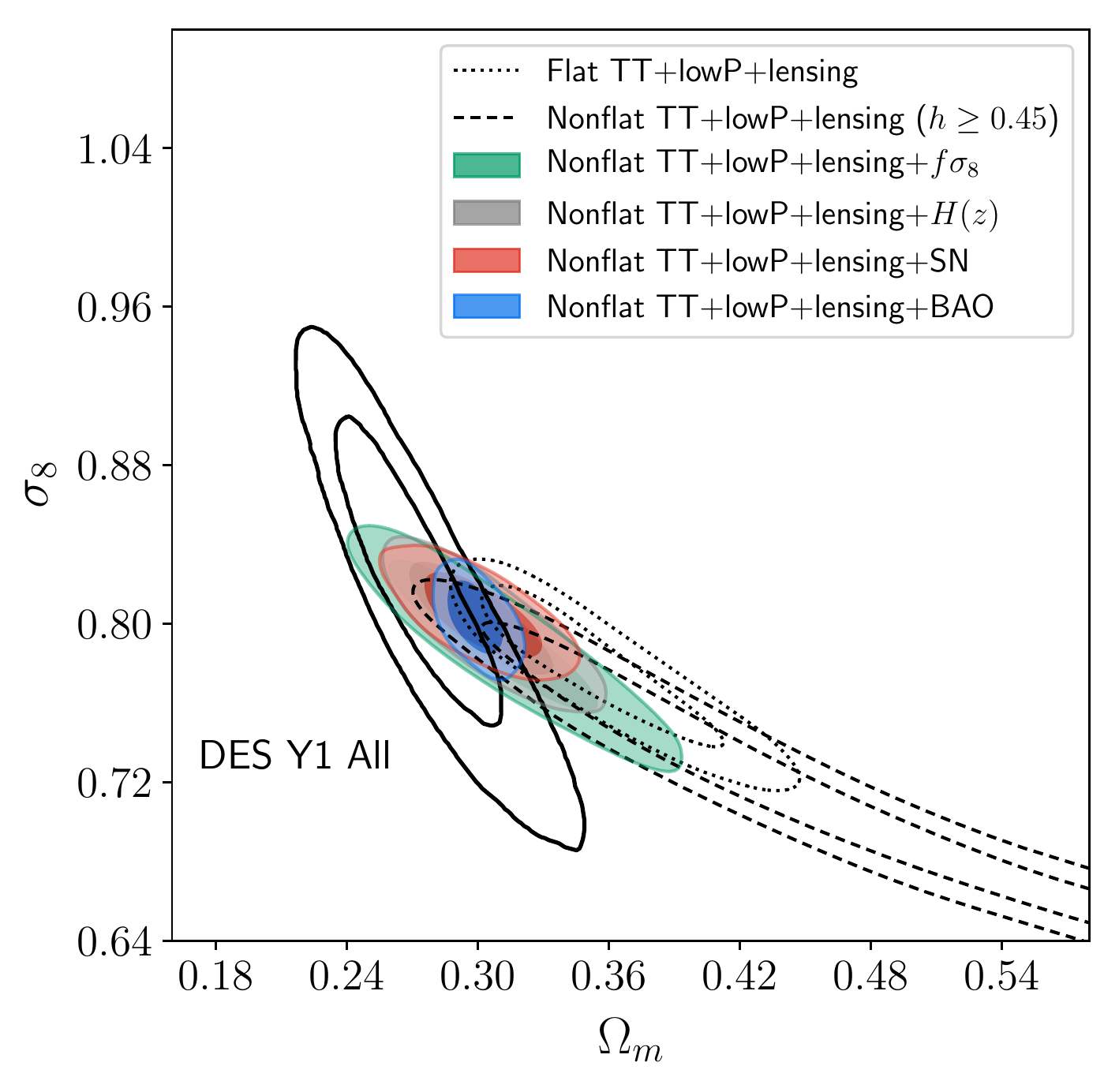}}
\mbox{\includegraphics[width=65mm]{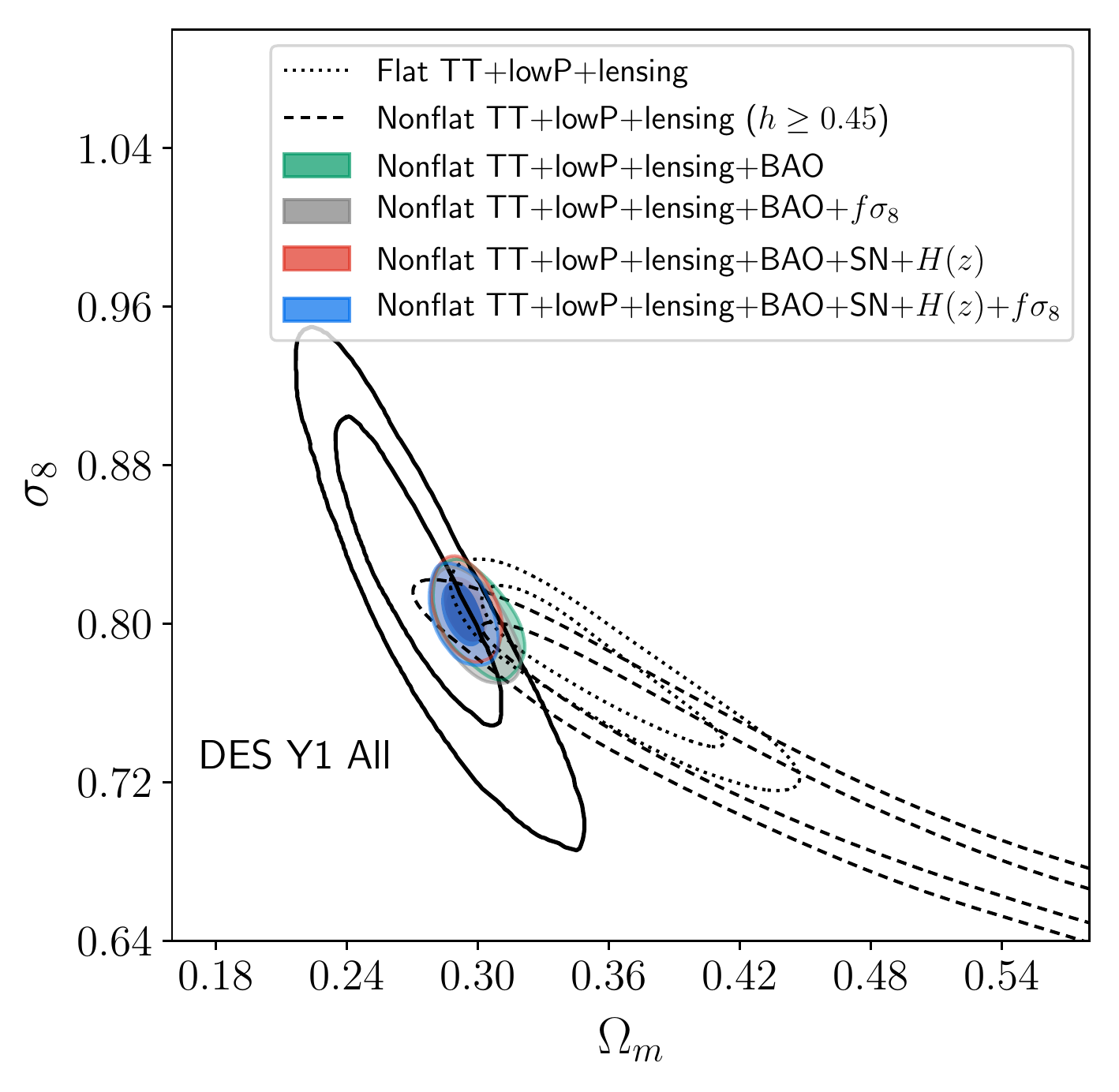}} \\
\caption{Same as Fig.\ \ref{fig:omm_sig8_flat} but for the untilted nonflat
$\phi\textrm{CDM}$ model.
}
\label{fig:omm_sig8_nonflat}
\end{figure*}

From the total data combination (also including CMB lensing data), $\sigma_8$'s 
measured using the two $\phi$CDM models, Tables 
\ref{tab:para_flat_lensing} and \ref{tab:para_nonflat_lensing}, 
agree to 0.034$\sigma$ (of the quadrature 
sum of both the error bars). Figures \ref{fig:omm_sig8_flat} and 
\ref{fig:omm_sig8_nonflat} show the marginalized two-dimensional 
$\Omega_m$--$\sigma_8$ likelihood 
contours for the 
flat tilted and nonflat untilted $\phi$CDM models constrained using the 
CMB and non-CMB data. In each panel we also show the $\Lambda\textrm{CDM}$ 
model constraints from a combined analysis of the first year galaxy 
clustering and weak lensing data of the Dark Energy Survey (DES Y1 All)
\citep{DESCollaboration2017a}, whose 1$\sigma$ confidence limits are 
$\Omega_m=0.264_{-0.019}^{+0.032}$ and $\sigma_8=0.807_{-0.041}^{+0.062}$.
The marginalized likelihood contours in the $\Omega_m$--$\sigma_8$ plane 
derived by adding each of the four sets of non-CMB data in turn to the CMB data 
are consistent with each other. Here 
the BAO data provide the most stringent constraints among the four sets of 
non-CMB data.

Although the $\sigma_8$ constraints from the flat tilted and nonflat untilted
$\phi$CDM analyses (ignoring and accounting for CMB lensing data) are 
consistent with the DES Y1 All measurements, the $\Omega_m$ bounds 
here prefer a 
larger value by about $1.3\sigma$ (of the quadrature sum of both the error 
bars) for the flat tilted $\phi$CDM case for the total data collection. 
We note that the best-fit point of the nonflat untilted $\phi$CDM model 
constrained by using the CMB data (also including lensing) combined with all 
non-CMB measurements enters inside
the 1$\sigma$ (68.3\%) region of the DES Y1 All likelihood distribution
(lower right panel of Fig.\ \ref{fig:omm_sig8_nonflat}), unlike the case 
for the 
flat tilted $\phi$CDM model (Fig.\ \ref{fig:omm_sig8_flat} lower right panel).


\begin{table*}
\caption{Untilted nonflat $\phi\textrm{CDM}$ model parameters constrained by using Planck TT + lowP, SN, BAO, $H(z)$, and $f\sigma_8$ data (mean and 68.3\% confidence limits).}
\begin{ruledtabular}
\begin{tabular}{lccc}
  Parameter               &  TT+lowP ($h\ge 0.45$)    &  TT+lowP+SN             &   TT+lowP+BAO    \\[+0mm]
 \hline \\[-2mm]
  $\Omega_b h^2$          &  $0.02329 \pm 0.00021$    &  $0.02313 \pm 0.00020$  &   $0.02306 \pm 0.00020$     \\[+1mm]
  $\Omega_c h^2$          &  $0.1095 \pm 0.0011$      &  $0.1094  \pm 0.0011$   &   $0.1096 \pm 0.0011$       \\[+1mm]
  $H_0$ [km s$^{-1}$ Mpc$^{-1}$] & $48.4 \pm 2.5$     &  $64.2 \pm 2.3$         &   $66.68 \pm 0.91$        \\[+1mm]
  $\tau$                  &  $0.108 \pm 0.021$        &  $0.132   \pm 0.018$    &   $0.138 \pm 0.016$       \\[+1mm]
  $\ln(10^{10} A_s)$      &  $3.126 \pm 0.042$        &  $3.174   \pm 0.036$    &   $3.185 \pm 0.033$        \\[+1mm]
  $\Omega_k$              &  $-0.074 \pm 0.016$       &  $-0.0162 \pm 0.0064$   &   $-0.0067 \pm 0.0023$     \\[+1mm]
  $\alpha$~[95.4\%~C.L.]   & $<1.81$                   &  $<0.20$                &   $<0.46$                  \\[+1mm]
 \hline \\[-2mm]
  $100\theta_\textrm{MC}$ &  $1.04217 \pm 0.00042$    &  $1.04209 \pm 0.00042$  &   $1.04204 \pm 0.00041$     \\[+1mm]
  $\Omega_m$              &  $0.573 \pm 0.057$        &  $0.324 \pm 0.023$      &   $0.2999 \pm 0.0086$     \\[+1mm]
  $\sigma_8$              &  $0.733 \pm 0.026$        &  $0.819 \pm 0.017$      &   $0.815 \pm 0.017$       \\[+1mm]
    \hline \hline \\[-2mm]
  Parameter               &  TT+lowP+$H(z)$           &  TT+lowP+SN+BAO         &  TT+lowP+SN+BAO+$H(z)$      \\[+0mm]
 \hline \\[-2mm]
  $\Omega_b h^2$          &  $0.02309 \pm 0.00021$    &  $0.02306 \pm 0.00019$  &  $0.02307  \pm 0.00019$       \\[+1mm]
  $\Omega_c h^2$          &  $0.1098 \pm 0.0011$      &  $0.1096  \pm 0.0011$   &  $0.1097   \pm 0.0011$      \\[+1mm]
  $H_0$ [km s$^{-1}$ Mpc$^{-1}$]  & $65.3 \pm 2.2$    &  $67.16 \pm 0.70$       &  $67.24 \pm 0.72$             \\[+1mm]
  $\tau$                  &  $0.136 \pm 0.017$        &  $0.134   \pm 0.016$    &  $0.136    \pm 0.016$         \\[+1mm]
  $\ln(10^{10} A_s)$      &  $3.182 \pm 0.035$        &  $3.178   \pm 0.033$    &  $3.182    \pm 0.033$         \\[+1mm]
  $\Omega_k$              &  $-0.0100 \pm 0.0052$     &  $-0.0073 \pm 0.0020$   &  $-0.0068  \pm 0.0019$        \\[+1mm]
  $\alpha$~[95.4\%~C.L.]   & $<0.60$                  &  $<0.28$                &  $<0.29$                     \\[+1mm]
 \hline \\[-2mm]
  $100\theta_\textrm{MC}$ &  $1.04204 \pm 0.00042$    &  $1.04202 \pm 0.00041$  &  $1.04206  \pm 0.00041$       \\[+1mm]
  $\Omega_m$              &  $0.314 \pm 0.022$        &  $0.2957 \pm 0.0063$    &  $0.2952 \pm 0.0066$            \\[+1mm]
  $\sigma_8$              &  $0.813 \pm 0.022$        &  $0.819 \pm 0.016$      &  $0.820 \pm 0.016$           \\[+1mm]
   \hline \hline \\[-2mm]
     Parameter            &  TT+lowP+$f\sigma_8$      & TT+lowP+BAO+$f\sigma_8$ &  TT+lowP+SN+BAO+$H(z)$+$f\sigma_8$      \\[+0mm]
 \hline \\[-2mm]
  $\Omega_b h^2$          &  $0.02309 \pm 0.00021$    &  $0.02304 \pm 0.00020$  &  $0.02307  \pm 0.00020$       \\[+1mm]
  $\Omega_c h^2$          &  $0.1091  \pm 0.0010$     &  $0.1093  \pm 0.0011$   &  $0.1092   \pm 0.0011$      \\[+1mm]
  $H_0$ [km s$^{-1}$ Mpc$^{-1}$]  & $63.8 \pm 4.0$    &  $66.91  \pm 0.91$      &  $67.37 \pm 0.71$             \\[+1mm]
  $\tau$                  &  $0.122 \pm 0.019$        &  $0.128   \pm 0.017$    &  $0.126    \pm 0.016$         \\[+1mm]
  $\ln(10^{10} A_s)$      &  $3.152 \pm 0.039$        &  $3.164   \pm 0.033$    &  $3.161    \pm 0.033$         \\[+1mm]
  $\Omega_k$              &  $-0.0136 \pm 0.0090$     &  $-0.0063 \pm 0.0023$   &  $-0.0065  \pm 0.0020$        \\[+1mm]
  $\alpha$~[95.4\%~C.L.]  &  $<0.86$                  &  $<0.46$                &  $<0.30$                     \\[+1mm]
 \hline \\[-2mm]
  $100\theta_\textrm{MC}$ &  $1.04206 \pm 0.00041$    &  $1.04204 \pm 0.00042$  &  $1.04206  \pm 0.00041$       \\[+1mm]
  $\Omega_m$              &  $0.330 \pm 0.042$        &  $0.2972 \pm 0.0084$    &  $0.2930 \pm 0.0062$            \\[+1mm]
  $\sigma_8$              &  $0.789 \pm 0.028$        &  $0.805 \pm 0.016$      &  $0.809 \pm 0.015$                \\[+0mm]
\end{tabular}
\end{ruledtabular}
\label{tab:para_nonflat}
\end{table*}

Table \ref{tab:chi2_phiCDM} lists $\chi^2$'s
for the best-fit flat tilted and nonflat untilted $\phi\textrm{CDM}$ 
models. The best-fit position in parameter space is found using Powell's 
minimization method, an efficient algorithm to locate the $\chi^2$ minimum.
We list the $\chi^2$ contribution of each data set. 
The total $\chi^2$ is the sum of the 
individual ones from the high-$\ell$ CMB TT 
likelihood ($\chi_{\textrm{PlikTT}}^2$),
the low-$\ell$ CMB power spectra of temperature and polarization
($\chi_{\textrm{lowTEB}}^2$), lensing ($\chi_{\textrm{lensing}}^2$),
SN ($\chi_{\textrm{SN}}^2$), BAO ($\chi_{\textrm{BAO}}^2$),
$f\sigma_8$ ($\chi_{f\sigma_8}^2$), $H(z)$ data ($\chi_{H(z)}^2$), 
and from the foreground nuisance parameters
($\chi_{\textrm{prior}}^2$). As a result of 
the nonstandard normalization of the Planck data likelihoods,
the number of CMB degrees of freedom is ambiguous. 
Thus, the absolute value of $\chi^2$ for the Planck CMB data is arbitrary, 
and only the relative difference between $\chi^2$ of one model and another 
is meaningful for the Planck data. For the non-CMB data, the 
degrees of freedom are 10, 15, 31, 1042\footnote{
This is the number of degrees of freedom for the flat $\Lambda$CDM model,
given by the number of data points (1048) minus the number of parameters such
as the matter density ($\Omega_m$) and the five internal nuisance
parameters.} for the $f\sigma_8$, BAO, $H(z)$, SN 
observations, respectively, resulting in 1098 degrees of freedom all together. 
The reduced $\chi^2$'s for the individual non-CMB data sets are 
$\chi^2 / \nu \lesssim 1$. There are 189 points in the Planck TT + lowP 
(binned) CMB data anisotropy angular power spectrum and 197 points when 
the CMB lensing measurements are included.


\begin{table*}
\caption{Untilted nonflat $\phi\textrm{CDM}$ model parameters constrained by using Planck TT + lowP + lensing, SN, BAO, $H(z)$, and $f\sigma_8$ data (mean and 68.3\% confidence limits).}
\begin{ruledtabular}
\begin{tabular}{lccc}
  Parameter               &  TT+lowP+lensing ($h\ge 0.45$)  &  TT+lowP+lensing+SN     &   TT+lowP+lensing+BAO    \\[+0mm]
 \hline \\[-2mm]
  $\Omega_b h^2$          &  $0.02303 \pm 0.00020$    &  $0.02305 \pm 0.00020$  &   $0.02302 \pm 0.00020$     \\[+1mm]
  $\Omega_c h^2$          &  $0.1091 \pm 0.0011$      &  $0.1092  \pm 0.0011$   &   $0.1095 \pm 0.0011$       \\[+1mm]
  $H_0$ [km s$^{-1}$ Mpc$^{-1}$] & $55.2 \pm 6.1$     &  $66.5 \pm 2.1$         &   $66.73 \pm 0.93$        \\[+1mm]
  $\tau$                  &  $0.105 \pm 0.020$        &  $0.117   \pm 0.015$    &   $0.126 \pm 0.014$       \\[+1mm]
  $\ln(10^{10} A_s)$      &  $3.118 \pm 0.039$        &  $3.142   \pm 0.031$    &   $3.161 \pm 0.026$        \\[+1mm]
  $\Omega_k$              &  $-0.032 \pm 0.017$       &  $-0.0098 \pm 0.0055$   &   $-0.0062 \pm 0.0023$     \\[+1mm]
  $\alpha$~[95.4\%~C.L.]  &  $<3.30$                  &  $<0.26$                &   $<0.46$                  \\[+1mm]
 \hline \\[-2mm]
  $100\theta_\textrm{MC}$ &  $1.04213 \pm 0.00041$    &  $1.04213 \pm 0.00041$  &   $1.04210 \pm 0.00041$     \\[+1mm]
  $\Omega_m$              &  $0.45 \pm 0.10$          &  $0.301 \pm 0.018$      &   $0.2992 \pm 0.0086$     \\[+1mm]
  $\sigma_8$              &  $0.716 \pm 0.049$        &  $0.804 \pm 0.014$      &   $0.803 \pm 0.012$       \\[+1mm]
    \hline \hline \\[-2mm]
  Parameter               &  TT+lowP+lensing+$H(z)$   &  TT+lowP+lensing+SN+BAO &  TT+lowP+lensing+SN+BAO+$H(z)$      \\[+0mm]
 \hline \\[-2mm]
  $\Omega_b h^2$          &  $0.02303 \pm 0.00020$    &  $0.02302 \pm 0.00020$  &  $0.02303  \pm 0.00020$       \\[+1mm]
  $\Omega_c h^2$          &  $0.1095 \pm 0.0010$      &  $0.1094  \pm 0.0011$   &  $0.1095   \pm 0.0010$      \\[+1mm]
  $H_0$ [km s$^{-1}$ Mpc$^{-1}$]  & $66.5 \pm 2.4$    &  $67.16 \pm 0.72$       &  $67.25 \pm 0.74$             \\[+1mm]
  $\tau$                  &  $0.125 \pm 0.015$        &  $0.122   \pm 0.013$    &  $0.123    \pm 0.012$         \\[+1mm]
  $\ln(10^{10} A_s)$      &  $3.158 \pm 0.029$        &  $3.153   \pm 0.025$    &  $3.155    \pm 0.024$         \\[+1mm]
  $\Omega_k$              &  $-0.0069 \pm 0.0045$     &  $-0.0071 \pm 0.0020$   &  $-0.0065  \pm 0.0020$        \\[+1mm]
  $\alpha$~[95.4\%~C.L.]   & $<0.60$                  &  $<0.31$                &  $<0.32$                    \\[+1mm]
 \hline \\[-2mm]
  $100\theta_\textrm{MC}$ &  $1.04210 \pm 0.00041$    &  $1.04209 \pm 0.00041$  &  $1.04209  \pm 0.00042$       \\[+1mm]
  $\Omega_m$              &  $0.302 \pm 0.022$        &  $0.2950 \pm 0.0064$    &  $0.2945 \pm 0.0065$            \\[+1mm]
  $\sigma_8$              &  $0.801 \pm 0.019$        &  $0.807 \pm 0.011$      &  $0.807 \pm 0.011$           \\[+1mm]
   \hline \hline \\[-2mm]
     Parameter            &  TT+lowP+lensing+$f\sigma_8$  & TT+lowP+lensing+BAO+$f\sigma_8$ &  TT+lowP+lensing+SN+BAO+$H(z)$+$f\sigma_8$      \\[+0mm]
 \hline \\[-2mm]
  $\Omega_b h^2$          &  $0.02304 \pm 0.00019$    &  $0.02303 \pm 0.00020$  &  $0.02304  \pm 0.00020$       \\[+1mm]
  $\Omega_c h^2$          &  $0.1091  \pm 0.0011$     &  $0.1091  \pm 0.0010$   &  $0.1093  \pm 0.0010$      \\[+1mm]
  $H_0$ [km s$^{-1}$ Mpc$^{-1}$]  & $65.8 \pm 3.4$    &  $66.77  \pm 0.91$      &  $67.36 \pm 0.72$             \\[+1mm]
  $\tau$                  &  $0.119 \pm 0.017$        &  $0.125   \pm 0.013$    &  $0.122    \pm 0.012$         \\[+1mm]
  $\ln(10^{10} A_s)$      &  $3.146 \pm 0.034$        &  $3.158   \pm 0.026$    &  $3.152    \pm 0.024$         \\[+1mm]
  $\Omega_k$              &  $-0.0087 \pm 0.0065$     &  $-0.0062 \pm 0.0023$   &  $-0.0063  \pm 0.0020$        \\[+1mm]
  $\alpha$~[95.4\%~C.L.]  &  $<0.79$                  &  $<0.48$                &  $<0.31$                      \\[+1mm]
 \hline \\[-2mm]
  $100\theta_\textrm{MC}$ &  $1.04209 \pm 0.00041$    &  $1.04208 \pm 0.00041$  &  $1.04210  \pm 0.00041$       \\[+1mm]
  $\Omega_m$              &  $0.309 \pm 0.032$        &  $0.2981 \pm 0.0084$    &  $0.2931 \pm 0.0064$            \\[+1mm]
  $\sigma_8$              &  $0.792 \pm 0.025$        &  $0.799 \pm 0.012$      &  $0.805 \pm 0.011$                \\[+0mm]
\end{tabular}
\end{ruledtabular}
\label{tab:para_nonflat_lensing}
\end{table*}

In the last column of Table \ref{tab:chi2_phiCDM}, we list $\Delta\chi^2$,
the excess $\chi^2$ of the best-fit seven parameter $\phi$CDM model relative 
to the $\chi^2$ of the related six parameter $\Lambda\textrm{CDM}$ 
model that is constrained by using
the same data combination. The minimum $\chi^2$ values for the 
$\Lambda\textrm{CDM}$ and XCDM models are presented in Tables 7 and 8 
of \citet{ParkRatra2018b}. These models are nested; the seven parameter 
flat tilted $\phi$CDM (nonflat untilted $\phi$CDM) model reduces to the 
six parameter flat tilted $\Lambda$CDM (nonflat untilted $\Lambda$CDM) model 
when $\alpha$ goes to zero.\footnote{This is also true of the XCDM 
parameterization when the equation of state parameter $w$ goes to $-1$.} 
Here the ambiguity in the number of Planck CMB data degrees of freedom is 
no longer an obstacle to converting the $\Delta\chi^2$ to 
a relative goodness of fit probability. From $\sqrt{-\Delta \chi^2}$, for 
the complete data (accounting for CMB lensing), for a single 
additional free parameter, 
we find that the flat tilted $\phi$CDM (nonflat untilted $\phi$CDM) model
is a 0.40$\sigma$ (0.93$\sigma$) better fit to the data than is the flat tilted 
$\Lambda$CDM (nonflat untilted $\Lambda$CDM) model.\footnote{XCDM does 
not do as well as $\phi$CDM, with the flat tilted XCDM (nonflat untilted 
XCDM) parameterization being a 0.28$\sigma$ (0.87$\sigma$) better fit to 
the data than is the flat tilted $\Lambda$CDM (nonflat untilted $\Lambda$CDM) 
model \citep{ParkRatra2018b}. We emphasize that nonflat untilted 
$\Lambda$CDM does not fit as well as flat tilted $\Lambda$CDM,
although as discussed in \citet{Oobaetal2018a, Oobaetal2017, Oobaetal2018b}
and \citet{ParkRatra2018a, ParkRatra2018b}, it is not known how to transform 
this into a relative probability because the Planck 2015 CMB data number 
of degrees of freedom is unavailable and the two six 
parameter models are not nested.}
These findings are compatible with those of \citet{Oobaetal2018c} and
\citet{Oobaetal2017}. 

Of all three flat cases, tilted flat $\phi$CDM best fits the 
combined data (although there is no significant difference between all
three cases), but at a lower level of significance than the 1.3$\sigma$ found 
by \citet{Oobaetal2018c} using a very small sample of non-CMB data compared
to what we have used here, and far from the 3 or 4$\sigma$ result 
found in earlier approximate analyses by \citet{Solaetal2017b, Solaetal2017c}.
While the tilted flat $\phi$CDM and XCDM cases do not provide a much
better fit to the data, available data  
allow for the possibility that dark energy is dynamical.

It is clear that relative to the flat models, in terms of 
$\Delta\chi^2$ values, the nonflat models do a worse job of fitting the 
higher-$\ell$ $C_\ell$'s than they do at fitting the lower-$\ell$ $C_\ell$'s.
However, the models are not nested so it is not possible to turn these 
differences into relative goodness of fit probabilities (as the number of 
degrees of freedom of the Planck 2015
data is ambiguous). We note that there have been 
studies on systematic differences between constraints determined from 
the higher-$\ell$ and the lower-$\ell$ Planck 2015 data \citep{Addisonetal2016,PlanckCollaboration2017}. Also, in the 
tilted flat $\Lambda$CDM model, there seem to be inconsistencies between the 
higher-$\ell$ Planck and the South Pole Telescope 
CMB data \citep{Ayloretal2017}. Possibly, if these differences are real, when
they are resolved this could result in a decrease of the 
$\Delta\chi^2$'s found here.

\begin{figure*}[htbp]
\centering
\mbox{\includegraphics[width=100mm,bb=15 230 570 750]{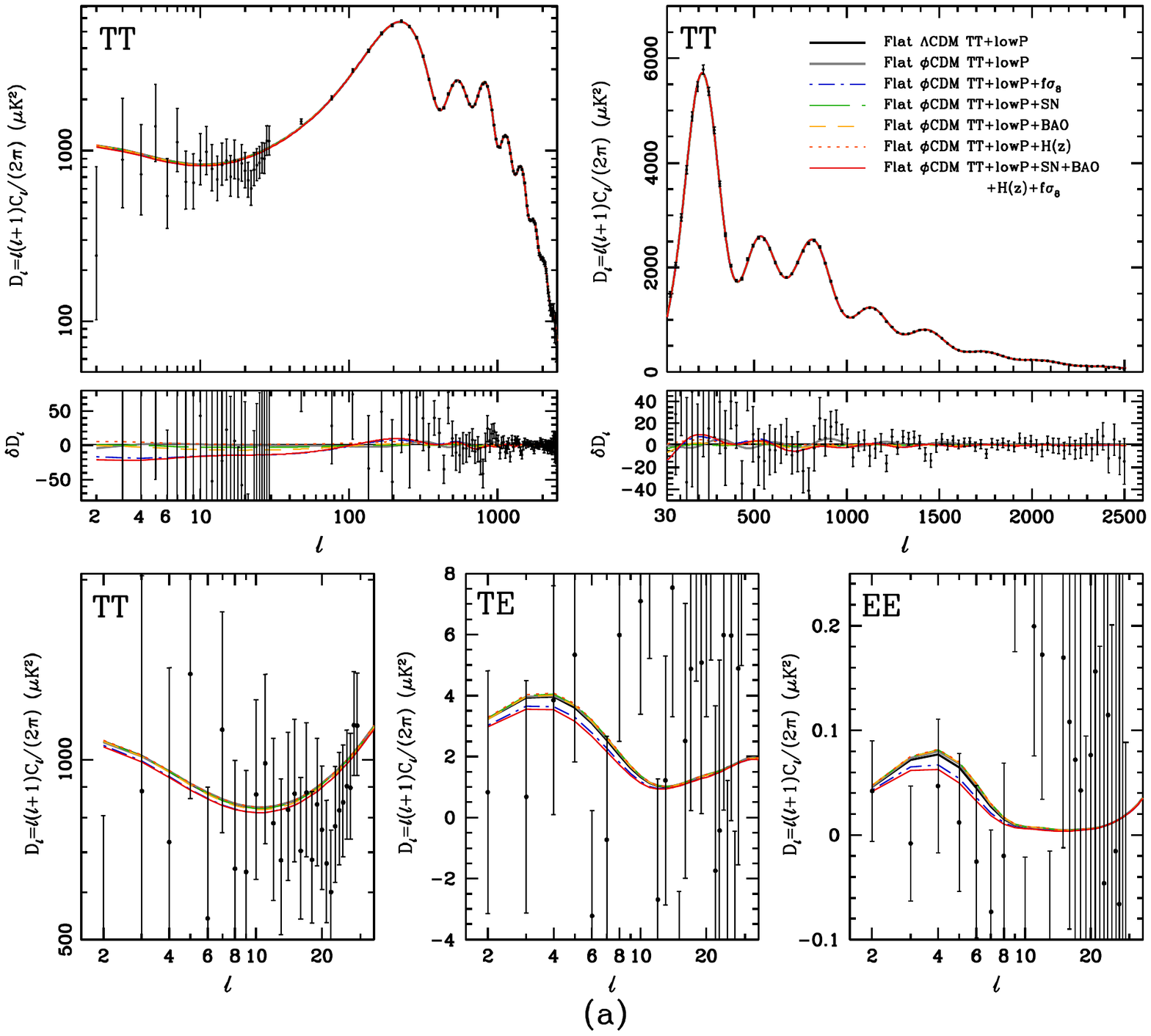}} \\
\mbox{\includegraphics[width=100mm,bb=15 210 570 750]{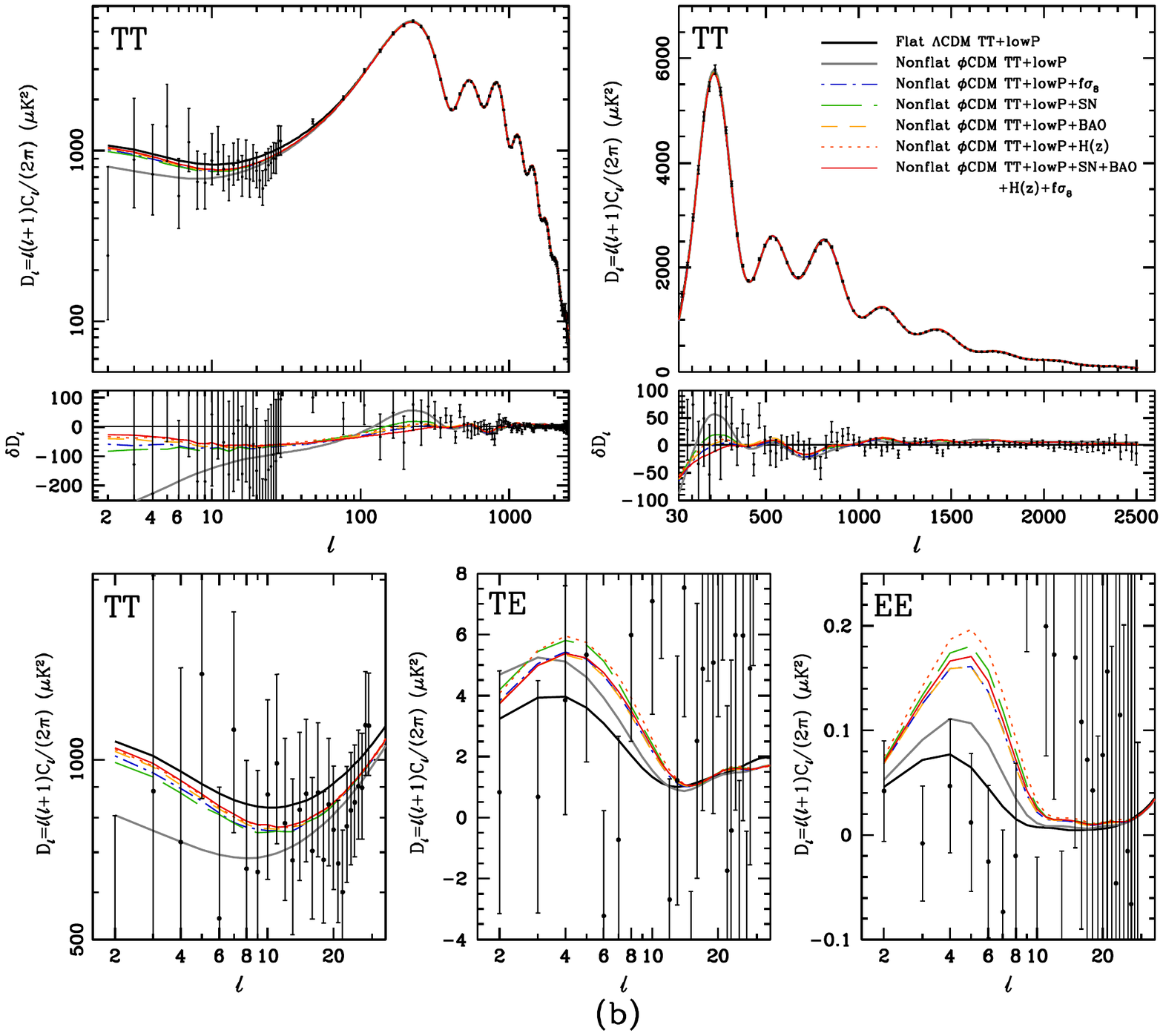}}
\caption{Best-fit CMB anisotropy angular power spectra of (a) flat tilted (top five panels) and (b) nonflat
untilted $\phi$CDM models (bottom five panels) constrained
by using the Planck 2015 CMB TT + lowP data (ignoring the lensing data) in conjunction with BAO, $H(z)$, SN, and $f\sigma_8$ data.
For comparison, the best-fit angular power spectra of the flat tilted $\Lambda\textrm{CDM}$ model are shown as
black curves. $\delta D_\ell$ residuals for the TT power spectra are shown 
with respect to the flat tilted $\Lambda\textrm{CDM}$ power spectrum that best fits the TT + lowP data.
}
\label{fig:ps_cmb}
\end{figure*}

\begin{figure*}[htbp]
\centering
\mbox{\includegraphics[width=100mm,bb=15 230 570 750]{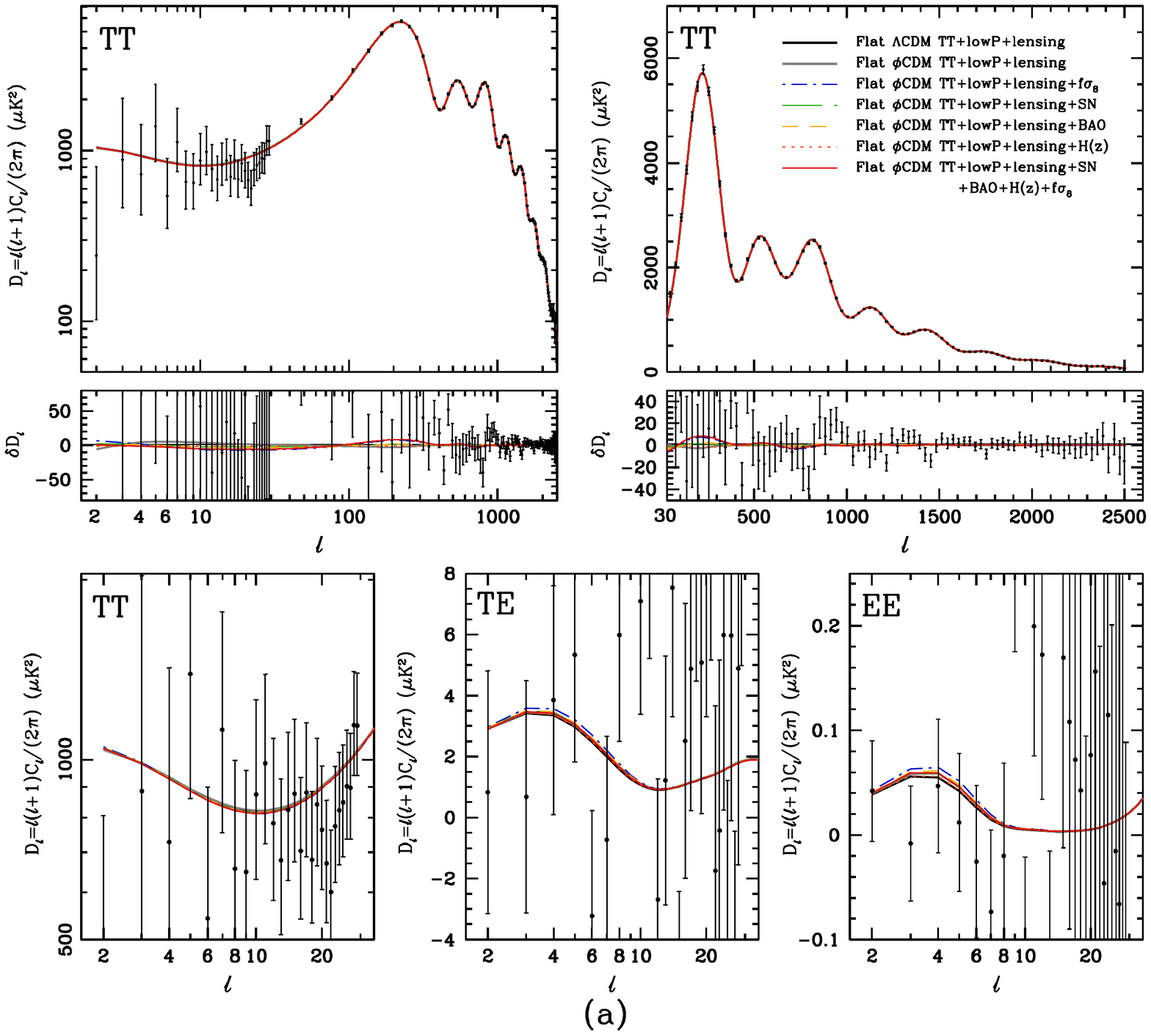}} \\
\mbox{\includegraphics[width=100mm,bb=15 210 570 750]{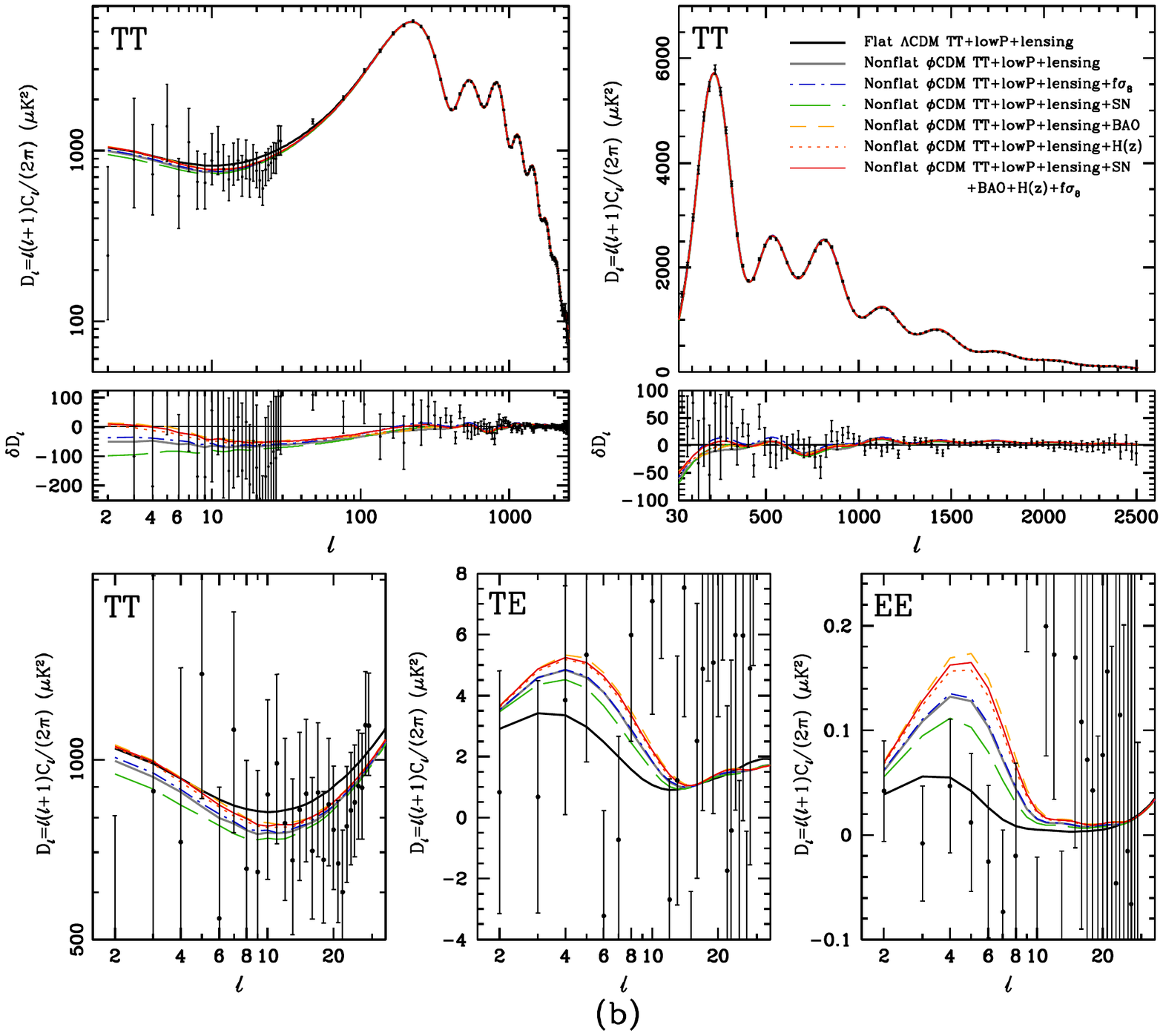}}
\caption{Same as Fig. \ref{fig:ps_cmb} but now accounting for the CMB lensing data. $\delta D_\ell$ residuals of the TT power spectra are shown with respect to the flat tilted $\Lambda\textrm{CDM}$ power spectrum that best fits the TT + lowP + lensing data.
}
\label{fig:ps_cmb_lensing}
\end{figure*}

Figures \ref{fig:ps_cmb} and \ref{fig:ps_cmb_lensing} plot the CMB 
high-$\ell$ TT, and the low-$\ell$ TT, TE, EE power spectra of the best-fit 
flat tilted and nonflat untilted $\phi$CDM dynamical dark energy inflation 
models, ignoring and accounting for the lensing data, respectively. 
The best-fit flat tilted $\phi$CDM models favored by the CMB and 
non-CMB data are in good agreement with the observed CMB power spectra at all $\ell$
\citep[this is also the case for the best-fit flat tilted XCDM parameterization,][]{ParkRatra2018b}.
However, similar to the nonflat $\Lambda\textrm{CDM}$ and XCDM cases
studied in \citet{ParkRatra2018a, ParkRatra2018b}, the nonflat untilted
$\phi$CDM 
model constrained with the Planck 2015 CMB anisotropy data and each non-CMB 
data set generally provides a poorer fit to the low-$\ell$ EE power spectrum 
while it provides a better fit to the low-$\ell$ TT power spectrum (see the 
bottom left 
panel of Figs.\ \ref{fig:ps_cmb} and \ref{fig:ps_cmb_lensing}). The 
best-fit $C_\ell$ model power spectra shapes relative to the Planck 
CMB data points are compatible with the $\chi^2$ values listed in 
Table \ref{tab:chi2_phiCDM}. For example, the best-fit untilted nonflat 
$\phi$CDM model constrained by using the TT + lowP + $H(z)$ data has 
an EE power spectrum shape at low-$\ell$ that is most deviant from the Planck 
data and a corresponding value of $\chi^2_{\textrm{lowTEB}}=10500.10$
that is larger by $3.69$ relative to the best-fit tilted
flat $\Lambda\textrm{CDM}$ model
$\chi^2_{\textrm{lowTEB}}=10496.41$ for the TT + lowP data \citep[see Table 
\ref{tab:chi2_phiCDM} here and Table 7 of][]{ParkRatra2018b}.

\begin{figure*}[h!]
\mbox{\includegraphics[width=85mm,bb=30 170 500 620]{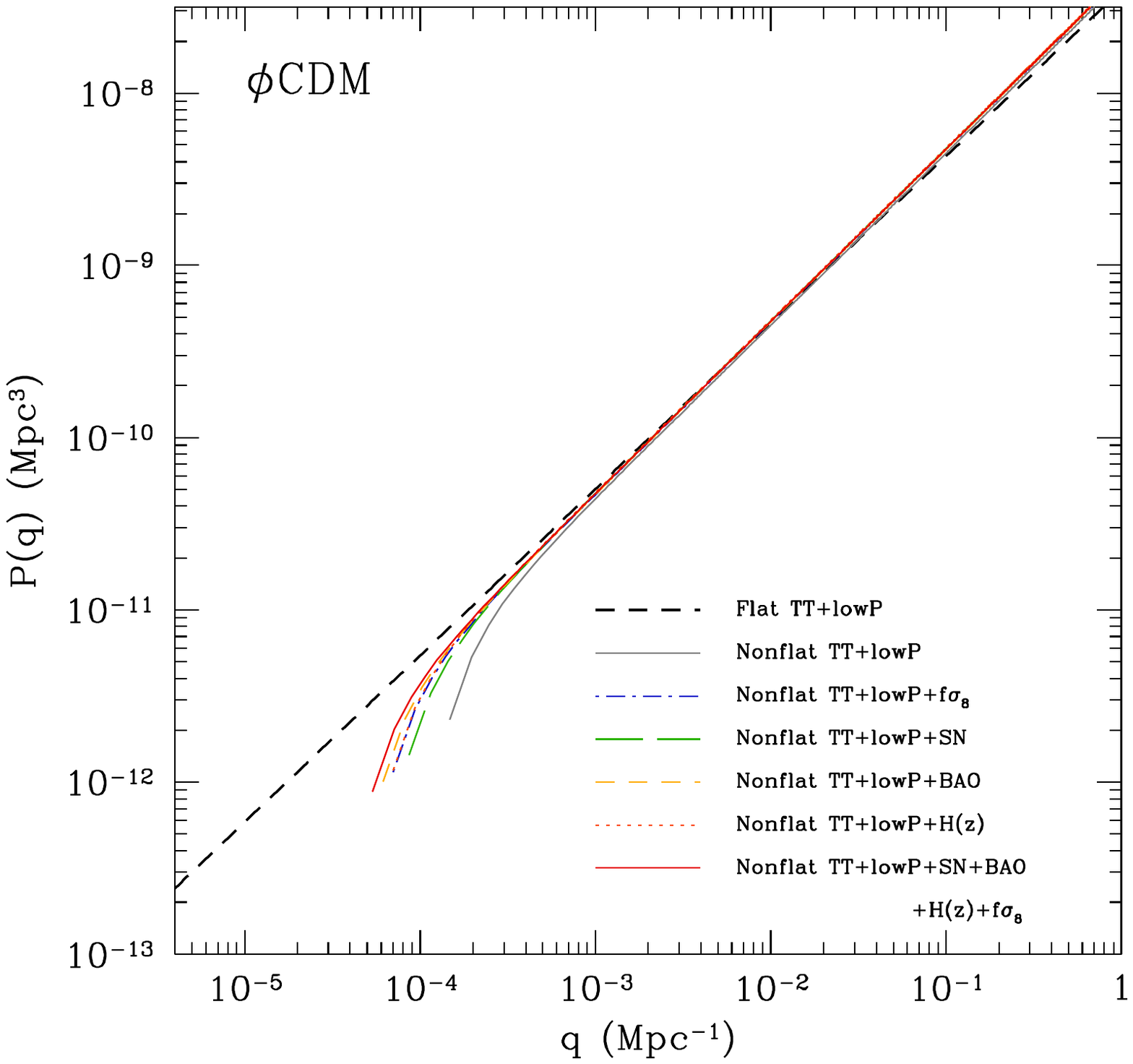}} 
\mbox{\includegraphics[width=85mm,bb=30 170 500 620]{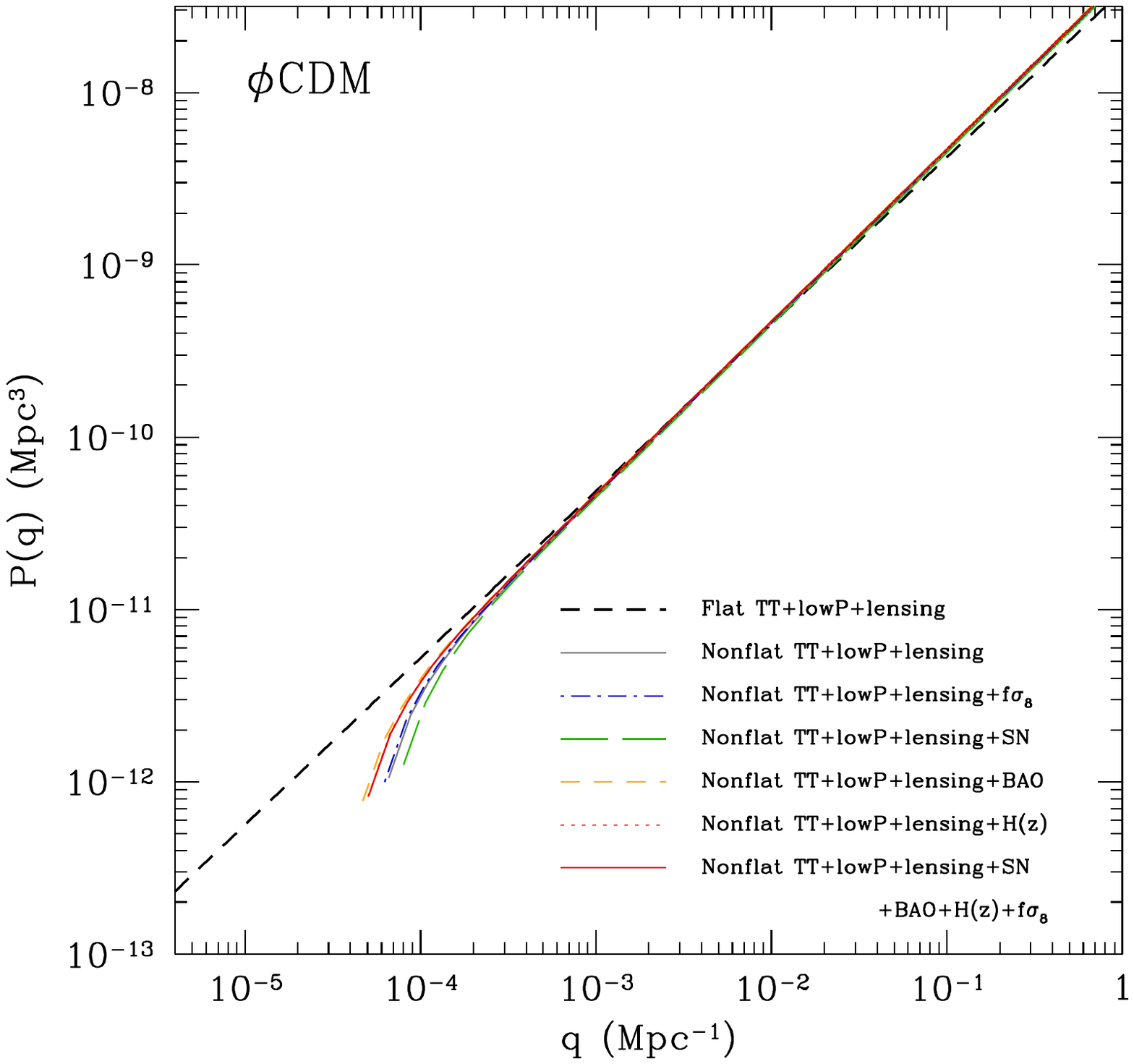}} 
\caption{Primordial scalar-type perturbation power spectra
with non-power-law form of the best-fit untilted nonflat $\phi$CDM models
constrained with Planck TT + lowP (left panel) and TT + lowP + lensing
data (right panel) together with SN, BAO, $H(z)$, $f\sigma_8$ non-CMB data
sets.
In both panels the primordial power spectrum of the best-fit tilted flat $\phi$CDM model is shown as dashed lines. See Sec.\ 3 for the definition of $q$. 
Note that all power spectra are normalized to $P(q)=A_s$ at 
$k_0=0.05~\textrm{Mpc}^{-1}$.
}
\label{fig:pq}
\end{figure*}

Figure \ref{fig:pq} shows the best-fit primordial power spectra of 
fractional energy density spatial inhomogeneity perturbations for the 
nonflat untilted $\phi$CDM 
model constrained by using the Planck TT + lowP (left) and TT + lowP + 
lensing (right panel) data together with the other non-CMB data sets. The 
low $q$ region reduction in power in the best-fit closed untilted 
$\phi$CDM inflation model power spectra in Fig.\ \ref{fig:pq} contributes 
to the TT power reduction at low-$\ell$ of the best-fit closed untilted model 
$C_\ell$'s (see Figs.\ \ref{fig:ps_cmb} and \ref{fig:ps_cmb_lensing} lower 
panels) relative to the best-fit flat tilted model $C_\ell$'s.\footnote{The usual and integrated Sachs-Wolfe effects, as well as other effects, also 
play a role in determining the shape of the low-$\ell$ $C_\ell$'s.} 
The most dramatic case is that of the best-fit untilted nonflat $\phi$CDM 
model for the TT + lowP data, consistent with the low-$\ell$ TT power reduction 
(Figs.\ \ref{fig:ps_cmb}$b$).\footnote{Figure 24 (bottom-right panel) of \citet{PlanckCollaboration2018} shows the primordial power spectrum derived from the Planck CMB data. (We note that their Fig.\ 24 has been derived under the assumption that space is flat, and consequently ignores the effect of the spatial curvature Sachs-Wolfe and Integrated Sachs-Wolfe effects on the CMB power spectra that were used in the derivation of this figure.) This power spectrum is a power law over wavenumbers in the interval $5 \times 10^{-3} \la k\ [{\rm Mpc}^{-1}] \la 2 \times 10^{-1}$ but at smaller wavenumbers their power spectrum amplitude errors are much larger and the Planck power spectrum is not inconsistent with our closed model power spectra plotted in Fig.\ \ref{fig:pq}.}

\section{Conclusion}

We have used the flat tilted and the nonflat untilted $\phi$CDM dynamical dark 
energy inflation models to measure cosmological parameters 
from a reliable, large compilation of observational data.

Our main findings, in summary, are:
\begin{itemize}
\item We confirm, but at a lower significance of 0.40$\sigma$, 
the result of 
\citet{Oobaetal2018c} that the flat tilted $\phi$CDM model better 
fits the data than does the standard flat tilted $\Lambda$CDM model. While
the improvement is not significant, it does mean that current data 
allow for the possibility that dark energy is dynamical.  
\item In the nonflat untilted $\phi$CDM case, we confirm, with greater 
significance, the \citet{Oobaetal2018b} result that cosmological data 
does not require flat spatial hypersurfaces for this model, and that 
the nonflat untilted $\phi$CDM model better fits (at 0.93$\sigma$) 
the data than does the nonflat untilted $\Lambda$CDM model (qualitatively 
the standard flat tilted $\Lambda$CDM model provides a better 
fit to the data than does the nonflat untilted $\Lambda$CDM model).  
In the nonflat untilted $\phi$CDM model, these data (including CMB lensing 
data) 
favor a closed model at more than 3.1$\sigma$ significance, in which spatial 
curvature contributes a little less than two-thirds of a percent of the 
cosmological energy budget now.
\item $H_0$ is measured here in an manner that is almost model-independent 
and is  
consistent with many other $H_0$ measurements. However, as is well known, an 
estimate of $H_0$ from the local expansion rate \citep{Riessetal2018} is about
3.3$\sigma$ larger.   
\item $\sigma_8$ here is measured in an almost model-independent manner and is 
consistent with the recent DES estimate \citep{DESCollaboration2017a}.
\item The value of $\Omega_m$ is more model dependent than the value of 
$\sigma_8$ and the $\Omega_m$ value measured using the nonflat untilted 
$\phi$CDM model is more consistent with the recent DES estimate \citep{DESCollaboration2017a}. 
\item $\Omega_c h^2$, $\tau$, and a few of the other 
cosmological parameter values are quite model dependent.
\end{itemize}

These results are very similar to those for the XCDM
dynamical dark energy parameterization presented in \cite{ParkRatra2018b}.

\begin{table*}
\caption{Individual and total $\chi^2$ values for the best-fit tilted flat and untilted nonflat $\phi\textrm{CDM}$ inflation models.}
\begin{ruledtabular}
\begin{tabular}{lcccccccccr}
    Data sets   & $\chi_{\textrm{Plik}}^2$  & $\chi_{\textrm{lowTEB}}^2$  &  $\chi_{\textrm{lensing}}^2$  &  $\chi_{\textrm{SN}}^2$  & $\chi_{\textrm{BAO}}^2$  &  $\chi_{H(z)}^2$   &  $\chi_{f\sigma_8}^2$ & $\chi_{\textrm{prior}}^2$  &  Total $\chi^2$      & $\Delta\chi^2$ \\[+0mm]
 \hline \\[-2mm]
 \multicolumn{11}{c}{Tilted flat $\phi\textrm{CDM}$ model} \\
  \hline \\[-2mm]
   TT+lowP                                & 766.02 & 10496.47 &      &         &       &       &       & 0.24 & 11262.25 & $+0.32$     \\[+1mm]
   \quad\quad +SN                         & 763.68 & 10496.29 &      & 1036.31 &       &       &       & 2.01 & 12298.30 & $-0.01$     \\[+1mm]
   \quad\quad +BAO                        & 764.40 & 10495.97 &      &         & 12.85 &       &       & 1.95 & 11275.18 & $-0.07$     \\[+1mm]
   \quad\quad +$H(z)$                     & 763.35 & 10496.70 &      &         &       & 14.93 &       & 1.88 & 11276.86 & $-0.07$     \\[+1mm]
   \quad\quad +$f\sigma_8$                & 766.75 & 10494.91 &      &         &       &       & 12.13 & 1.96 & 11275.75 & $-0.05$     \\[+1mm]
   \quad\quad +BAO+$f\sigma_8$            & 767.30 & 10494.94 &      &         & 12.04 &       & 11.99 & 1.98 & 11288.25 & $-0.25$     \\[+1mm]
   \quad\quad +SN+BAO                     & 764.23 & 10496.03 &      & 1036.12 & 12.94 &       &       & 2.03 & 12311.36 & $-0.05$     \\[+1mm]
   \quad\quad +SN+BAO+$H(z)$              & 764.23 & 10496.04 &      & 1036.16 & 12.96 & 14.82 &       & 2.01 & 12326.21 & $+0.00$     \\[+1mm]
   \quad\quad +SN+BAO+$H(z)$+$f\sigma_8$  & 766.85 & 10494.80 &      & 1036.07 & 12.57 & 14.79 & 12.05 & 2.11 & 12339.24 & $-0.12$     \\[+1mm]
 \hline\\[-2mm]
   TT+lowP+lensing                        & 766.47 & 10495.03 & 9.24 &         &       &       &       & 2.05 & 11272.79 & $+0.35$     \\[+1mm]
   \quad\quad +SN                         & 766.31 & 10494.80 & 9.21 & 1036.07 &       &       &       & 2.13 & 12308.51 & $-0.08$     \\[+1mm]
   \quad\quad +BAO                        & 767.08 & 10494.69 & 9.06 &         & 12.29 &       &       & 1.88 & 11285.00 & $-0.07$     \\[+1mm]
   \quad\quad +$H(z)$                     & 766.28 & 10494.84 & 9.21 &         &       & 14.84 &       & 2.09 & 11287.26 & $-0.01$     \\[+1mm]
   \quad\quad +$f\sigma_8$                & 767.53 & 10494.57 & 8.74 &         &       &       & 11.54 & 2.20 & 11284.58 & $-0.04$     \\[+1mm]
   \quad\quad +BAO+$f\sigma_8$            & 767.84 & 10494.56 & 8.69 &         & 12.05 &       & 11.63 & 2.18 & 11296.95 & $-0.37$     \\[+1mm]
   \quad\quad +SN+BAO                     & 766.37 & 10494.80 & 9.16 & 1036.10 & 12.60 &       &       & 2.10 & 12321.14 & $-0.03$     \\[+1mm]
   \quad\quad +SN+BAO+$H(z)$              & 766.71 & 10494.74 & 9.17 & 1036.21 & 12.49 & 14.83 &       & 1.94 & 12336.08 & $+0.07$     \\[+1mm]
   \quad\quad +SN+BAO+$H(z)$+$f\sigma_8$  & 767.76 & 10494.53 & 8.70 & 1036.13 & 12.44 & 14.79 & 11.60 & 2.08 & 12348.04 & $-0.16$     \\[+1mm]
   \hline \\[-2mm]
    \multicolumn{11}{c}{Untilted nonflat $\phi\textrm{CDM}$ model}  \\
   \hline \\[-2mm]
   TT+lowP                                & 773.80 & 10496.68 &      &         &       &       &       & 1.88 & 11272.36 & $+0.26$     \\[+1mm]
   \quad\quad +SN                         & 776.52 & 10498.66 &      & 1036.99 &       &       &       & 1.86 & 12314.03 & $-0.91$     \\[+1mm]
   \quad\quad +BAO                        & 783.17 & 10497.33 &      &         & 13.52 &       &       & 1.86 & 11295.88 & $-0.20$     \\[+1mm]
   \quad\quad +$H(z)$                     & 778.01 & 10500.10 &      &         &       & 17.30 &       & 1.83 & 11297.23 & $+0.08$     \\[+1mm]
   \quad\quad +$f\sigma_8$                & 781.56 & 10497.32 &      &         &       &       & 12.66 & 1.85 & 11293.39 & $-1.40$     \\[+1mm]
   \quad\quad +BAO+$f\sigma_8$            & 782.60 & 10499.20 &      &         & 12.24 &       & 10.94 & 1.74 & 11306.72 & $-0.69$     \\[+1mm]
   \quad\quad +SN+BAO                     & 782.64 & 10498.87 &      & 1036.17 & 12.54 &       &       & 1.91 & 12332.13 & $-0.17$     \\[+1mm]
   \quad\quad +SN+BAO+$H(z)$              & 782.77 & 10498.26 &      & 1036.05 & 12.63 & 15.73 &       & 2.07 & 12347.52 & $-1.05$     \\[+1mm]
   \quad\quad +SN+BAO+$H(z)$+$f\sigma_8$  & 783.46 & 10498.06 &      & 1036.24 & 12.22 & 15.62 & 11.02 & 2.18 & 12358.81 & $-1.17$     \\[+1mm]
  \hline\\[-2mm]
   TT+lowP+lensing                        & 789.84 & 10495.16 & 8.60 &         &       &       &       & 1.22 & 11292.37 & $+0.08$     \\[+1mm]
   \quad\quad +SN                         & 786.79 & 10493.86 & 9.83 & 1036.12 &       &       &       & 1.72 & 12328.32 & $+0.01$     \\[+1mm]
   \quad\quad +BAO                        & 784.96 & 10498.48 & 8.82 &         & 11.39 &       &       & 2.05 & 11305.69 & $-1.71$     \\[+1mm]
   \quad\quad +$H(z)$                     & 785.54 & 10497.28 & 8.84 &         &       & 15.35 &       & 1.80 & 11308.81 & $-1.29$     \\[+1mm]
   \quad\quad +$f\sigma_8$                & 787.33 & 10495.34 & 8.64 &         &       &       & 10.05 & 1.71 & 11303.07 & $+0.14$     \\[+1mm]
   \quad\quad +BAO+$f\sigma_8$            & 785.85 & 10498.06 & 8.53 &         & 11.50 &       &  9.77 & 2.02 & 11315.73 & $-1.58$     \\[+1mm]
   \quad\quad +SN+BAO                     & 783.34 & 10499.33 & 9.65 & 1036.67 & 11.77 &       &       & 1.84 & 12342.59 & $-0.20$     \\[+1mm]
   \quad\quad +SN+BAO+$H(z)$              & 785.03 & 10498.28 & 8.99 & 1036.79 & 11.78 & 15.38 &       & 1.80 & 12358.05 & $-0.82$     \\[+1mm]
   \quad\quad +SN+BAO+$H(z)$+$f\sigma_8$  & 786.41 & 10497.91 & 8.46 & 1036.65 & 11.98 & 15.36 &  9.38 & 1.84 & 12368.00 & $-0.86$     \\[+1mm]
\end{tabular}
\\[+1mm]
Note: $\Delta\chi^2$ of tilted flat or untilted nonflat $\phi\textrm{CDM}$ model represents the excess value relative to $\chi^2$
of the tilted flat or untilted nonflat $\Lambda\textrm{CDM}$ model estimated for the same combination of data sets \citep[listed in Table 7 of][]{ParkRatra2018b}.
\end{ruledtabular}
\label{tab:chi2_phiCDM}
\end{table*}

%
%
\acknowledgements{
We acknowledge valuable discussions with J.\ Ooba.
C.-G.P.\ was supported by research funds of Chonbuk National University in 2017 and
the Basic Science Research Program through the National Research Foundation of Korea (NRF)
funded by the Ministry of Education (No. 2017R1D1A1B03028384). B.R.\ was supported in part by DOE
grant DE-SC0019038.
}

%
%

\def\and{{and }}
\bibliographystyle{yahapj}


\end{document}